\documentclass[a4paper,11pt]{article}
\usepackage[numbers]{natbib}   
\usepackage{amsmath}
\usepackage{amsfonts}
\usepackage{amssymb}
\usepackage{amsthm}
\usepackage{graphicx}
\usepackage{hyperref,setspace}
\usepackage{bm}
\usepackage{xcolor}
\usepackage{graphicx}
\usepackage{epstopdf}
\usepackage{mathtools}
\usepackage{multicol}
\usepackage{booktabs}
\usepackage{blindtext}
\usepackage{pdftexcmds}
\usepackage{array}
\usepackage{authblk}

\newtheorem{proposition}{Proposition}
\newtheorem{lemma}{Lemma}
\newtheorem{corollary}{Corollary}
\newtheorem{outcome}{Outcome}
\newcommand{\tmmathbf}[1]{\ensuremath{\boldsymbol{#1}}}

\newcommand*{\mline}[1]{%
\begingroup
    \renewcommand*{\arraystretch}{1.1}%
   \begin{tabular}[c]{@{}>{\raggedright\arraybackslash}p{2cm}@{}}#1\end{tabular}%
  \endgroup
}

\title{Technology innovation in evolutionary green transition:
environmental quality and economic sustainability}
\date{}

\author[1]{Fausto Cavalli \thanks{fausto.cavalli@unimib.it} }
\author[2]{Alessandra Mainini \thanks{alessandra.mainini@unicatt.it}}
\author[3]{Enrico Moretto \thanks{enrico.moretto@unimib.it}}
\author[4]{Ahmad Naimzada \thanks{ahmad.naimzada@unimib.it}}

\affil[1,3,4]{Department of Economics, Management and Statistics, University of Milano-Bicocca, Milano, Italy}

\affil[2]{Department of Economics and Social Science, Catholic University, Piacenza, Italy}

\begin{document}
\maketitle
\begin{abstract}
We propose an evolutionary model to study the transition toward green technology under the influence of innovation. Clean and dirty technologies are selected according to their profitability under an environmental tax, which depends on the overall pollution level. Pollution itself evolves dynamically:
it results from the emissions of the two types of producers, naturally decays, and is reduced through the implementation of the current abatement technology.
The regulator collects tax revenues and allocates them between the implementation of the existing abatement technology and its innovation, which increases the stock of knowledge and thereby enhances abatement effectiveness.
From a static perspective, we show the existence of steady states, both with homogeneous populations of clean or dirty producers and with heterogeneous populations where both technologies coexist. We discuss the mechanisms through
which these steady states emerge and how they may evolve into one another.
From a dynamical perspective, we characterize the resulting scenarios, showing how innovation can foster a green transition if coupled with a suitable level of taxation. At the same time, we investigate how improper environmental
policies may also produce unintended outcomes, such as environmental
deterioration, reversion to dirty technology, or economic unsustainability.
\end{abstract}

\section{Introduction}

In recent decades the exacerbation of climate change, extreme natural
events, ecosystem degradation, and rising pollution have placed
environmental concerns at the core of public debate and policy
agendas. Addressing these challenges clearly requires an integrated
approach that accounts for economic, environmental, and social
dimensions ({\citet{Ros00}}). The overarching goal is to achieve
sustainable development objectives that are environmentally and
economically viable, fostering a green transition from
fossil-fuel-based technologies systems to clean ones,
without undermining economic growth.  Recent disruptions to global
supply chains, driven by the pandemic and geopolitical tensions, have
reinforced the urgency of rethinking the energy foundations of
production systems.

A variety of instruments have been proposed and evaluated to encourage
the green transition, involving either private actors or public
authorities.  Private efforts focus on investments in research and
innovation and on the adoption of sustainable practices, such as
resource efficiency and circular economy principles. One of the main
challenges lies in the introduction and diffusion of clean
technologies, which are designed to reduce the negative environmental
impacts of production and promote sustainability. Like many other
innovations, adopting clean technologies results in additional
economic expenses when compared to conventional ones, which are both
more polluting and less costly also thanks to years of
development. Consequently, a transition towards clean technologies
requires substantial effort and investment, particularly if the goal
is to catch up with established conventional technologies. This
technological gap may also discourage research efforts directed toward
clean innovations.

Research contributions in the ``eco-innovation'' field grew
considerably in the last decade. If, according to \citet{diaz2015eco}
in 2014 ``eco-innovation is still a young area of research'', a
relevant number of contributions can be now found. Among many, it is
worth mentioning research articles by \citet{arilla2024quantifying}
and \citet{ding2025supply}. If the first paper deals with
``sustainable development goals (SDGs)'' at an European Union level
and claims that there exists a positive correlation between
eco-innovation indicator and employment levels, the second focuses on
France. The authors, here, aim at determining if eco-innovation is
capable of ``mitigating environmental degradation'', a topic that this
paper tackles in two ways: the first is by devoting a fraction of the
amount of money levied from dirty producers directly to reduce
pollution while the second is by directing the remaining part of money
to research establishing the amelioration of the existing
technology. The literature highlights that even environmentally
oriented innovation pathways may generate unintended drawbacks. For
example, the Jevons paradox, dating back to {\citet{Jev865}},
illustrates how improvements in energy efficiency can, by lowering
usage costs, increase energy demand and emissions
({\citet{Sor09,Sau13}}). Likewise, {\citet{AABH12}} highlighted how
innovation may become locked into fossil-intensive trajectories,
delaying a genuine green transition, while clean technologies
themselves can entail  a different evolution according to
  the initial state, giving rise to path dependency (see also
{\citet{AHTZ19}}), as well as {\citet{SABRNOM20}} remarked how the
extraction and processing of critical materials needed for clean
technologies production often harm ecosystems and the environment.

In order to contain and channel innovation trends towards the original
objectives, issues described above make it necessary to pursue
appropriate public measures, including environmental taxation,
penalties for polluting or not correctly disclosing emissions and
incentives for cleaner production systems. The impact of taxation and
technological innovation on renewable energy is the main focus of
\citet{ebaidalla2024impact}. This author performs a statistical
analysis on data from 37 countries spanning 25 years and concludes
that, if, on one hand, taxation carries a negative effect on
investments on renewable energy, on the other technological innovation
has a `positive and significant' impact. An empirical analysis on how
taxation can provide resources useful for improving green technology
can be found in \citet{sharif2023demystifying}. There, a panel
analysis exploiting data collected from 6 Asian countries on a time
span of 23 years pinpoints that ``environmental taxes have a positive
effect on green technology innovation''. However, also environmental
taxation or penalties have their own drawbacks. Using carbon or
pollution tax revenues for end-of-pipe technologies risks prolonging
polluting plant operation rather than promoting structural changes
({\citet{FN08}}). High taxes or penalties may also prompt firms to
engage in greenwashing or corrupt practices ({\citet{LM11}}). Finally,
as discussed in a report of the European Environment
agency\footnote{The role of (environmental) taxation in supporting
  sustainability transitions, 2022,
  \url{https://www.eea.europa.eu/en/analysis/publications/the-role-of-environmental-taxation-in-supporting-sustainability-transitions}},
the amount of resources collected for taxation progressively reduces
as the green transition advances, if taxation mainly charges dirty
technologies.

Literature presented so far is, essentially, empirical. This proves
the interest of the scientific community towards environment
protection and research for technological innovation, but if the
analysis aims at providing a prescriptive contribution, more
theoretical studies should be considered. Concerning
this literature strand, we can mention a first class of contributions,
which has focused on studying the impact of taxation when it is
determined by the very companies engaged in  research and
  development (R\&D), as in the contribution by \citet{hall2019tax},
in a book by \citet{GoolsbeeJones+2021}. Here, two tax policies
related to innovation are analyzed. The first one is based on tax
credits and superdeductions for  R\&D  while the second
one goes under the name of ``patent box''. This is a reduction of
corporate tax rates granted to revenues that come from patents or
intellectual properties developed by companies. Even if this is a well
established policy in, as of 2017, 42 countries, as we already
remarked, to prevent private action from leading to outcomes different
from those desired, a well-designed public regulatory framework is
necessary. Following what \citet{rodrik2014green} claims, at an
institutional level, a properly designed industrial policy can help
creating `green growth' that ``can be defined as a trajectory of
economic development based on sustainable use of non-renewable
resources and that fully internalizes environmental costs''.
According to this author, such policy can counter well established
skepticism that sees difficulties in obtaining effective public
intervention capable of correctly and fully implement green
technologies.

Building on all these considerations, this contribution develops a
theoretical model to investigate under which conditions the regulator
policy choices, taxation and innovation investments, can foster a
green transition that is both environmentally effective and
economically sustainable. The work draws on the contribution by
\citet{zeppini2015discrete} and its extension by
{\citet{CMN24}}\footnote{The resulting model we propose is a nonlinear
  evolutionary dynamical system, consistent with
  {\citet{zeppini2015discrete}} and {\citet{CMN24}}. Among other
  applications of nonlinear dynamics to the same topic, we refer the
  interested reader to {\citet{LI2025107215}}, in which carbon
  abatement in a Cournot duopoly with a carbon tax reducing profits
  through abatement technologies is studied. The focus is on firms'
  ``green reputation,'' defined as consumer willingness to purchase
  their products. At equilibrium, green reputation decreases with
  ``green efficiency'' and increases with costs and the tax rate,
  while the dynamic analysis reveals equilibria characterized by
  complex bifurcations across parameter regions.}. Both contributions
examine how pollution taxes drive the transition from dirty to clean
technologies, based on the idea that a properly designed levy, by
reducing polluters' profitability, can push firms toward cleaner production. While Zeppini does not
consider at all the use of tax revenues, Cavalli et al. focus on the
case where all revenues finance pollution abatement through
already existing, and with no improvement, technologies.  Here, we investigate whether resources could also be
effectively allocated to research for developing new methods of
environmental preservation.

As in {\citet{CMN24}}, producers choose between a ``dirty'' and a
``clean'' technology, being the former one more polluting but also
more profitable. Technology adoption follows an evolutionary
mechanism, where fitness depends on profitability and taxation
penalties, proportional to the pollution level and more burdensome for
the dirty option. The accumulation of pollution is caused by emissions
from both types of producers, and it is mitigated by natural decay and
abatement. The regulator allocates tax revenues between innovation in
abatement technologies and their implementation. Unlike
{\citet{CMN24}}, abatement effectiveness is now endogenous, evolving
with the stock of knowledge generated, which is modeled as a
cumulative learning process, building on past achievements in existing
technologies and expanding through new investments.

The main results are grounded on both static and dynamical analyses,
complemented by several simulation case studies, linking theoretical
and numerical outcomes to the empirical evidence in the
literature. The aim is to classify the resulting scenarios according
to whether a green transition occurs, the environmental situation
improves, and policy choices prove economically sustainable. The
analysis shows, in particular, that the key driver of the green
transition is represented by an adequate level of taxation, which
combined with an optimal resource allocation between innovation and
implementation can reduce pollution levels. Conversely, when taxation
does not incentivize clean technologies, careful allocation of
resources to innovation may still trigger a green transition, though
without environmental improvements. The study also highlights cases
where misguided policies lead to transitions back to dirty
technologies or to a deterioration of environmental conditions.  In
addition, the dynamical investigations demonstrate the potential for
multiple coexisting steady states to emerge, with a path dependence
that leads similar frameworks to evolve differently depending on
policy choices. Finally, the results indicate that even situations
characterized by a fully realized static green transition can be
unstable, giving rise to out-of-equilibrium dynamics in which agents
erratically switch between technologies.

The remainder of the contribution is organized as follows. In Section
\ref{sec:model} we present the model, whose static properties are
analyzed and discussed in Section \ref{sec:static}. The dynamical
investigation is carried on in Section \ref{sec:dyn}, in which we also
provide overall interpretation of the results and of their policy
implications. Section \ref{sec:concl} collects final insights and
possible future research steps.

\section{The model}\label{sec:model}

This model aims to explore how research
and innovation contribute to an evolutionary green
transition. The dynamics evolve at discrete time
  $t,$ and with a unit-mass population of producers 
that can switch between two technologies, the clean ($C$) and the dirty ($D$)
ones, characterized by different
emission levels.
Clean producers adopt a technology characterized by lower emission
levels, in contrast to dirty producers, whose technology
  has higher emissions, and then a more harmful impact on
  environment.  A regulator introduces an environmental
  tax proportional to the pollution
  level,
 and allocates the collected resources between
  technological innovation for abatement and its implementation.


The model has three core components, namely the share of
    population adopting clean technologies, the level of pollutants
    and the stock of knowledge accumulated through investments in
    research and innovation.
 In what follows, we describe all these mechanisms in detail.

\textit{Evolutionary selection of technologies}

We denote with $x_{C, t} = x_t \in [0, 1]$ and $x_{D, t} = 1 - x_t$ the fractions of clean and dirty producers, respectively. We assume that transition from dirty to clean producers, or vice versa, occurs according to a replicator-adapted evolutionary selection mechanism\footnote{In contrast to {\citet{zeppini2015discrete}} and {\citet{CMN24}}, where
  the evolution of $x_t$ is described by means of the well-known
  {\citet{brock1997rational}} recursive expression, in the present contribution we adopt the replicator framework, which offers greater analytical tractability, leading to clearer and more interpretable results. While this modelling choice limits the possibility of a direct comparison with some of the results in {\citet{CMN24}} and    {\citet{zeppini2015discrete}}, the focus of our analysis is
  different, and meaningful qualitative comparisons can still be
  drawn. Moreover, we stress that the evolutionary selection based on the replicator mechanism admits, as steady states, those characterized  by populations of no clean and of clean only producers, where this latter one accounts for a complete green transition.} (see, for instance, {\citet{cressman2003evolutionary}}). In other words, $x_t$
evolves over time according to
\begin{equation}
  x_{C, t + 1} = x_{t + 1} = \frac{x_t}{x_t + (1 - x_t) e^{\beta (\lambda_D -
  p_t \tau_D - (\lambda_C - p_t \tau_C))}} = \frac{x_t}{x_t + (1 - x_t)
  e^{\beta (\lambda_D - \lambda_C - p_t (\tau_D - \tau_C))}} . \label{eq:x2}
\end{equation}
 Parameter
  $\beta \ge 0$ represents the evolutionary pressure (or intensity of choice) of the selection mechanism, whereas
  $\lambda_D,\,\lambda_C \in \mathbb R$ represent profitability of dirty and clean technologies, respectively.\footnote{We stress
  that, as in {\citet{CMN24}} and {\citet{zeppini2015discrete}},
  $\lambda_i$ represents the component of profits intrinsically
   related to the production process, net of the regulator
  intervention. For a complete economic interpretation and details
  about their micro-foundation, we refer the interested reader to the
  supplementary material in {\citet{CMN24}}.}. We assume
$\lambda_D > \lambda_C$ to focus on a framework in which
the presence of a conventional,  vastly adopted, and
well-known dirty technology allows for profitability advantage over the clean one\footnote{Indeed, as shown in   {\citet{zeppini2015discrete}}, the opposite scenario may also occur,  where the green transition is driven by the  profitability advantage of clean technologies. Conversely, one of the aims of this work is to explore whether a transition towards clean technologies can be triggered even under less favorable conditions --- specifically, when clean technologies do not enjoy a profitability advantage.}. This effect is counterbalanced by the charged green tax, 
which is proportional to the pollution stock $p_t$.
The amount of taxation  depends on parameters $\tau_D > 0$ and $0 \leq \tau_C < \tau_D$, 
  which denote per-unit pollution tax on clean and dirty agents,
  respectively, so that $\tau_C p_t$ and $\tau_D p_t$
represent tax burden on each type of agent. 
This implies that the environmental tax affects to a    greater extent profits realized by dirty technologies rather than 
those achieved by clean producers. The resulting fitness measure for each technology
is then    equal to $\lambda_i - \tau_ip_t $, $i = D,\,C$. A direct consequence is that,  when the 
 population is homogeneous, namely it is 
 composed only by dirty (resp. clean) producers, that is
 $x_t = 0$, (resp. $x_t = 1$), no transition toward different population of producers is possible, 
 that is 
 $x_{t + 1} = 0$   (resp. $x_{t + 1} = 1$). This occurs no
 matter how a technology is profitable or not, and regardless of the  intervention of the regulator, that is independently of the fitness differential
   $\lambda_0 - (\tau_D - \tau_C)p_t$, where
   $\lambda_0 = \lambda_D - \lambda_C > 0$ is the profitability
   advantage of the dirty technology over the clean one. Conversely, when  both types of producers coexist,
(i.e., $x_t \in (0, 1)$) without any evolutionary pressure (i.e., $\beta = 0$), all agents remain locked into their currently adopted technology. 
 As a result, the recursive expression \eqref{eq:x2} reduces to $x_{t + 1} = x_t$. On the other side, when
   population is non-homogeneous and $\beta > 0$, agents are pushed
   towards the choice with the largest fitness.  To describe this mechanism, we introduce threshold
\begin{equation}
  \bar{r} = \frac{\lambda_0}{\tau_D - \tau_C}, \label{eq:r}
\end{equation}
which is the level of pollution with respect to which there is no an
advantage in being clean or dirty.  Then, if the current
  level of pollution is above this threshold ($p_t > \bar r$), fitness
  coming from being clean is greater than that of being dirty, because
  of the reduced amount of taxation, and an increasing number of
  agents becomes clean. The opposite situation occurs
  when the current level of pollution is below this threshold
  ($p_t < \bar r$), and agents switch to the more convenient dirty
  technology. In presence of an extremely high intensity of choice,
(i.e. $\beta\to+\infty$), 
under condition $p_t < \bar r$ (resp. $p_t > \bar r$),
  all agents adopt dirty (resp. clean) technologies, that is,
  $x_{t + 1} = 0$ (resp. $x_{t + 1} = 1$).
Amount $\bar r$ encompasses the relative benefit of the
  (higher) profitability of a dirty technology taking into account the
  (higher) taxation it incurs, The amount $\bar r$ then
identifies a particular cut-off point in terms of pollution, at which
the profitability advantage enjoyed by dirty producers is exactly
offset by the burden of higher taxation. 
As discussed above, threshold $\bar r$ plays a key role
in the transition toward clean technologies\footnote{To avoid
  misinterpretations, we stress that $\bar r$ does not encompass the   function of convincing agents to take seriously into account environmental sustainability and deal with fundamental issues such   as climate, atmosphere and ecosystem. In other words, $\bar{r}$ does not embody an `ethical' purpose, aimed at convincing agents to confront environmental issues. Indeed, $\bar{r}$ has to do only with profitability, production achievements, and economic advantages, though it may allow to a transition that is, undoubtedly,
  environmentally
  friendly.}, 
 and its implications should be carefully taken into account by the policymaker when setting the taxation level.

\textit{Environmental dynamics}

The environmental sphere is described in terms of the stock of
pollution, for which we assume the following dynamical adjustment
\begin{equation}
  p_{t + 1} = \max \{p_t - \alpha p_t + \varepsilon_C x_t + \varepsilon_D (1 -
  x_t) - \theta_{t + 1} (1 - \omega) p_t (\tau_C x_t + \tau_D (1 - x_t)) ; 0\}
  . \label{eq:p2}
\end{equation}
Equation \eqref{eq:p2} shows that the current level
  of pollution $p_t$ evolves according to the (normalized) amount of
  new emissions of pollutants
  $\varepsilon_C x_t + \varepsilon_D (1 - x_t)$, with parameters
  $\varepsilon_D > \varepsilon_C \geq 0$ denoting pollution emitted   due to dirty and clean technologies respectively, to natural decay $- \alpha p_t$, with $\alpha \in [0, 1]$ and to abatement
  $- \theta_{t + 1} (1 - \omega) p_t (\tau_C x_t + \tau_D (1 -
  x_t))$. Quantity $\theta_{t + 1}$ represents pollution abatement technology effectiveness, and $\tau_C x_t + \tau_D (1 - x_t)$ the (normalized) total revenues from taxation per unit of pollutant at time $t$; the latter, multiplied by $\omega$, gives the share of resources allocated to the implementation of current technology, being $\omega \in[0, 1]$ the portion of resources destined for research. 
According to this, we can define
\begin{equation}
  \bar{\tau} (x_t) = \tau_C x_t + \tau_D  (1 - x_t) \label{eq:taum}
\end{equation}
and
\begin{equation}
  \bar{\varepsilon} (x_t) = \varepsilon_C x_t + \varepsilon_D  (1 - x_t).
  \label{eq:varepsilonm}
\end{equation}

Mechanism \eqref{eq:p2} is essentially the same adopted in 
{\citet{CMN24}} (to which we refer for the related literature), but
with two significant differences. The first one regards the effectiveness of
pollution abatement, which is exogenous in {\citet{CMN24}}, while in the present work evolves over time according to the amount of resources destined for innovation. The second difference is that in {\citet{CMN24}} taxation is entirely addressed to abatement technology, whereas in the present contribution the regulator decides how to split resources between innovation and abatement, by allocating an amount
$\omega \bar{\tau}_t$ for innovation, and the remaining amount $(1 - \omega) \bar{\tau}_t$ for the implementation of the abatement technology.

\textit{Endogenous technology innovation}

The effectiveness of the abatement technology 
depends on the accumulated stock of knowledge, which evolves over time according to the following
process\footnote{A mechanism describing the evolution of the stock of
knowledge, similar to the one introduced in (\ref{eq:k2}), but in a continuous
time setting, can be found, for example, in 
{\citet{la2010endogenous}}.}:
\begin{equation}
  k_{t + 1} = \sigma k_t + dk^{\gamma}_t  (\omega p_t  \bar{\tau} (x_t))^{1 -
  \gamma} . \label{eq:k2}
\end{equation}
The next period knowledge, $k_{t + 1}$, is the result of the    current stock of knowledge, $k_t$, which shrinks, due to its obsolescence, at rate 
  $\left(1-\sigma\right) \in (0, 1]$ and newly produced stock of knowledge, which is described by a Cobb-Douglas function with constant returns to scale.
The latter depends on the existing
  level of knowledge $k_t$ and the amount of research investments $\omega p_t \bar{\tau} (x_t)$, 
  with parameter $\gamma \in [0, 1]$  being output elasticity of $k_t$  and $d > 0$ the total
  factor productivity.
Assuming that the effectiveness of pollution abatement technology linearly depends on the current stock of knowledge,  then $\theta_t$ evolves over time according to
\begin{equation}
  \theta_{t + 1} = c_1 k_t + c_2, \label{eq:th2}
\end{equation}
where $c_1 > 0$ and $c_2 \geq 0$. Condition $c_1 > 0$ ensures a
positive marginal effect of knowledge on abatement effectiveness, thus
excluding the uninteresting case in which knowledge has no impact on
technological improvement. Parameter $c_2$ captures the ex-ante effectiveness of
pollution-reducing technologies at the initial time $t = 0$,
represents the well-established technological level before any knowledge-driven enhancement and may, in
principle, be zero. We emphasize that we consider a single abatement
technology, whose effectiveness $\theta_t$ results from the
combination of both ex-ante and innovation-driven components. For
interpretative purposes, it is useful to distinguish between contribution of $c_1$ and $c_2$ to effectiveness
$\theta_{t + 1}$, and discuss them as if they were two distinct
components that could, in principle, be observed separately. The policy maker, therefore, faces a trade-off
between  improving the baseline level of abatement effectiveness $c_2$ and implementing it at its current
  level. 
The analysis carried on in the next sections shows 
  that this choice depends on the relative magnitude of $c_2$ compared
  to $c_1$.
 Specifically, the greater $c_2$ with respect to $c_1$ is,
  the less significant the marginal impact of the share of resources
  allocated to research, $\omega$, becomes. Putting together equations \eqref{eq:x2}, \eqref{eq:p2}, \eqref{eq:k2} and
\eqref{eq:th2} we obtain system
\begin{equation}
  \left\{ \begin{array}{l}
    x_{t + 1} = \dfrac{x_t}{x_t + (1 - x_t) e^{\beta (\lambda_0 - p_t (\tau_D
    - \tau_C))}},\\
    \\
    p_{t + 1} = \max \{p_t (1 - \alpha) + \bar{\varepsilon}_t - \theta_{t + 1}
    (1 - \omega) p_t  \bar{\tau}_t ; 0\},\\
    \\
    k_{t + 1} = \sigma k_t + dk^{\gamma}_t (\omega p_t  \bar{\tau}_t)^{1 -
    \gamma},\\
    \\
    \theta_{t + 1} = c_1 k_t + c_2,
  \end{array} \right. \label{eq:model4}
\end{equation}
which is reduced to the following three-dimensional system by substituting the expression for $\theta_{t + 1}$ in the last equation into the second one, thus obtaining
\begin{equation}
  M : \left\{ \begin{array}{l}
    x_{t + 1} = \dfrac{x_t}{x_t + (1 - x_t) e^{\beta (\lambda_0 - p_t (\tau_D
    - \tau_C))}},\\
    \\
    p_{t + 1} = \max \{p_t (1 - \alpha) + \bar{\varepsilon}_t - (c_1 k_t +
    c_2) (1 - \omega) p_t  \bar{\tau}_t ; 0\},\\
    \\
    k_{t + 1} = \sigma k_t + dk^{\gamma}_t (\omega p_t  \bar{\tau}_t)^{1 -
    \gamma} .
  \end{array} \right. \label{eq:model}
\end{equation}
We then define function
\[ M : \hspace{0.17em} [0, 1] \times [0, + \infty)^2 \rightarrow [0, 1] \times
   [0, + \infty)^2, \hspace{0.27em} \hspace{0.27em} \hspace{0.27em} (x, p, k)
   \mapsto M (x, p, k), \]
whose components are described by the right-hand sides (from now on rhs) of each equation in \eqref{eq:model}. In the analysis of \eqref{eq:model}, we focus on the policy parameter
  $\omega$, which represents the main novelty of the model. We examine its role both from a static and a dynamical perspective, with the aim of understanding its contribution to a green transition achievement and an environmental quality improvement. 
To this end, we now study the effects of extreme values of $\omega$. If $\omega = 0$, the third equation in \eqref{eq:model} becomes \[ k_{t + 1} = \sigma k_t . \]
Then, for any initial level of knowledge $k_0 > 0$, the solution for $k_t$ is 
\[ k_t = \sigma^t k_0, \] 
which implies, for any $\sigma \in [0, 1)$,
\[ \lim_{t \to \infty} \hspace{0.17em} k_t = 0. \]
This means that the long run value of $\theta_{t+1}$ is constant and equal to $c_2$, and, as a consequence, equation \eqref{eq:p2} becomes
\[ p_{t + 1} = \max \{p_t (1 - \alpha) + \bar{\varepsilon}_t - c_2 p_t \bar{\tau}_t ; 0\}, \]
Then, in this special case, the first two equations in
\eqref{eq:model} are very similar to those in {\citet{CMN24}}. On the
other hand, the case $\omega = 1$ represents the limit scenario in which tax revenues are exclusively allocated to
innovation. As a result, 
 no resource is available for the implementation of abatement technologies, and the second equation in \eqref{eq:model} takes the form
\[ p_{t + 1} = \max \{p_t (1 - \alpha) + \bar{\varepsilon}_t ; 0\}.\] This latter case is somewhat extreme and unrealistic, and taken into consideration only as a limit benchmark. In what follows, most attention is given to values
  of $\omega$ belonging to the open interval $(0,1)$.
In the next sections, we perform static and dynamical analysis of the model. It should be kept in mind that the aim is to assess whether an appropriate allocation of resources between innovation in abatement technologies and their implementation can promote a green transition that is both environmentally effective and economically sustainable. In interpreting the results, we consider three main aspects of the regulator's policy design: 

1) the scale of the green transition: configurations in which a higher
proportion of producers adopt clean technologies are more desirable. A broader
diffusion of green awareness can have positive spillover effects on firms'
internal production choices as well as on consumers' behavior;

2) the environmental quality: policy
measures leading to lower pollution levels are
preferable;

3) the per-unit pollution taxation:
outcomes in which environmental quality improvement and green
transition are achieved with lower $\tau_D$ are preferable. An excessive level of $\tau_D$ may result in a scenario that facilitates widespread corruption and/or encourages greenwashing, rather than achieving significant environmental
  improvements\footnote{It should be noted that some of the
  dimensions considered in the interpretation of the results (such as
  the social impacts of a high presence of green firms, or the
  implications of excessive taxation for corruption) are not yet
  embedded in the model. These elements are currently assessed as
  possible exogenous consequences. Future research will seek to
  incorporate selected aspects into the model, thereby enabling their
  examination from an endogenous perspective.}.

Furthermore, to facilitate the discussion, we focus on three simulative case
studies, 
 each of which characterized by the same parameter configuration: $\lambda_0 = 0.5,
\tau_C = 1, \varepsilon_D = 0.6, \sigma = 0.3, d = 10, \alpha = 0.2, \gamma =
0.1$. These three
  cases differ only in the clean technology emission level,
  $\varepsilon_C$, compared to the dirty technology one,
  $\varepsilon_D$; in particular, it can be low
  ($\varepsilon_C = 0.002$), intermediate ($\varepsilon_C = 0.2$), or
  high ($\varepsilon_C = 0.55$). 
Each case is analyzed as a function of the policy
parameters $\omega$ and $\tau_D$ and for different values of
parameters $c_1$ and $c_2$, which
endogenously govern abatement effectiveness. Evolutionary pressure
$\beta$ has no impact on the static analysis, so
 its value is specified in the dynamic analysis section\footnote{We emphasize that when choosing the
  parameter setting, we normalized the per-unit taxation level for
  green producers, while the remaining parameters are selected to
  yield, as will be shown, a minimal degree of dynamic complexity,
  enabling a clearer interpretation of the results.  Nonetheless, the
  analytical results allow the conclusions to be extended to any
  parameter configuration.}. Finally, in presenting the results, we
focus on the dynamics of the share 
 of clean agents and the pollution level $p$, but we omit the behavior of the knowledge stock $k$, which is less relevant for the interpretation of the results. We note that 
 the three case studies can represent initial situation that are
connected to some scenarios we discussed in the review of the
empirical literature in the Introduction.
In particular, the first case may represent a situation
  where, at the beginning, private investment efforts as well as
  government support in providing financial support for production system innovation, leads to a significant technology improvement.
In the second case, such improvement is still observed, though to a lesser extent. The third case, by contrast, depicts the
  possibility of greenwashing, as clean technologies differ only
  marginally from dirty ones.
  
\section{Static analysis}\label{sec:static}

In this section, we study model \eqref{eq:model} from  a
static perspective, focusing on the possible existing steady states,
how they emerge and disappear and and how they depend on parameters.
 
Steady states are generically denoted by
$\tmmathbf{\xi}^{\ast} = (x^{\ast}, p^{\ast}, k^{\ast})$.
In what follows, we widely make use of parameter
  $\bar r$, defined in \eqref{eq:r}, and parameter
\begin{equation}
  \chi = \left( \frac{d}{1 - \sigma} \right)^{\frac{1}{1 - \gamma}},
\label{eq:chi}
\end{equation}
which 
encompasses all parameters dealing with knowledge production mechanism \eqref{eq:k2}.

\subsection{Steady states}\label{sec:static_ss}

Steady state quantities
  $p_0^{\ast},\,p_1^{\ast},\,x_a^{\ast},\,x_b^{\ast},\, k_a^{\ast}$ and $k_b^{\ast}$, which appear in next Proposition, have a lengthy analytical expression, which is then relegated to the proof in Appendix.

\begin{proposition}
  \label{th:ss}Let
  \[ \tmmathbf{\xi}_0^{\ast} = (0, p_0^{\ast}, \chi \omega \tau_D
    p_0^{\ast}), \hspace{0.27em} \hspace{0.27em} \hspace{0.27em}
    \hspace{0.27em} \tmmathbf{\xi}_1^{\ast} = (1, p_1^{\ast}, \chi
    \omega \tau_C p_1^{\ast}), \hspace{0.27em} \hspace{0.27em}
    \hspace{0.27em} \hspace{0.27em} \tmmathbf{\xi}_i^{\ast} =
    (x_i^{\ast}, \bar{r}, k_i^{\ast}), \hspace{0.27em} i \in \{a, b\}
    . \] Then, model \eqref{eq:model} always admits
    steady states $\tmmathbf{\xi}_0^{\ast}$ and
    $\tmmathbf{\xi}_1^{\ast}$, for which the share $x^*$ of clean
    producers is $0$ or $1$, respectively. On the contrary, steady
    states $\tmmathbf{\xi}_a^{\ast}$ and $\tmmathbf{\xi}_b^{\ast}$,
    for which $x_a^{\ast}, \hspace{0.17em} x_b^{\ast} \in (0, 1)$,
    only exist for suitable parameter configurations and, in this
    case, $x_a^{\ast} \leq x_b^{\ast}$.
\end{proposition}

Proposition \eqref{th:ss} suggests the existence of four possible sets of
steady states, which we denote by
\[ \mathcal{S}_0 = \{\tmmathbf{\xi}_0^{\ast}, \tmmathbf{\xi}_1^{\ast} \},
   \hspace{0.27em} \hspace{0.27em} \hspace{0.27em} \hspace{0.27em}
   \mathcal{S}_a = \{\tmmathbf{\xi}_0^{\ast}, \tmmathbf{\xi}^{\ast}_a,
   \tmmathbf{\xi}_1^{\ast} \}, \hspace{0.27em} \hspace{0.27em} \mathcal{S}_b =
   \{\tmmathbf{\xi}_0^{\ast}, \tmmathbf{\xi}^{\ast}_b, \tmmathbf{\xi}_1^{\ast}
   \}, \hspace{0.27em} \hspace{0.27em} \hspace{0.27em} \hspace{0.27em}
   \mathcal{S}_2 = \{\tmmathbf{\xi}_0^{\ast}, \tmmathbf{\xi}^{\ast}_a,
   \tmmathbf{\xi}^{\ast}_b, \tmmathbf{\xi}_1^{\ast} \} . \]
We note that we introduced distinct sets $S_a$ and $S_b$ even if they actually
contain the same number (and, seemingly, the same kinds) of steady states. In
what follows, it will become clearer that the internal steady states
$\tmmathbf{\xi}_a^{\ast}$ and $\tmmathbf{\xi}_b^{\ast}$ possess specific
properties that allow them to be identified and distinguished from one
another.

As already noted, a quick look at (\ref{eq:x2}) permits to conclude that $x_t
= 0$ and $x_t = 1$ are always solutions, leading to steady states
$\tmmathbf{\xi}_0^{\ast}$ and $\tmmathbf{\xi}_1^{\ast}$, and direct
substitution in (\ref{eq:p2}) yields the corresponding level of pollution, and
knowledge. In this case the steady state is characterized by
\textit{homogeneous} populations of agents, namely by a population
consisting of either all clean or all dirty agents.

As one expects, knowledge is positively affected by resources
collected through taxation: the higher the share of resources devoted
to technology innovation (measured by $\omega$) and the higher the
per-unit taxation charged on profits (measured by $\tau_D$ or
$\tau_C$) the higher the steady stock of knowledge. For the same
reason, $p^{\ast}$ also contributes to the accumulation of knowledge,
as collected resources are proportional to the level of
pollution. Effects of $d$, the total productivity of existing
technologies and investments in innovation, and $1 - \sigma$, the
level of technological obsolescence, affect
knowledge 
as suggested by common sense, having, respectively, a positive and
negative effect on $k^{\ast}$.

Steady states $\tmmathbf{\xi}_0^{\ast}$ and $\tmmathbf{\xi}_1^{\ast}$ represent, respectively, the scenarios of a green transition that has not yet begun and one that is fully completed.

With regard to $\tmmathbf{\xi}_a^{\ast}$ and $\tmmathbf{\xi}_b^{\ast}$, Proposition \ref{th:ss} shows that these steady states exist only under specific additional conditions. They are \textit{internal} steady states, meaning that they are characterized by an intermediate share $x_i^{\ast} \in (0, 1)$ of clean producers, and thus by a {\textit{heterogeneous}} population consisting of both clean and dirty producers. These steady states may represent
transitional phases between $\tmmathbf{\xi}_0^{\ast}$ and
$\tmmathbf{\xi}_1^{\ast}$, and can therefore play a key role in capturing the dynamics of the green transition.

A crucial point to note is that, regardless of the composition of population, these internal steady states are associated with a
\textit{pollution level that is unaffected by $\omega$}, namely by the allocation of resources between innovation and implementation of abatement technologies. This implies that, even if an increase in $\omega$ leads to a higher share of clean producers, this may not result in an improvement in environmental quality\footnote{Conversely,  when $\tau_D$ increases, the corresponding steady-state pollution
  level decreases. In {\citet{CMN24}}, it was shown and discussed that  an increase in the per-unit taxation of dirty producers may actually lead to a deterioration in environmental quality. This contrasting outcome stems from the different evolutionary selection mechanisms adopted in {\citet{CMN24}} and in the present work. Since our
  primary focus here is on the effects of $\omega$, the predictable and unambiguous response of $p^{\ast} = \bar{r}$ to changes in $\tau_D$ allows for a clearer attribution of any counterintuitive results to the role of $\omega$. This, in turn, further supports the choice of adopting in the present contribution a replicator-based  evolutionary mechanism.}. The underlying economic rationale is that in model \eqref{eq:model}, as in {\citet{CMN24,zeppini2015discrete}},
the economic sphere is essentially exogenous, with environmental
taxation, determined by the level of pollution, being the only
endogenous component. Threshold $\bar r$, that determines whether the 
evolutionary transition favors clean or dirty technologies, is thus essentially a pollution level  making a heterogeneous
population sustainable. Changes in $\omega$ may therefore affect the proportion of agents choosing one technology over the other, provided environmental quality stays the same\footnote{As we will
  discuss in Section \ref{sec:concl}, the modelling of the economic sphere, in order to include also direct spillovers of the policy choices, will be addressed in future research.}.

\begin{outcome}
  \label{out:pconstw}In a heterogeneous population of producers,
  increasing investments in innovation do not affect steady state pollution level, while increasing per-unit taxation of dirty producers makes the corresponding pollution level lower.
\end{outcome}

Moreover, the actual relevance of $\tmmathbf{\xi}_a^{\ast}$ and
$\tmmathbf{\xi}_b^{\ast}$ is related to their dynamical stability, as already remarked for $\tmmathbf{\xi}_0^{\ast}$ and $\tmmathbf{\xi}_1^{\ast}$.

In what follows, we study in details the resulting possible scenarios, in particular analyzing the way a transition from a steady state configuration to another one may occur, and highlighting   the possibility of emerging, or, on the contrary, disappearing of $\tmmathbf{\xi}_a^{\ast}$ and
$\tmmathbf{\xi}_b^{\ast}$.

\subsection{Steady scenarios evolutions}\label{sec:static_evol}

We now describe how both steady states $\tmmathbf{\xi}^{\ast}_a$
and/or $\tmmathbf{\xi}^{\ast}_b$ may emerge or disappear\footnote{In
  particular, we emphasize that model \eqref{eq:model} may admit
  steady state solutions even for infeasible values of the share,
  i.e. when $x < 0$ or $x > 1$. This observation allows us to say that $\tmmathbf{\xi}_a^{\ast}$ and/or $\tmmathbf{\xi}_b^{\ast}$ enter(s) (respectively, leave(s)) the feasibility region when a parameter  change causes a steady state solution characterized by an inconsistent share $x \not\in (0, 1)$ (respectively, a consistent share $x \in (0, 1)$) to become feasible (respectively, unfeasible).} upon changing
$\omega$. To do that, we set ourselves in a situation characterized by
\begin{equation}
  p^{\ast} = \bar r, \qquad k^{\ast} =
  \omega \chi \bar r \bar{\tau} (x), \label{eq:pkequi}
\end{equation}
that is
\begin{itemize}
\item the pollution level corresponds to the cut-off
    point defined in \eqref{eq:r}, at which clean and dirty
    technologies yield identical fitness levels and no producer
    benefits from a change in its current technology; 
  \item  the stock of knowledge corresponds to its long run value assuming a 
  pollution level $p^{\ast}$, and a share $x$ of clean producers.
\end{itemize}
Under condition \eqref{eq:pkequi} a steady-state scenario for \eqref{eq:model}
can occur only for specific population distributions, for which the emitted
pollution is exactly offset by natural decay and abatement efforts. To this
end, we define
\begin{equation}
  g_1 (x, \omega) = \frac{c_1 \lambda_0^2 \omega \chi (1 - \omega)  (\tau_D (1
  - x) + \tau_C x)^2}{(\tau_D - \tau_C)^2} = c_1 \chi \omega (1 - \omega)
  (\bar{r}  \bar{\tau} (x))^2, \label{eq:g1}
\end{equation}
which represents the pollution removed thanks to innovation in
  abatement technology only (whose effect is encompassed in $c_1$),
and
\begin{equation}
  g_2 (x, \omega) = \frac{c_2 \lambda_0  (1 - \omega)  (\tau_D (1 - x) +
  \tau_C x)}{\tau_D - \tau_C} = c_2  (1 - \omega) \bar{r}  \bar{\tau} (x).
  \label{eq:g2}
\end{equation}
Observe that the overall effect
  on $g_1$ in \eqref{eq:g1} depends on $c_1 \chi$, which is the
  marginal effect of knowledge on abatement effectiveness, and
  $\omega \bar r \bar{\tau} (x)$ and
  $(1 - \omega) \bar r \bar{\tau} (x)$, which are the amounts of
  resources addressed to technology innovation and implementation,
  respectively. 
Consequently, quantity $g_1$ is the impact of implementing the component of effectiveness by means of new technology.

Similarly, in \eqref{eq:g2}, $c_2$ is the ex-ante effectiveness of technology while $(1 - \omega) \bar r \bar{\tau}$ represents the impact of resources addressed to pollution abatement induced by 
 the implementation of the baseline component of 
technology.
Therefore, the existence of an internal steady state for model \eqref{eq:model} requires 
\begin{equation}
  g_1(x,\omega) + g_2(x,\omega) + \bar{r}\alpha = g(x,\omega) = \bar{\varepsilon}(x)
  \label{eq:bal}
\end{equation}
where, for any $(x, \omega) \in [0, 1]^2$, function
  $g (x, \omega)$ represents the entire amount of removed pollution, including the amount $\bar r \alpha$ of naturally absorbed pollution.

We observe that the left-hand side (lhs from now on) of
\eqref{eq:bal} decreases in $x$: indeed, when the share of clean
producers increases then emission of pollutants diminishes. 
Limit cases occur when the population consists exclusively of
  dirty ($x = 0$) or clean ($x = 1$) producers, and thus removed
  pollution level is $\varepsilon_D$ or
  $\varepsilon_C < \varepsilon_D$, respectively. We now delve into studying some properties of $g (x,\omega)$.

\begin{proposition}
  \label{th:g}Consider function
  $g : \hspace{0.17em} (x, \omega) \in [0, 1]^2 \to [0, + \infty)$,
  where $g$ is defined in \eqref{eq:bal}. Then the graph of
  $x \mapsto g (x, \omega), \hspace{0.27em} x \in [0, 1],$ is a convex and decreasing parabola for any $\omega \in (0, 1)$, and a straight line with negative slope when either $\omega = 0$ or $\omega =
  1$. On the other hand, the graph of
  $\omega \mapsto g (x, \omega), \hspace{0.27em} \omega \in [0, 1], $
  is a concave parabola for any $x \in [0,1]$. In the particular case
  of $c_2 = 0$, $g (x, \omega) = g (x, 1 - \omega)$ and hence it is
  symmetric with respect to $\omega = \frac{1}{2}$.
\end{proposition} 
Indeed, assume that steady state
  $\tmmathbf{\xi}_a^{\ast}$ exists. A geometrical consequence of
  \eqref{eq:bal} and Proposition \ref{th:g} is that, as the number of
  clean producers increases, $\tmmathbf{\xi}_a^{\ast}$ describes a
  transition from a situation in which the removed pollution is higher
  than that produced to the inverse situation, in which removed
  pollution is lower than that produced. Conversely, steady state
  $\tmmathbf{\xi}_b^{\ast}$, when it exists, describes the opposite
  transition.

Next outcome explains in detail the effect of the share
  $x$ of clean producers, as described in Proposition \ref{th:g}. 

{\outcome{\label{out:share_ss} Assume that condition \eqref{eq:pkequi}
    holds.  Then, share  $x$ of clean producers has a
    negative marginal effect on the amount of removed pollution
     $g(x,\omega) + \bar r\alpha$. Indeed, both its
    components, $g_1$, which is related to innovation, and $g_2$,
    which has to do with ex-ante effectiveness, decrease in $x$.}}

According to Outcome \ref{out:share_ss}, for a given $\omega$,
function $x \mapsto g (x, \omega)$ is decreasing. Indeed, an increase
of  $x$ reduces the amount 
  $\bar{\tau} (x) = \tau_D - (\tau_D - \tau_C)x$ of resources
collected through taxation. Clearly, this results in two effects,
\begin{itemize}
\item[a)] the reduction of the share of resources $\omega \bar{\tau} (x)$ destined for
  innovation;
  \item[b)] the  reduction of the share of resources $(1 - \omega)  \bar{\tau} (x)$ destined for implementation of
  abatement technology.
\end{itemize}  When $c_2 >0$, the effect described in b)
explains why pollution quantity eliminated through ex-ante
effectiveness of abatement linearly decreases.  Conversely, the
additional stock of pollution removed by means of innovation is a consequence of both a) and b). Indeed, on the one hand resources destined for innovation decrease, thus inducing a reduction
in the level of knowledge $k^{\ast}$, and, then, a 
  limited improvement $c_1 k^{\ast}$ in the effectiveness of
technology abatement. On the other hand, less resources are addressed
to implementation of technology abatement. These
  overlapping effects explain why the dependence of $g$ on $x$ is
  quadratic.

To investigate, for each $x$, the effect of $\omega$ on the eliminated
stock of pollution, we start from the
  limit case $\omega = 0$, where no resources are destined for  innovation. By direct substitution, condition \eqref{eq:bal}
  becomes
\begin{equation}
  \bar{\varepsilon} (x) = c_2  \bar{r}  \bar{\tau} (x) + \bar{r} \alpha,
  \label{eq:bal0}
\end{equation}
in which the rhs includes only those
  contributions to pollution removal coming from ex-ante abatement
  effectiveness and natural decay. Next result points out conditions
under which internal steady states exist in the case $\omega = 0$.

\begin{figure}[h]
  \begin{center}
    \includegraphics[width=0.9\textwidth]{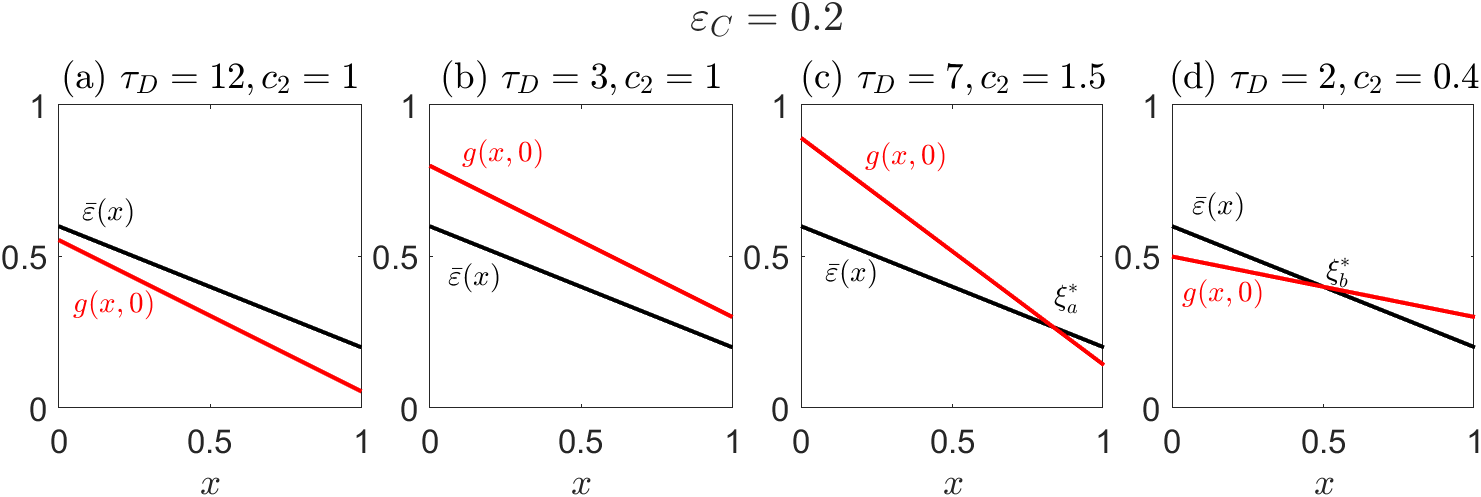}
  \end{center}
  \caption{Graphical representation of different outcomes of
      Proposition \ref{th:ss0} for no investments in innovation $(\omega = 0)$. Red line denotes the level of pollution abatement, the black one the stock of emitted pollution.  Internal steady state occurs when they intersect.\label{fig:bal0}}
\end{figure}

\begin{proposition}
  \label{th:ss0}Assume $\omega = 0$, and define
  \[ l_i = \frac{\lambda_0  (\alpha + \tau_i c_2)}{\tau_D - \tau_C},
     \hspace{0.27em} \hspace{0.27em} \hspace{0.27em} i \in \{C, D\} . \]
  Then, steady states are
  \begin{itemize}
    \item[a)] $\tmmathbf{\xi}_0^{\ast}$, $\tmmathbf{\xi}_1^{\ast}$, together
    with $\tmmathbf{\xi}^{\ast}$ characterized by the share
    \[ x^{\ast} = \frac{l_D - \varepsilon_D}{\lambda_0 c_2 - \varepsilon_D +
       \varepsilon_C} \in (0, 1) \]whenever $l_C < \varepsilon_C < \varepsilon_D < l_D$ or $\varepsilon_C < l_C <
    l_D < \varepsilon_D$;
    
  \item[b)] $\tmmathbf{\xi}_0^{\ast}$ and $\tmmathbf{\xi}_1^{\ast}$
    only, whenever either $\varepsilon_D \geq l_D$
      and $\varepsilon_C \geq l_C$, or  $\varepsilon_D \leq l_D$ and
    $\varepsilon_C \leq l_C$.
  \end{itemize}
\end{proposition}
We first note that $l_D$ and $l_C$ represent the pollution eliminated
in the two extreme situations of fully dirty ($x = 0$) and fully clean
($x = 1$) populations, corresponding to the rhs of \eqref{eq:bal0}
in each of these situations.   Proposition \ref{th:ss0} states
  that the existence of an internal steady state, in addition to
  $\tmmathbf{\xi}_0^{\ast}$ and $\tmmathbf{\xi}_1^{\ast}$, is related
  to the pollution of clean and dirty producers. We refer to Figure
  \ref{fig:bal0}.  Specifically,
  consider case (b) of Proposition \ref{th:ss0}, where both
  technologies are either particularly harmful to environment (see
  panel (a) of Figure \ref{fig:bal0}) or sufficiently clean (see panel
  (b) of Figure \ref{fig:bal0}). Then, only steady states
  characterized by a homogeneous population exist, suggesting that a
  green transition has already occurred or not yet started.
Conversely, consider case (a) of Proposition
  \ref{th:ss0}, where one technology produces pollution below a
  certain threshold. Then an internal steady state arises.\\
Specifically, if emissions of dirty producers are low
($\varepsilon_D < l_D$), this additional steady state  $\tmmathbf{\xi}_a^{\ast}$ is characterized by a favorable \textit{ex ante} scenario 
($c_2 > (\varepsilon_D - \varepsilon_C) / \lambda_0$) (see
  panel (c) of Figure \ref{fig:bal0}); if instead
$\varepsilon_D > l_D$, the new steady state is
  $\tmmathbf{\xi}_b^{\ast}$ with a poor \textit{ex ante} scenario
($c_2 < (\varepsilon_D - \varepsilon_C) / \lambda_0$, see
  panel (d) of Figure \ref{fig:bal0}). Overall, the dynamics can be
interpreted by comparing emissions from clean and dirty producers with
the pollution removed in the two extreme  benchmark cases
of homogeneous population.

In Figure \ref{fig:bal0}, we report four panels describing the content
of Proposition \ref{th:ss0}, focusing on the case study with
intermediate emissions for clean producers ($\varepsilon_C = 0.2$) and
for different values of $\tau_D$ and $c_2$. In each plot
we represent, on varying $x$, two straight lines, the black one
representing produced pollution, and the red one the eliminated
pollution.

We now concentrate on effects of a change in resources destined for innovation, $\omega$, on the level of eliminated pollutants. To this end, we note that
\[ \frac{\partial g_1}{\partial \omega} = c_1 \chi (1 - 2 \omega)  (\bar{r} 
   \bar{\tau})^2 \hspace{0.27em} \hspace{0.27em} \hspace{0.27em}
   \hspace{0.27em} \hspace{0.27em} \text{and} \hspace{0.27em} \hspace{0.27em}
   \hspace{0.27em} \hspace{0.27em} \hspace{0.27em} \frac{\partial
   g_2}{\partial \omega} = - c_2  \bar{r}  \bar{\tau} < 0. \]
We observe that the sign of $\frac{\partial g_1}{\partial \omega}$ depends on the steady state marginal effect of knowledge on technological effectiveness ($c_1 \chi$) and on the marginal impact of resources allocated to pollution reduction through innovation $(1 - 2 \omega) (\bar{r} \bar{\tau})^2$, which varies with $\omega$. When $\omega < 1 / 2$, increasing $\omega$ has the positive effect
to raise $g_1$, while for $\omega > 1 / 2$ the effect is the opposite. This follows from the dual role of $\omega$: on the one hand, a higher $\omega$ enhances abatement through innovation; on the other, it reduces resources for implementing existing technologies. Thus, with a small share of resources devoted to innovation, the positive effect on $g_1$ dominates, whereas for large $\omega$ the negative effect prevails. In contrast, $\frac{\partial g_2}{\partial \omega}$ depends on the ex-ante abatement effectiveness ($c_2$) and on the marginal effect of resources used for existing technologies ($-
\bar{r} \bar{\tau}$), which is always negative regardless of $\omega$. Hence, raising $\omega$ only decreases resources available for implementation.
Consequently, when $\omega \geq 1 / 2$ the stock of pollution removed via
innovation declines, whereas for $\omega < 1 / 2$ multiple outcomes are possible. The main insights are summarized in the following outcome.

{\outcome{\label{out:w_ss} Under condition \eqref{eq:pkequi}, amount of pollution removed as a result of innovation effectiveness increases with investments in innovation when $\omega < 1 / 2$, and then decreases. On the
other side, quantity of pollution being removed for ex-ante effectiveness decreases. Marginal benefits on pollution reduction by investing in innovation
as $\omega$ increases are decreasing.}}

We now focus on the case $\omega < 1 / 2$, and investigate how monotonicity of $g$ is affected by parameters and variables. Since
\begin{equation}
  \frac{\partial g}{\partial \omega} = \frac{\partial g_1}{\partial \omega} +
  \frac{\partial g_2}{\partial \omega} = \bar{r}  \bar{\tau} (x)  (c_1 \chi (1
  - 2 \omega) \bar{r}  \bar{\tau} (x) - c_2), \label{eq:dgw}
\end{equation}
the sign of $\frac{\partial g}{\partial \omega}$ depends on that of $c_1 \chi
(1 - 2 \omega)  \bar{r}  \bar{\tau} (x) - c_2,$ which is positive if and only if
\begin{equation}
  R_{\tau} (\omega, x) = (1 - 2 \omega)  \bar{r}  \bar{\tau} (x) >
  \frac{c_2}{c_1 \chi} = R_c . \label{eq:g_sign}
\end{equation}

We stress that in \eqref{eq:g_sign}, the rhs $R_c$ is the ratio between ex-ante effectiveness and the steady state marginal effect of knowledge on technological effectiveness, depending solely on technological aspects, independently of taxation. Conversely, the rhs $R_{\tau} (\omega, x)$ reflects only taxation effects, excluding baseline technology and innovation capability.

Condition \eqref{eq:g_sign} implies that investing in innovation is
beneficial either because the marginal impact of resources devoted to
innovation is sufficiently large compared to that for the baseline
technology, or because ex-ante effectiveness is small relative to the
marginal effect of knowledge.  If condition \eqref{eq:g_sign} fails,
the conclusion is reversed. Before analyzing the behavior of
$\frac{\partial g}{\partial \omega}$ as $\omega$ increases, we note
that $R_{\tau} (\omega, x)$ decreases in $x$, $\tau_D$, and
$\omega$. The following result is a direct consequence of Outcome
\ref{out:share_ss}.

{\outcome{\label{out:w_x_ss}Ceteris paribus, under condition
\eqref{eq:pkequi}, the impact of taxation is mostly beneficial for innovation when the industry is populated by dirty producers, while it reaches its minimum when it is populated by clean producers.}}

We have the following result, for which we introduce quantities:
\[ \omega_i = \frac{1}{2} - \frac{c_2  (\tau_D - \tau_C)}{2 c_1 \lambda_0
   \tau_i \chi}, \hspace{0.27em} \hspace{0.27em} \hspace{0.27em} i \in \{C,
   D\}, \hspace{0.27em} \hspace{0.27em} \hspace{0.27em} \hspace{0.27em}
   \hspace{0.27em} \text{and} \hspace{0.27em} \hspace{0.27em} \hspace{0.27em}
   \hspace{0.27em} \hspace{0.27em} \tilde{x} (\omega) = \frac{\tau_D}{\tau_D -
   \tau_C} - \frac{c_2}{c_1 \lambda_0 \chi (1 - 2 \omega)} . \]
\begin{proposition}
  \label{th:incw} The monotonicity of $g$ depends on parameters as follows:
  \begin{itemize}
    \item[a)] if $R_{\tau} (0, 0) < R_c,$ then $\frac{\partial g}{\partial
    \omega} < 0$ for any $x \in (0, 1)$ and any $\omega \in (0, 1)$;
    
    \item[b)] if $R_{\tau} (0, 1) < R_c < R_{\tau} (0, 0)$ then
    \begin{enumerate}
      \item $\frac{\partial g}{\partial \omega} > 0$ for $x \in (0, \tilde{x}
      (\omega))$ and $\frac{\partial g}{\partial \omega} < 0$ for $x \in
      (\tilde{x} (\omega), 1)$ whenever $\omega \in (0, \omega_D)$;
      
      \item $\frac{\partial g}{\partial \omega} < 0$ for any $x \in (0, 1)$
      whenever $\omega \in (\omega_D, 1)$;
    \end{enumerate}
    \item[c)] if $R_{\tau} (0, 1) > R_c$ then
    \begin{enumerate}
      \item $\frac{\partial g}{\partial \omega} > 0$ for any $x \in (0, 1)$
      whenever $\omega \in (0, \omega_C)$;
      
      \item $\frac{\partial g}{\partial \omega} > 0$ for $x \in (0, \tilde{x}
      (\omega))$ and $\frac{\partial g}{\partial \omega} < 0$ for $x \in
      (\tilde{x} (\omega), 1)$ whenever $\omega \in (\omega_C, \omega_D)$;
      
      \item $\frac{\partial g}{\partial \omega} < 0$ for any $x \in (0, 1)$
      whenever $\omega \in (\omega_D, 1)$.
    \end{enumerate}
  \end{itemize}
\end{proposition}

Each case described in Proposition \ref{th:incw} is characterized in
terms of condition \eqref{eq:g_sign}. Case a) corresponds to
$R_{\tau} (0, 0) < R_c$, namely investing in innovation is not
convenient, even in a scenario where population consists only of dirty
producers ($x = 0$), and thus we would expect the wider benefits from
doing so (see Outcome \ref{out:w_x_ss}). These potential advantages
decrease along with the increase in either $\omega$ or $x$ (see
Outcomes \ref{out:share_ss} and \ref{out:w_ss}), therefore we conclude
that innovation, at least in this setting, cannot bring any benefit
($\frac{\partial g}{\partial \omega} < 0$ for any $\omega$). In the
opposite situation, we have case c), occurring when
$R_{\tau} (0, 1) > R_c$, namely when investing in innovation would be
beneficial, even in a scenario where population consists only of clean
producers ($x = 1$), and thus we would expect to gain minimum benefits
from doing so (see Outcome \ref{out:share_ss}). Consequently, we
expect significant positive effects for any $x < 1$, and investing in
innovation increases the amount of pollution that can be
eliminated. However, as $\omega$ increases, marginal (decreasing)
benefits of investing in innovation (see Outcome \ref{out:w_ss}) are
overtaken by marginal (constant) advantages of implementing
technology. This effect is significant when $x$ is larger (see Outcome
\ref{out:share_ss}), so investing in innovation is appropriate if the
number of dirty producers is sufficiently large, and no longer when
this number is small enough. A further increase in $\omega$ rules out
the possibility to gain positive effects from innovation, whatever
the population distribution is. Finally, case b),
i.e. for $R_{\tau} (0, 1) < R_c < R_{\tau} (0, 0)$ is intermediate
between limit situations a) and c), so that it can be easily
interpreted along the lines of cases a) and c).


Proposition \ref{th:incw} identifies the necessary conditions for the
emergence or disappearance of steady states. We summarize in what follows the possible cases, reminding that 
conditions, under which each case occurs can be inferred from Proposition \ref{th:incw} and geometrical considerations.
\begin{corollary}
  \label{th:static_trans}A new (couple of) steady state(s) can enter
  or leave the feasible region from $x^{\ast} = 0,\,x^{\ast} = 1$ or $x^{\ast} \in(0,1).$
\end{corollary}

\begin{table}[t!]
  {\small{
      \begin{tabular}{lp{2.7cm}p{2.5cm}p{2.1cm}p{2.0cm}p{1.5cm}l}
     & If, for $\bar{\omega}$, removed
    
    and emitted pollution
    
    coincide for a population of & and on a neigh.
    
    of $\bar{x}$, removed pollution & and, on a
    
    neigh. of $\bar{\omega}$,
    
    increasing $\omega,$ removed
    
    pollution on a
    
    neighb. of $\bar{x}$ & then  & feasible region through & Figure \ \\
    \hline
    a) & dirty producers ($\bar{x} = 0$) & is greater than
    
    emissions & decreases & $\tmmathbf{\xi}_b^{\ast}$ enters  & $\tmmathbf{\xi}_0^{\ast}$ &
    \ref{fig:ss_internal_transitions1} (a)\\
    \hline
    b) & dirty producers ($\bar{x} = 0$) & is greater than
    
    emissions & increases & $\tmmathbf{\xi}_b^{\ast}$ leaves   & $\tmmathbf{\xi}_0^{\ast}$ &
    \ref{fig:ss_internal_transitions1} (b)\\
    \hline
    c) & dirty producers ($\bar{x} = 0$) & is smaller than
    
    emissions & decreases & $\tmmathbf{\xi}_a^{\ast}$ leaves   & $\tmmathbf{\xi}_0^{\ast}$ &
    \ref{fig:ss_internal_transitions1} (c),(d)\\
    \hline
    d)  & dirty producers ($\bar{x} = 0$) & is smaller than
    
    emissions & increases & $\tmmathbf{\xi}_a^{\ast}$ enters   & $\tmmathbf{\xi}_0^{\ast}$ &
    \ref{fig:ss_internal_transitions1} (e),(f)\\
    \hline
    e) & mixed producers ($\bar{x} \in (0, 1)$) & is greater than
    
    emissions  & decreases & $\tmmathbf{\xi}_a^{\ast},
    \tmmathbf{\xi}_b^{\ast}$ enter   &  & \ref{fig:ss_internal_transitions2} (a)\\
    \hline
    f) & mixed producers ($\bar{x} \in (0, 1)$) & is greater than
    
    emissions  & increases & $\tmmathbf{\xi}_a^{\ast},
    \tmmathbf{\xi}_b^{\ast}$ leaves   &  & \ref{fig:ss_internal_transitions2} (b)\\
    \hline
    g) & clean producers ($\bar{x} = 1$) & is greater than
    
    emissions  & decreases & $\tmmathbf{\xi}_a^{\ast}$ enters   & $\tmmathbf{\xi}_1^{\ast}$ &
    \ref{fig:ss_internal_transitions3} (a)\\
    \hline
    h) & clean producers ($\bar{x} = 1$) & is greater than
    
    emissions  & increases & $\tmmathbf{\xi}_a^{\ast}$ leaves   & $\tmmathbf{\xi}_1^{\ast}$ &
    \ref{fig:ss_internal_transitions3} (b)\\
    \hline
    i) & clean producers ($\bar{x} = 1$) & is smaller than
    
    emissions  & decreases & $\tmmathbf{\xi}_b^{\ast}$ leaves & $\tmmathbf{\xi}_1^{\ast}$ &
    \ref{fig:ss_internal_transitions3} (c)-(d)\\
    \hline
    l) & clean producers ($\bar{x} = 1$) & is greater than
    
    emissions  & increases & $\tmmathbf{\xi}_b^{\ast}$ enters  & $\tmmathbf{\xi}_1^{\ast}$ &
    \ref{fig:ss_internal_transitions3} (e)-(f)\\
    \hline
  \end{tabular}
  
  \ }}
  
  \
  \caption{Possible situations giving rise to emergence or disappearance of
  steady states.\label{tab:emdis}}
\end{table}

\begin{figure}[h]
  \begin{center}
    \includegraphics[width=0.75\textwidth]{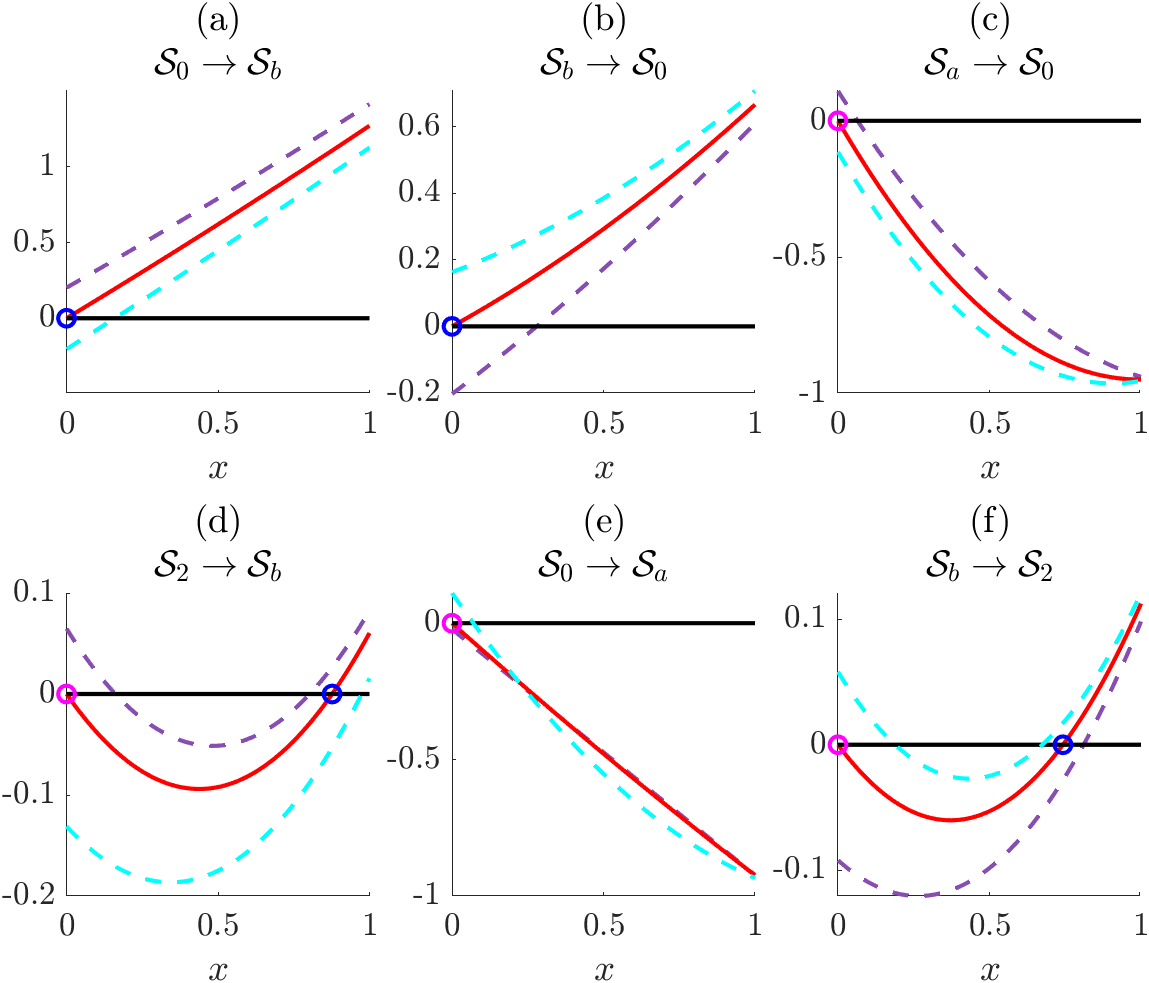}
    
  \end{center}
  \caption{Possible emergence/disappearance of a steady state entering/leaving
  the feasible region from
  $\tmmathbf{\xi}_0^{\ast}$.\label{fig:ss_internal_transitions1}}
\end{figure}

\begin{figure}[h]
  \begin{center}
    \includegraphics[width=0.5\textwidth]{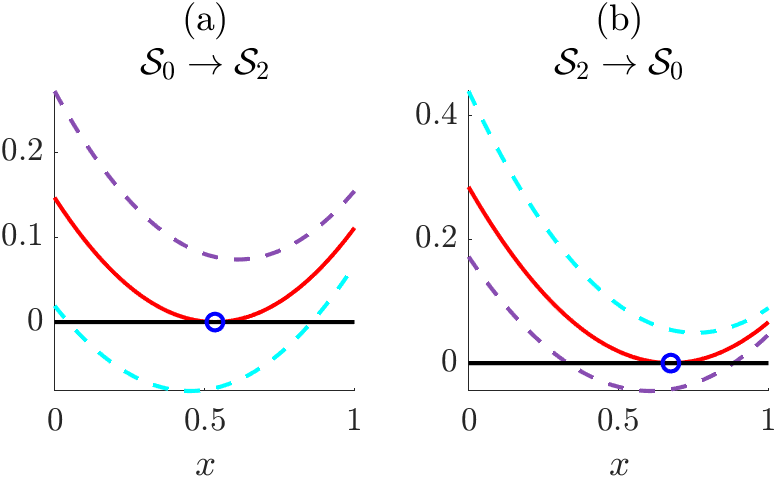}
    
  \end{center}
  \caption{Possible simultaneous emergence/disappearance of couple of steady
  states entering/leaving the feasible
  region.\label{fig:ss_internal_transitions2}}
\end{figure}

\begin{figure}[h]
  \begin{center}
    \includegraphics[width=0.75\textwidth]{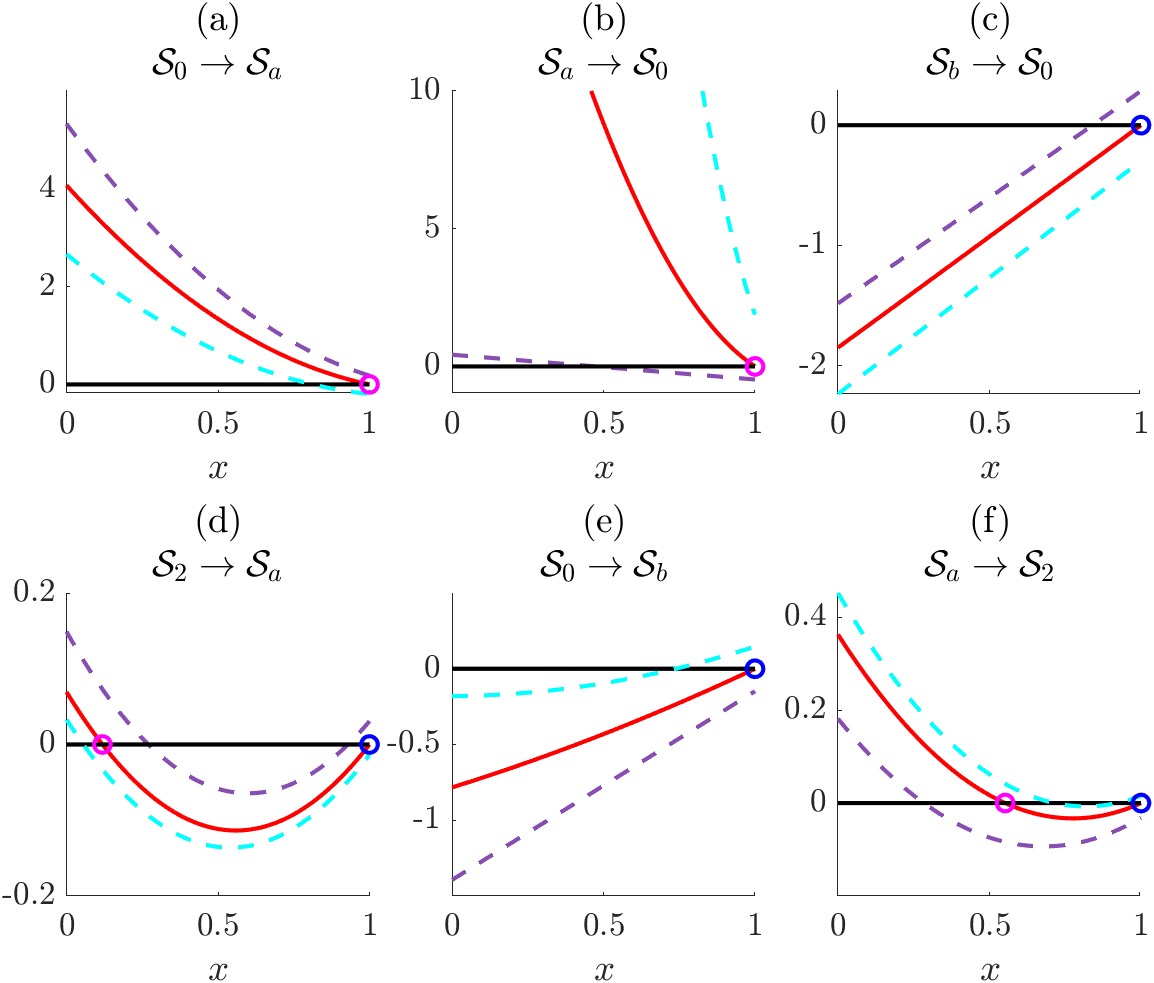}
    
  \end{center}
  \caption{Possible emergence/disappearance of a steady state entering/leaving
  the feasible region from
  $\tmmathbf{\xi}_1^{\ast}$.\label{fig:ss_internal_transitions3}}
\end{figure}

All different ways new steady states can either appear or disappear
are reported in Table \ref{tab:emdis}, which reads as follows. To
describe the first mechanism, focus on first row. Then read the
heading of first column, afterwards the content of the leftmost box in
first row, again the heading of second column, and so on. To describe
the second mechanism, focus on second row, and do the same. And so
forth. Last column of Table \ref{tab:emdis} reports the panel(s) of
Figures \ref{fig:ss_internal_transitions1},
\ref{fig:ss_internal_transitions2} and
\ref{fig:ss_internal_transitions3} by which the case at hand is
illustrated. To simplify readability, we define
\[ \Delta (x, \omega) = g (x, \omega) + \bar{r} \alpha -
  \bar{\varepsilon} (x), \] which is the difference between the amount
of eliminated   and produced pollution. When
$\Delta (x, \omega)$ is positive (resp. negative) then amount of
eliminated pollution is more (resp. less) than that
produced. Monotonicity of $\Delta (x, \omega)$ in $\omega$ is that of
$g (x, \omega)$, so Proposition \ref{th:incw} can be applied to
$\Delta (x, \omega)$ as well. As indicated in Table \ref{tab:emdis},
let $\bar{\omega}$ be the share of resources devoted to innovation at
which removed and emitted pollution coincide for a particular
distribution of shares. In each panel of Figures
\ref{fig:ss_internal_transitions1}, \ref{fig:ss_internal_transitions2}
and \ref{fig:ss_internal_transitions3}, the red curve represents
$\Delta (x, \bar{\omega})$, which is consistent with the occurrence of
a change in the internal steady state configuration, whereas purple
and light-blue dashed curves describe, respectively, what happens when
$\omega$ is suitably smaller or larger than $\bar{\omega}$. Further,
the black horizontal line corresponds to the horizontal axis,
therefore, when the red curve intersects this line, the amount of
eliminated and produced pollutants is the same. According to Outcome
\ref{out:share_ss}, $\Delta (x, \omega)$, and red curve as well,
increases if and only if marginal effects of an increase in clean
producers reduce quantity of removed pollution less than a decrease in
emissions.  Finally, blue and bright-pink circles highlight conditions
corresponding to steady states $\tmmathbf{\xi}_b^{\ast}$ and
$\tmmathbf{\xi}_a^{\ast}$, respectively; when they coincide, a blue
circle is used. All cases listed in Table \ref{tab:emdis} can be
described in terms of the mechanisms in Outcome \ref{out:share_ss} and
the conditions in Proposition \ref{th:incw}, which, in turn, are
defined with reference to Outcomes \ref{out:w_ss} and
\ref{out:w_x_ss}. We begin with cases a), b), c), and d) in Table
\ref{tab:emdis} (plotted in Figure \ref{fig:ss_internal_transitions1})
which are characterized by a transition in which a new steady state
emerges or vanishes as $\tmmathbf{\xi}_0^{\ast}$ is crossed. According
to occurrences a)\footnote{As an example, for occurrence a), Table
  \ref{tab:emdis} reads as follows: ``If, for some $\bar{\omega}$,
  removed and emitted pollution coincide for a population distribution
  of dirty producers ($\bar{x} = 0$) and on a neighborhood of
  $\bar{x}$, pollution removal is greater than emissions and if, on a
  neighborhood, of $\bar{\omega}$, increasing $\omega$ pollution
  removal on a neighborhood of $\bar{x}$ decreases, then
  $\tmmathbf{\xi}_b^{\ast}$ enters the feasible region from
  $\tmmathbf{\xi}_0^{\ast}$.''} and b), the amount of emissions is
smaller than eliminated pollutants when the population consists of a
large enough share of clean producers.

According to Outcome \ref{out:share_ss}, this holds true independently
of population distribution. If investing in innovation does not bring
any benefit in diminishing pollution when population consists of a
sufficiently large number of dirty producers in a neighborhood of
$\bar{x} = 0$, we are either in case a), c3), or b2) in Proposition
\ref{th:incw}. In increasing $\omega$, if proportion of dirty
producers is large enough, the amount of removed pollution exceeds
that emitted; on the other side, if most of population consists of
clean producers, the opposite occurs. This takes the form of
configuration $x_b$, according to which the two amounts coincide and a
new steady state $\tmmathbf{\xi}_b^{\ast}$ appears (panel (a), Figure
\ref{fig:ss_internal_transitions1}).

Case b) differs from a) only in that behavior resulting from an increased
investment in innovation has now positive effects, since it may occur
consistently to c1), c2) or b1) in Proposition \ref{th:incw}, which are
exactly situations that cannot arise in case a). Here, allocating resources in
innovation turns out to be a good strategy in terms of amount of removed
pollution; in particular, for $\omega < \bar{\omega}$, there exists a
threshold $x_b$ of clean producers such that removed pollution is less
(respectively, more) than that emitted if population consists of a share of
clean producers below (respectively, above) $x_b$. Threshold $x_b$ is
decreasing as $\omega$ increases, and this means that steady state
$\tmmathbf{\xi}_b^{\ast}$, existing whenever $\omega < \bar{\omega}$, leaves,
at some point, the feasible region (Figure \ref{fig:ss_internal_transitions1},
panel (b)).

Cases c) and d) are characterized by a level of emissions greater than that of
abatement, given a population entirely consisting of clean producers. However,
according to Outcome \ref{out:share_ss}, this may happen even in presence of
any population distribution (see Figure \ref{fig:ss_internal_transitions1},
panels (c) and (e)), or up to a certain threshold of clean producers (see
Figure \ref{fig:ss_internal_transitions1}, panels (d) and (f)). Now, assume
that emission is greater than abatement, independently of population
distribution, and that increasing investments in innovation does not bring any
benefit to reduce pollution (Table \ref{tab:emdis}, case c)). Then, given
$\omega < \bar{\omega}$, there exists a threshold $x_a$ of clean producers,
such that removed pollution is more (respectively, less) than that emitted if
population consists of a share of clean producers below (respectively, above)
$x_a$. Threshold $x_a$ is decreasing as $\omega$ increases, and this means
that steady state $\tmmathbf{\xi}_a^{\ast}$, existing whenever $\omega <
\bar{\omega}$, leaves, at some point, the feasible region (Figure
\ref{fig:ss_internal_transitions1}, panel (c)). This description is still
roughly the same when emissions exceeds abatement only up to a certain
threshold of clean producers; the only difference is that now there is an
additional share $x_b$ of clean producers at which $\Delta (x, \omega) = 0$,
and this corresponds to steady state $\tmmathbf{\xi}_b^{\ast}$ which is
consistent to levels of investment in innovation sufficiently close to
$\bar{\omega}$ (Figure \ref{fig:ss_internal_transitions1}, panel (d)). In case
that increasing $\omega$ has an opposite effect, the same way of proceeding
allows to understand all possible behaviors concerning case d) in Table
\ref{tab:emdis}, and reported in Figure \ref{fig:ss_internal_transitions1},
panels (e) and (f).

With regard to both cases e) and f) in Table \ref{tab:emdis} (Figure
\ref{fig:ss_internal_transitions2}), transition occurs through either the
appearance or disappearance of a couple of steady states. This follows
increasing marginal effects on abatement due to an increase in the share of
clean producers (Outcome \ref{out:share_ss}), which allows to be in a
situation such that, for a particular population distribution, $\Delta (x,
\omega)$ reaches its minimum.

Finally, the description of cases g), h), i) and l) (Figure
\ref{fig:ss_internal_transitions3}) is very similar to that of cases a), b),
c) and d), with the only difference that now transitions involve a unique
steady state $\tmmathbf{\xi}_1^{\ast}$ either entering or leaving the feasible
region.

We may in principle obtain all possible sequences of scenarios
transitions with the increasing of $\omega$ by simply combining all
cases of either appearance of disappearance of steady states mentioned
in Table \ref{tab:emdis} and depicted in Figures
\ref{fig:ss_internal_transitions1}, \ref{fig:ss_internal_transitions2}
and \ref{fig:ss_internal_transitions3}. Even if we do
  not mean to make a list, we, however, want to stress that,
according to Proposition \ref{th:incw}, amount of eliminated pollution
monotonicity may change at most once.  This, in
  particular, occurs at the beginning when, for any population
  distribution it increases and then decreases.

\begin{figure}[h]
  \begin{center}
    $\varepsilon_C = 0.002$
  \end{center}
  
  \begin{center}
    \includegraphics[width=0.9\textwidth]{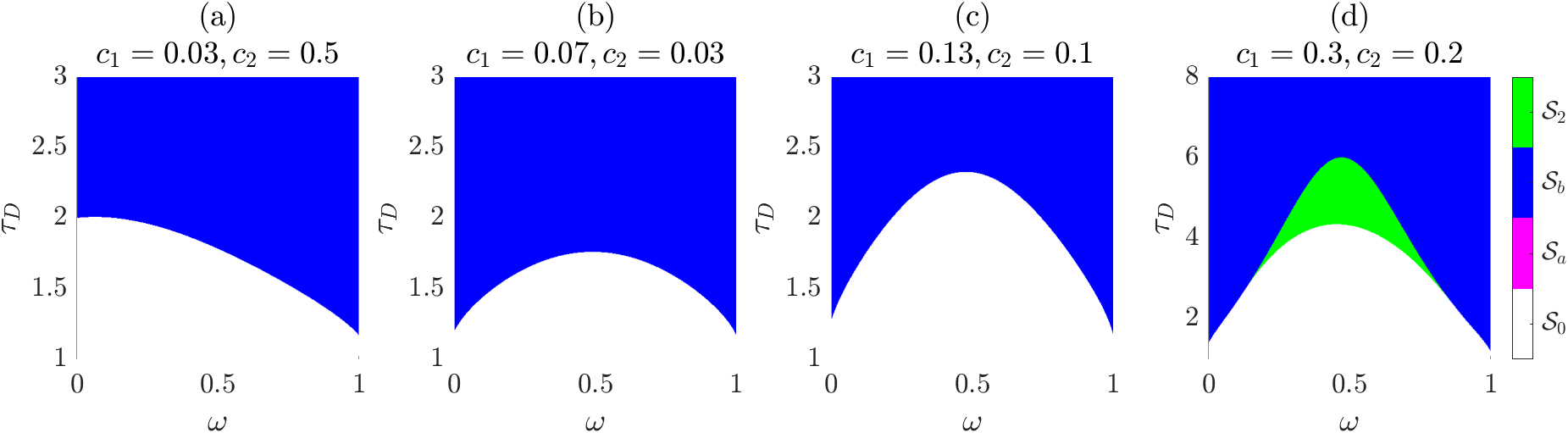}
  \end{center}
  \caption{Sequences of steady state configurations on varying $\omega$ and
  $\tau_D$ when the clean technology has low emission levels. White, blue,
  magenta and green colors are respectively used to represent steady state
  sets $\mathcal{S}_0, \mathcal{S}_b, \mathcal{S}_a$ and
  $\mathcal{S}_2$.\label{fig:ss_changes_small}}
\end{figure}

\begin{figure}[h]
  \begin{center}
    $\varepsilon_C = 0.2$
  \end{center}
  
  \begin{center}
    \includegraphics[width=0.9\textwidth]{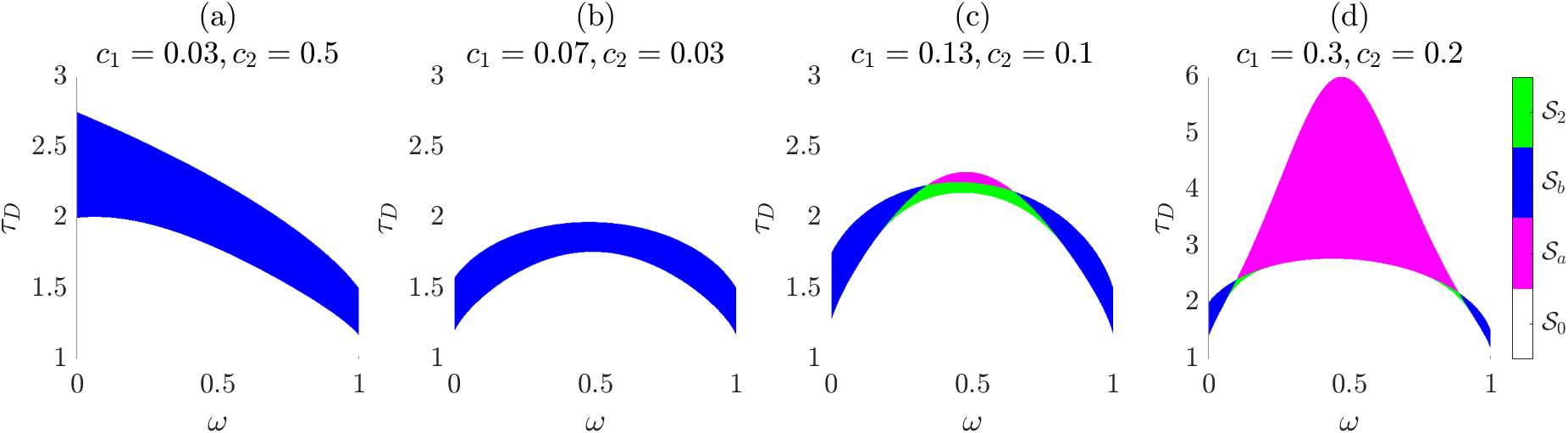}
  \end{center}
  \caption{Sequences of steady state configurations on varying $\omega$ and
  $\tau_D$ when the clean technology has intermediate emission levels. White,
  blue, magenta and green colors are respectively used to represent steady
  state sets $\mathcal{S}_0, \mathcal{S}_b, \mathcal{S}_a$ and
  $\mathcal{S}_2$.\label{fig:ss_changes_med}}
\end{figure}

\begin{figure}[h]
  \begin{center}
    $\varepsilon_C = 0.55$
  \end{center}
  
  \begin{center}
    \includegraphics[width=0.9\textwidth]{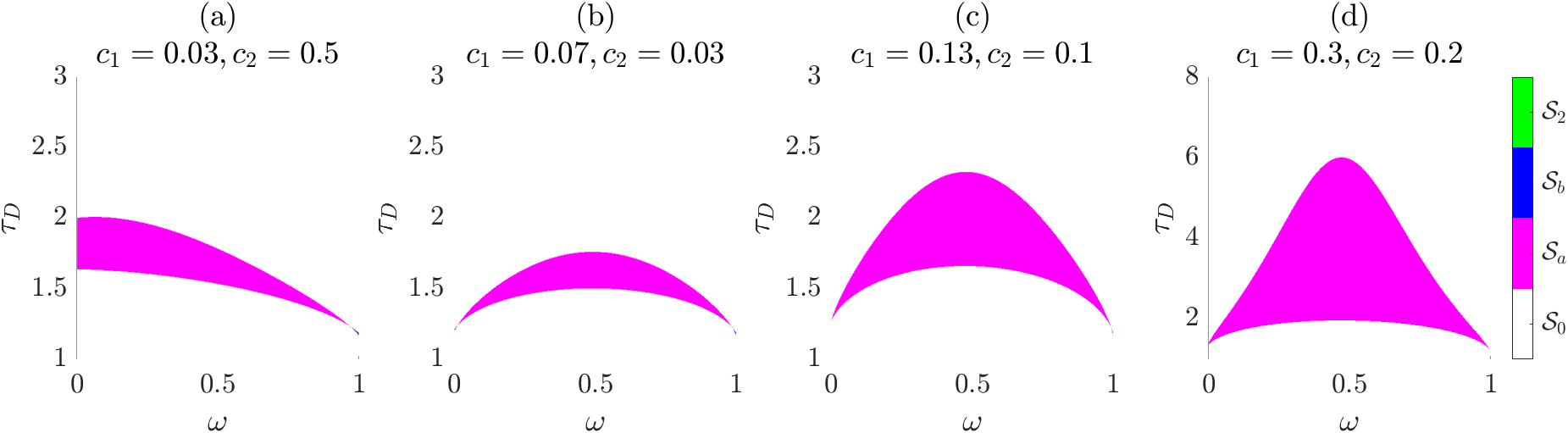}
  \end{center}
  \caption{Sequences of steady state configurations on varying $\omega$ and
  $\tau_D$ when the clean technology has high emission levels. White, blue,
  magenta and green colors are respectively used to represent steady state
  sets $\mathcal{S}_0, \mathcal{S}_b, \mathcal{S}_a$ and
  $\mathcal{S}_2$.\label{fig:ss_changes_large}}
\end{figure}

This ensures that $\tmmathbf{\xi}_a^{\ast}$
and/or $\tmmathbf{\xi}_b^{\ast}$ can enter/leave the feasible region at most twice. We depict possible sequences, on varying $\omega$ and $\tau_D$, in Figures
\ref{fig:ss_changes_small}--\ref{fig:ss_changes_large}, respectively
related to case studies with low, intermediate and high emissions for
the clean producers. In order to represent those regions of pairs
$(\omega, c_1)$ for which steady state sets are, respectively,
$\mathcal{S}_0, \mathcal{S}_b, \mathcal{S}_a$ and $\mathcal{S}_2$, we
use white, blue, bright-pink and green.  Any change regarding the set
of steady states reported in Figures
\ref{fig:ss_internal_transitions1}, \ref{fig:ss_internal_transitions2}
and \ref{fig:ss_internal_transitions3} is represented in Figures
\ref{fig:ss_changes_small}-\ref{fig:ss_changes_large} by the
transition from a colored region to another one with increasing
$\omega$, i.e., through an horizontal shift. Note that a vertical
shift describes a change in a steady state configuration due to an
increase of $\tau_D$. In each of Figures \ref{fig:ss_changes_small},
\ref{fig:ss_changes_med}, and \ref{fig:ss_changes_large}, four
different scenarios are presented, each characterized by a distinct
pairing of the parameters $c_1$ and $c_2$.  Specifically, in panels
(a) we have $c_1 < c_2$, while in the remaining panels (b), (c), and
(d), the ordering is reversed. In particular, in the scenarios
reported in

$\bullet$ panels (a), the effectiveness of ex-ante abatement significantly
exceeds the marginal effect of knowledge, which is small;

$\bullet$ panels (b), both the effectiveness of ex-ante abatement and the
marginal effect of knowledge are relatively low, though $c_1$ is more than
twice as large as $c_2$;

$\bullet$ panels (c), both the effectiveness of ex-ante abatement and the
marginal effect of knowledge take intermediate and roughly comparable values;

$\bullet$ panels (d), both the effectiveness of ex-ante abatement and the
marginal effect of knowledge have relevant size, with $c_1$ substantially
exceeding $c_2$.

Related to Figures \ref{fig:ss_changes_small},
\ref{fig:ss_changes_med}, and \ref{fig:ss_changes_large},
 we note that with low clean-technology emissions (Figure
\ref{fig:ss_changes_small}), sets $\mathcal{S}_b$ (blue) prevail,
whereas with high $\varepsilon_C$ (Figure \ref{fig:ss_changes_large}),
sets $\mathcal{S}_a$ (magenta) dominate. For intermediate emission
levels, steady sets exhibiting internal steady states shift from
$\mathcal{S}_b$ (panels (a)--(b) in Figure \ref{fig:ss_changes_med})
to $\mathcal{S}_a$ (panel (d))\footnote{For intermediate emission
  levels, the distribution shifts from $\mathcal{S}_b$ (panels
  (a)--(b) in Figure \ref{fig:ss_changes_med}) to $\mathcal{S}_a$
  (panel (d)). This pattern is not due to the specific parameter
  choices of the case studies but follows from relation
  \eqref{eq:bal}, which defines the internal steady states. The lhs of
  the equation becomes more strongly decreasing in $x$ as
  $\varepsilon_C$ decreases, while the rhs is convex and decreasing in
  $x$.  If $\varepsilon_C$ is low, emissions fall rapidly as the share
  of clean agents rises, so that, as $x$ grows, the system is more
  likely to move from emissions exceeding absorbed stocks to the
  opposite case, characterizing $\tmmathbf{\xi}_b^{\ast}$. Conversely,
  if $\varepsilon_C$ is high, a larger share of clean agents does not
  reduce emissions quickly; in this case, as $x$ increases, the system
  more often shifts from emissions below absorbed stocks to the
  opposite case, characterizing $\tmmathbf{\xi}_a^{\ast}$. Hence, for
  intermediate emission levels, both situations may arise, and
  coexistence of the two internal steady states is possible.

Moreover, for suitably small or large $\tau_D$, internal steady states
disappear (see Figure \ref{fig:ss_changes_small}; if $\tau_D$ increased
further, the white region would reappear). Finally, for further remarks, we
refer to the dynamics section, since the coexistence of steady states makes it
crucial to identify which ones are dynamically relevant, i.e., the stable
ones.}. Finally, for suitably small or large $\tau_D$, internal steady states
disappear (see Figure \ref{fig:ss_changes_small}; if $\tau_D$ increased
further, the white region would reappear).

\subsection{Comparative statics}\label{sec:static_cs}

Having examined how stationary configurations change with variations in policy parameters $\omega$ and $\tau_D$, we now turn our attention to their effects on each individual steady state. We start 
 with 
 those characterized by homogeneous populations of agents. Starting from $\tmmathbf{\xi}_0^{\ast}$ we focus on the behavior of the steady state levels of pollution, as indeed the population distributions are constant.

\begin{proposition}
  \label{th:csxi0} Consider the steady state $\tmmathbf{\xi}_0^{\ast} = (0,
  p_0^{\ast}, k_0^{\ast})$. Condition
  \begin{equation}
    \varepsilon_D > \frac{c_2^2 \tau_D + \alpha c_2}{c_1 \chi \tau_D},
    \label{eq:xi0sc}
  \end{equation}
  guarantees the existence of $\tilde{\omega} \in (0, 1 / 2)$ such that
  $p^{\ast}_0$ decreases for $\omega \in [0, \tilde{\omega})$ and increases
  for $\omega \in (\tilde{\omega}, 1]$. If condition \eqref{eq:xi0sc} is not
  fulfilled, then $p^{\ast}_0$ increases for any $\omega \in [0, 1]$.  
\end{proposition}
From Proposition \eqref{th:csxi0}, we see that improving environmental
quality in a population of dirty producers is possible through
increased investment in innovation, but only under specific conditions
as specified in \eqref{eq:xi0sc}. This condition introduces a
threshold related to emissions $\varepsilon_D,$ below
  which an increase in the share of resources devoted to innovation
is ineffective for the reduction of pollution, and $p_0^{\ast}$
increases with $\omega$, regardless  of the amount of
available resources.  The threshold increases with the efficiency of
ex-ante technologies, measured by $c_2$, which makes condition
\eqref{eq:xi0sc} harder
to be satisfied. This reflects the intuition that if the technology is
already sufficiently effective, particularly in terms of reducing
emissions from dirty producers, further investment in its improvement
is not worthwhile.  The same conclusion applies when $\tau_D$, the
per-unit tax on dirty producers, is high, and when the marginal
contribution of new knowledge, measured by $c_1$, is low. Similarly,
low levels of overall productivity $d$ and high technology
obsolescence $\sigma$ both raise this threshold. In all these cases,
the most effective choice is to invest in the implementation of
existing abatement systems and techniques, rather than diverting
resources to new research that provides only marginal improvements in
the quality of current technologies. Conversely, when $\varepsilon_D$
is sufficiently high, while ex-ante effectiveness $c_2$ is poor, the
impact of new knowledge on the effectiveness of new technologies $c_1$
is strong, the profits of dirty technologies are heavily reduced by
$\tau_D$, or when total productivity $d$ as well as technology
obsolescence $\sigma$ are low, condition \eqref{eq:xi0sc} is most
likely to hold. In such a case, investing in new research is
reasonable, at least initially: for low values of $\omega$, increasing
resources devoted to research and new knowledge reduces pollution.
This benefit continues until $\omega$ reaches $\tilde{\omega}$, the
share of resources devoted to research at which pollution reaches its
minimum. Beyond $\tilde{\omega}$, however, the marginal benefit of
allocating additional resources to new technologies becomes smaller
than the disadvantage arising from the lack of investment in existing
systems and methods. As a result, the initial benefit vanishes, making
it more convenient to redirect resources to existing technologies.


We stress that the threshold in \eqref{eq:xi0sc} is negatively affected by
$c_1 \tau_D$, which measures how innovation can benefit from resources raised
through taxation, since $c_1 \tau_D$ represents the potential impact of each
taxed unit of pollutant on new abatement technology. The role of $c_1$ is
clear, as investing in ineffective innovation is detrimental, but $\tau_D$
also has a crucial policy implication: effective innovation is possible only
with a suitable taxation level. This becomes explicit by rewriting
\eqref{eq:xi0sc} in a form that highlights the role of taxation. If $c_1
\varepsilon_D \chi - c_2^2 > 0,$then \eqref{eq:xi0sc} is equivalent to
\[ \tau_D > \frac{\alpha c_2}{c_1 \varepsilon_D \chi - c_2^2} . \]
These conditions confirm that if ex ante effectiveness is already high or the
marginal contribution of innovation is small, environmental improvement for a
homogeneous population of dirty producers cannot be achieved through
innovation alone. Conversely, improvement is possible if taxation on dirty
producers is sufficiently severe, allowing the regulator to reduce pollution
while increasing resources allocated to innovation. Now we focus on
$\tmmathbf{\xi}_1^{\ast}$.

\begin{proposition}
  \label{th:csxi1} Consider the steady state $\tmmathbf{\xi}_1^{\ast} = (1,
  p_1^{\ast}, k_1^{\ast})$. Condition
  \begin{equation}
    \varepsilon_C > \frac{c_2^2 \tau_C + \alpha c_2}{c_1 \chi \tau_C}
    \label{eq:xi1sc}
  \end{equation}
  guarantees the existence of $\tilde{\omega} \in (0, 1 / 2)$ such that
  $p^{\ast}_1$ decreases for $\omega \in [0, \tilde{\omega})$ and increases
  for $\omega \in (\tilde{\omega}, 1]$. If condition \eqref{eq:xi1sc} is not
  fulfilled, then $p^{\ast}_1$ increases for any $\omega \in [0, 1]$.
\end{proposition}

Proposition \eqref{th:csxi1} is analogous to \eqref{th:csxi0}, with the
difference that we now consider a scenario in which only clean producers
exist. Therefore, considerations about the role of the parameters, which now
are $\varepsilon_C$ and $\tau_C$ instead of $\varepsilon_D$ and $\tau_D$, are
the same. For the same reason, condition \eqref{eq:xi1sc} can be written in an
equivalent form. In particular, if we require that $c_1 \varepsilon_C \chi -
c_2^2 > 0,$ then \eqref{eq:xi1sc} can be written as
\[ \tau_C > \frac{\alpha c_2}{c_1 \varepsilon_C \chi - c_2^2} . \]
Concerning internal steady states $\tmmathbf{\xi}_a^{\ast}$ and
$\tmmathbf{\xi}_b^{\ast}$, we already remarked that they are characterized by
a pollution level $p^{\ast}$ that is not influenced by the choices on
investment/implementation, while on increasing per unit-taxation, we have that
$p^{\ast}$ decreases. Moreover, it appears evident that an higher per-unit
taxation on dirty producers has the aim of forcing a transition toward clean
technologies. An higher $\tau_D$ pushes down threshold $\bar{r}$, so that the
amount of pollution goes beyond it more likely, and this may have the effect
of convincing to shift towards behaviors that promote sustainable
technologies, in order to avoid to cut profitability excessively. So we focus
on the role of $\omega$ on $x^{\ast}$.
\begin{figure}[ht!]
  \begin{tabular}{|b{1.8cm} b{13cm}|}
    \hline
    \multicolumn{2}{|c|}{Low emission levels of clean
    producers }  \\
    \hline
    \rule{0pt}{55pt}
    \begin{tabular}{c}
    (a)\\    
    $c_1 = 0.03$\\    
      $c_2 = 0.5$
    \end{tabular} &            
    \mline{\includegraphics[width=0.8\columnwidth]{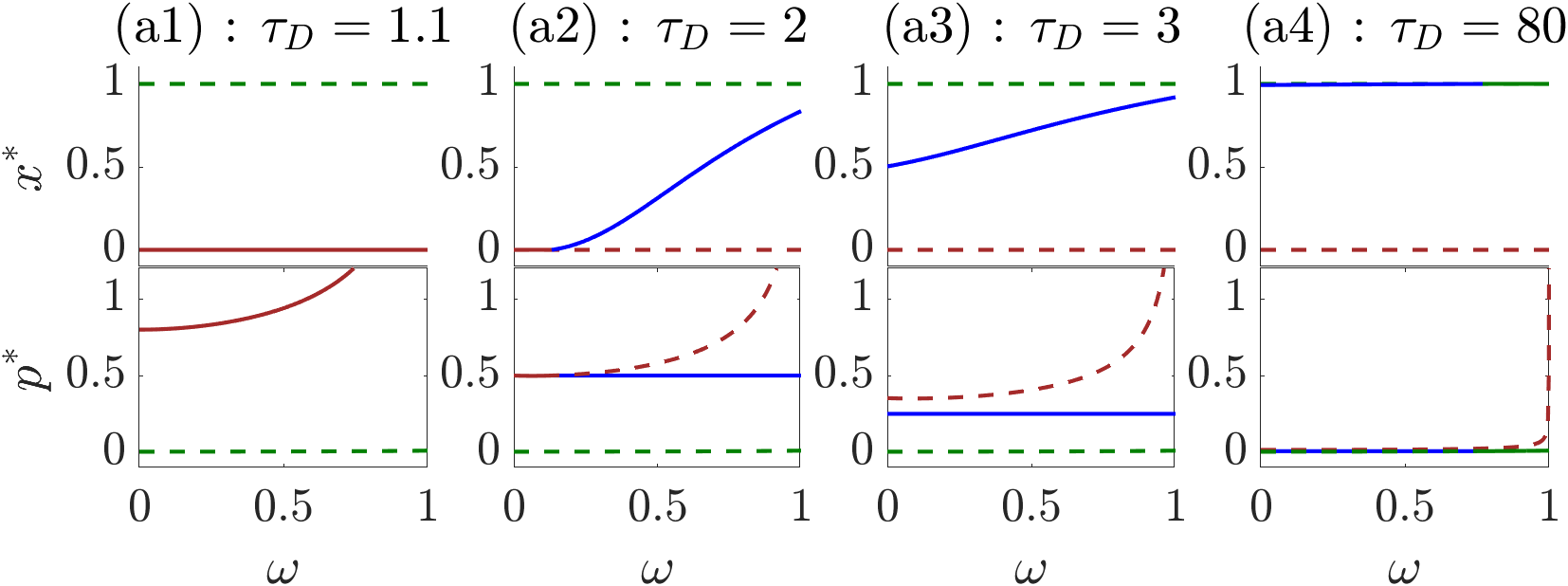}}\\
    \hline
    \rule{0pt}{55pt}
    \begin{tabular}{c}
      (b)\\
      $c_1 = 0.07$\\
      $c_2 = 0.03$\\
    \end{tabular}
    &            
    \mline{\includegraphics[width=0.8\columnwidth]{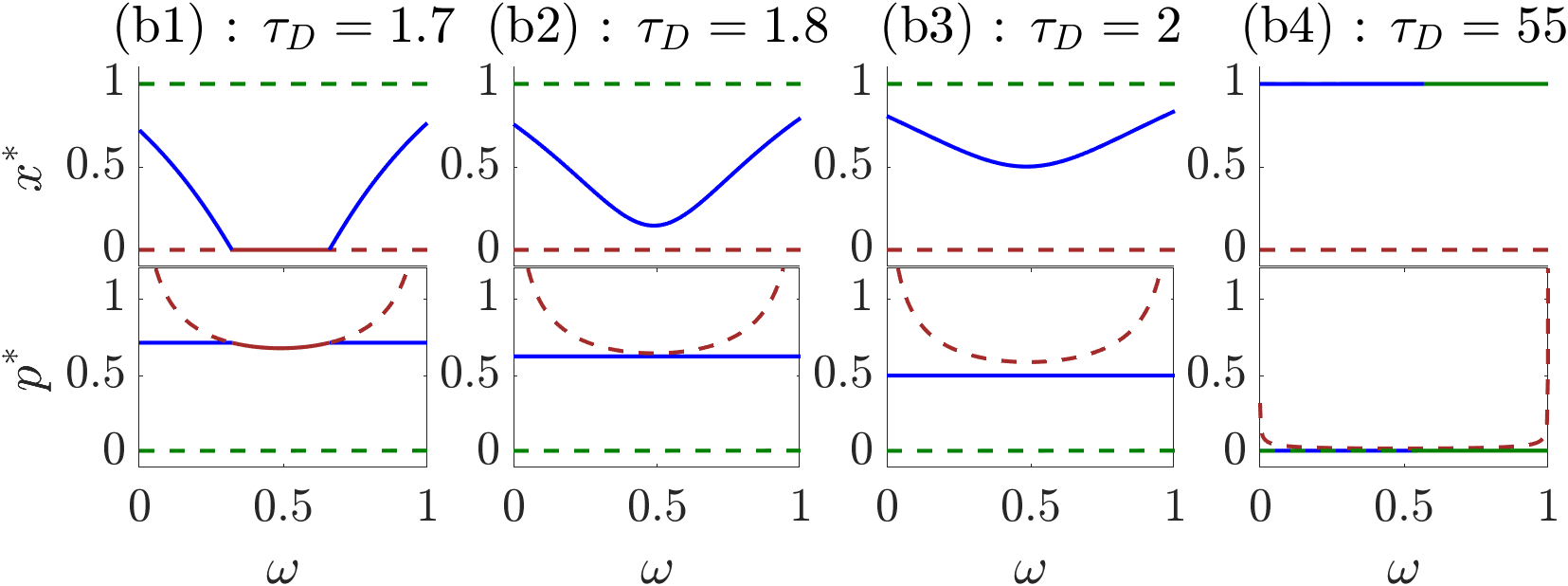}}\\
    \hline
    \rule{0pt}{55pt}
    \begin{tabular}{c}
     (c)\\
     $c_1 = 0.13$\\
     $c_2 = 0.1$
      \end{tabular}&
    \mline{\includegraphics[width=0.8\columnwidth]{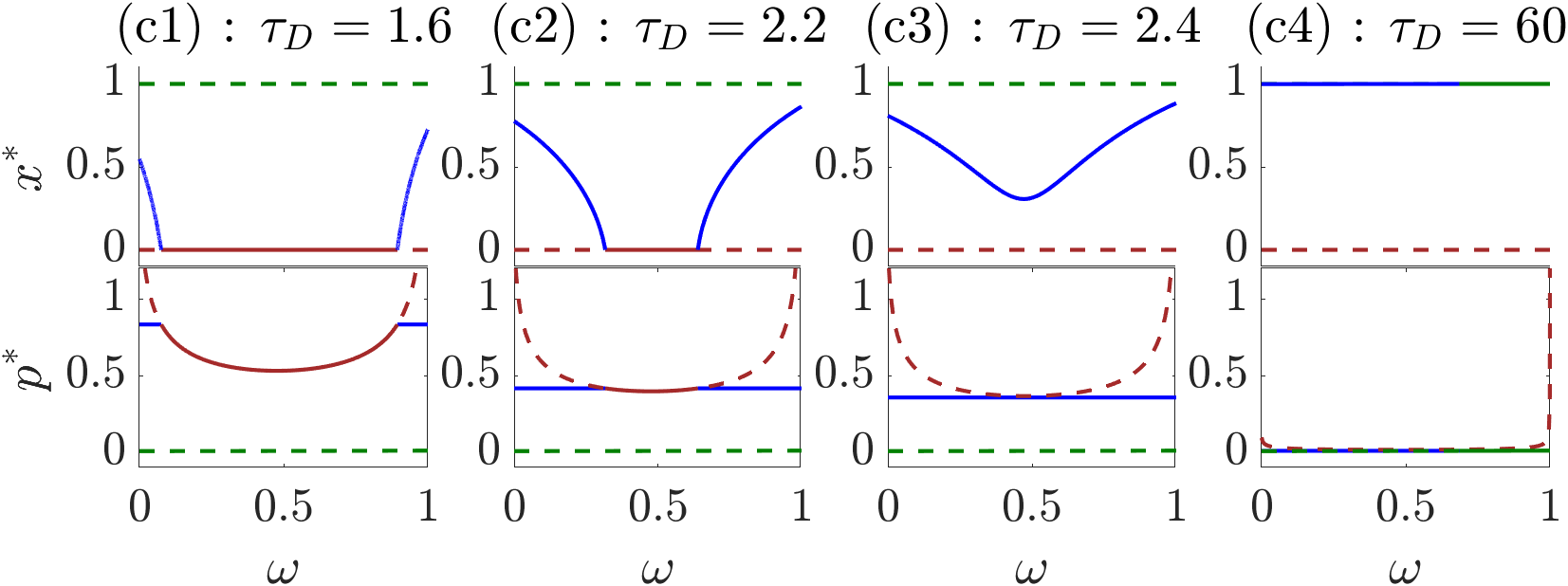}}\\
    \hline
    \rule{0pt}{55pt}
    \begin{tabular}{c}
    (d)\\
    $c_1 = 0.3$\\
     $c_2 = 0.2$
    \end{tabular}
    &
    \mline{\includegraphics[width=0.8\columnwidth]{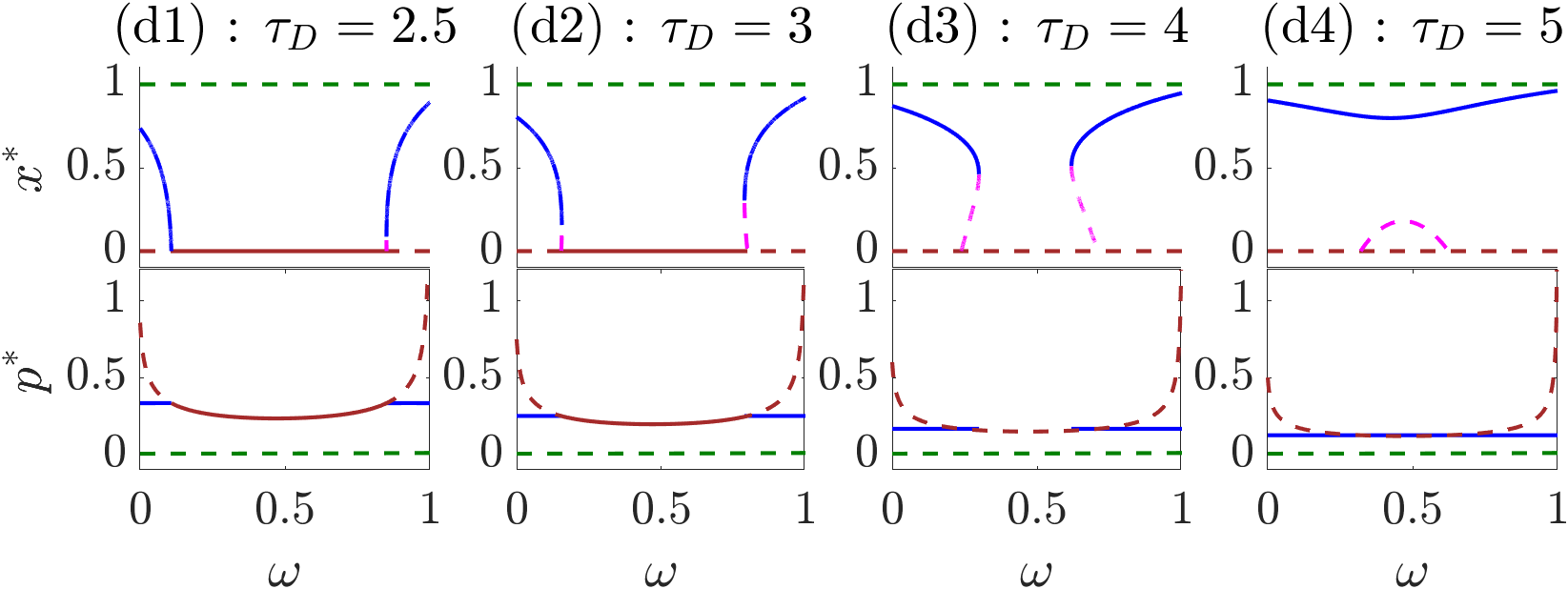}}\\
    \hline
  \end{tabular}
  \caption{Shares $x^{\ast}$ and pollution level $p^{\ast}$ as
    $\omega$ increases and for different values of $\tau_D$ in the
    case of different ex-ante and marginal effectiveness of abatement,
    for low clean technology emission level $\varepsilon_C = 0.002.$
    Brown, green, blue and magenta are respectively related to steady
    states
    $\tmmathbf{\xi}_0^{\ast}, \tmmathbf{\xi}_1^{\ast},
    \tmmathbf{\xi}_b^{\ast}$ and $\tmmathbf{\xi}_a^{\ast}$. Solid and
    dashed lines respectively represent stable and unstable steady
    states. \label{fig:sc_small} }
\end{figure}

\begin{proposition}
  \label{th:csab}Consider the steady state $\tmmathbf{\xi}_a^{\ast} =
  (x_a^{\ast}, p^{\ast}, k_a^{\ast})$, and denote with $I_a^{\omega}$ (resp.
  $I_a^{\tau_D}$) an interval of values of $\omega$ (resp. $\tau_D$) that
  guarantees the existence of $\tmmathbf{\xi}_a^{\ast}$. Then
  \begin{itemize}
    \item[(i)] $x_a^{\ast}$ can be either decreasing, increasing or
    increasing-decreasing on $I_a^{\omega}$;
    
    \item[(ii)] $x_a^{\ast}$ decreases on $I_a^{\tau_D}$.
  \end{itemize}
  Consider the steady state $\tmmathbf{\xi}_b^{\ast} = (x_b^{\ast}, p^{\ast},
  k_b^{\ast})$, and denote with $I_b^{\omega}$ (resp. $I_b^{\tau_D}$) an
  interval of values of $\omega$ (resp. $\tau_D$) that guarantees the
  existence of $\tmmathbf{\xi}_b^{\ast}$. Then
  \begin{itemize}
    \item[(i)] $x_b^{\ast}$ can be either decreasing, increasing or
    decreasing-increasing on $I_b^{\omega}$;
    
    \item[(ii)] $x_b^{\ast}$ decreases on $I_b^{\tau_D}$.
  \end{itemize}
\end{proposition}
Possible behaviors highlighted by Proposition \ref{th:incw} and the
definition of $x$ implicitly given by $\eqref{eq:bal}$ imply that
$x_b^{\ast}$ can be either increasing, decreasing or
decreasing-increasing as the share of resources devoted to innovation
grows, while $x_a^{\ast}$ can be either increasing, decreasing or
increasing-decreasing. Let us focus on $x_b^{\ast}$. Proposition
\ref{th:incw} opens to the possibility of both a green transition
driven by $\omega$ and scenarios in which the number of producers
adopting the clean technology decreases as investment in innovation
increases. A possible explanation for a backsliding toward dirty
technology is: if pollution is efficiently removed from the
environment, there is less incentive to shift towards cleaner
technologies, as the environmental tax burden $\tau_D p$
reduces. Hence, the same pollution level can be then achieved even in
the presence of a large share of dirty procurers, as their emissions
are more efficiently counteracted by the improved abatement, and this
can lead more agents to revert to the dirty technology. In other
words, a `clean' enough world lets some producers behave in a `dirty'
way. Conversely, an increase of the share occurs in the opposite
situation, namely when increasing investments in innovation is
ineffective for the improvement of the environmental quality, and
hence to keep constant the steady state pollution an increasingly
large share of clean procurers is required. Finally, both situations
can take place as $\omega$ increases. According to Outcome
\ref{out:w_ss}, an increase in $\omega$ is beneficial for the
environmental quality when $\omega$ is small, and detrimental when
$\omega$ is large, leading to the decreasing-increasing behavior of
$x_b^{\ast}$.

\begin{outcome}
  \label{out:xbchange}The increase (decrease) in abatement
  effectiveness as resources devoted to innovation rise makes the
  steady-state level of pollution $\bar{r}$ 
  sustainable in the presence of a share $x_b^{\ast} \in (0, 1)$
  greater (smaller) than that of agents adopting the dirty technology.
\end{outcome}

We refrain from discussing $x_a^{\ast}$ in detail, since its
interpretation can only be fully understood in light of the dynamic
properties. This consideration, to varying degrees, also applies to
the other results presented in this section. In fact, we remark that,
although up to four steady states may exist, they are not necessarily
dynamically relevant, as instability can make them unreachable except
under very specific conditions. Whether a transition involving the
emergence or disappearance of steady states benefits or harms the
regulator's objectives depends not only on the states involved but
also on their stability before and after the transition. The impact of
policies on steady states must therefore be evaluated also in terms of
their dynamic effects. To gain a deeper understanding of the
comparative statics results, it is necessary to complete the analysis
reported in Sections \ref{sec:static_ss}, \ref{sec:static_evol} and
\ref{sec:static_cs} by also considering the model from a dynamical
perspective. For now, we limit ourselves to referring to Figures
\ref{fig:sc_small}--\ref{fig:sc_large}, which correspond to the three
case studies\footnote{In addition to the parameters related to the
  three case of studies, the diagrams in Figures
  \ref{fig:sc_small}-\ref{fig:sc_large} are obtained setting
  $\beta = 2$, which has no influence on the steady state values.}
examined and illustrate the realization of all the possible scenarios
predicted by the model. We stress that each panel of Figures
\ref{fig:sc_small}--\ref{fig:sc_large} is in one-to-one
correspondence with a simulation related panels in Figures
\ref{fig:ss_changes_small}-\ref{fig:ss_changes_large}. Brown, green,
blue and magenta are respectively used for steady states
$\tmmathbf{\xi}_0^{\ast}, \tmmathbf{\xi}_1^{\ast},
\tmmathbf{\xi}_b^{\ast}$ and $\tmmathbf{\xi}_a^{\ast}$ and all the
possible monotonicity behaviors for $x$ and $p$ highlighted by the
comparative statics propositions can be straightforwardly
identified. We stress that in addition to the comparative statics
information, in Figures \ref{fig:sc_small}--\ref{fig:sc_large} also
the stability (solid lines) and instability (dashed lines) of the
steady states is represented. Comments on these figures will be
completed in the next section.

\section{Dynamical analysis}\label{sec:dyn}
In this section, we carry out a local stability analysis for the
steady states $\tmmathbf{\xi}^{\ast}_0$ and $\tmmathbf{\xi}^{\ast}_1$,
and provide simulations for $\tmmathbf{\xi}^{\ast}_a$
  and $\tmmathbf{\xi}^{\ast}_b$  because of lack of
analytical tractability. Moreover, we discuss the possible emergence
of complex dynamics.

\subsection{Stability analysis}\label{sec:dyn_stab}

Stability conditions for $\tmmathbf{\xi}_0^{\ast}$ are reported in the next proposition.

\begin{figure}[ht!]
  \begin{tabular}{|b{1.8cm} b{13cm}|}
    \hline
    \multicolumn{2}{|c|}{Intermediate emission levels of clean
    producers }  \\
    \hline
    \rule{0pt}{55pt}
    \begin{tabular}{c}
    (a)\\
    $c_1 = 0.03$\\
      $c_2 = 0.5$ \\
    \end{tabular} &            
    \mline{\includegraphics[width=0.8\columnwidth]{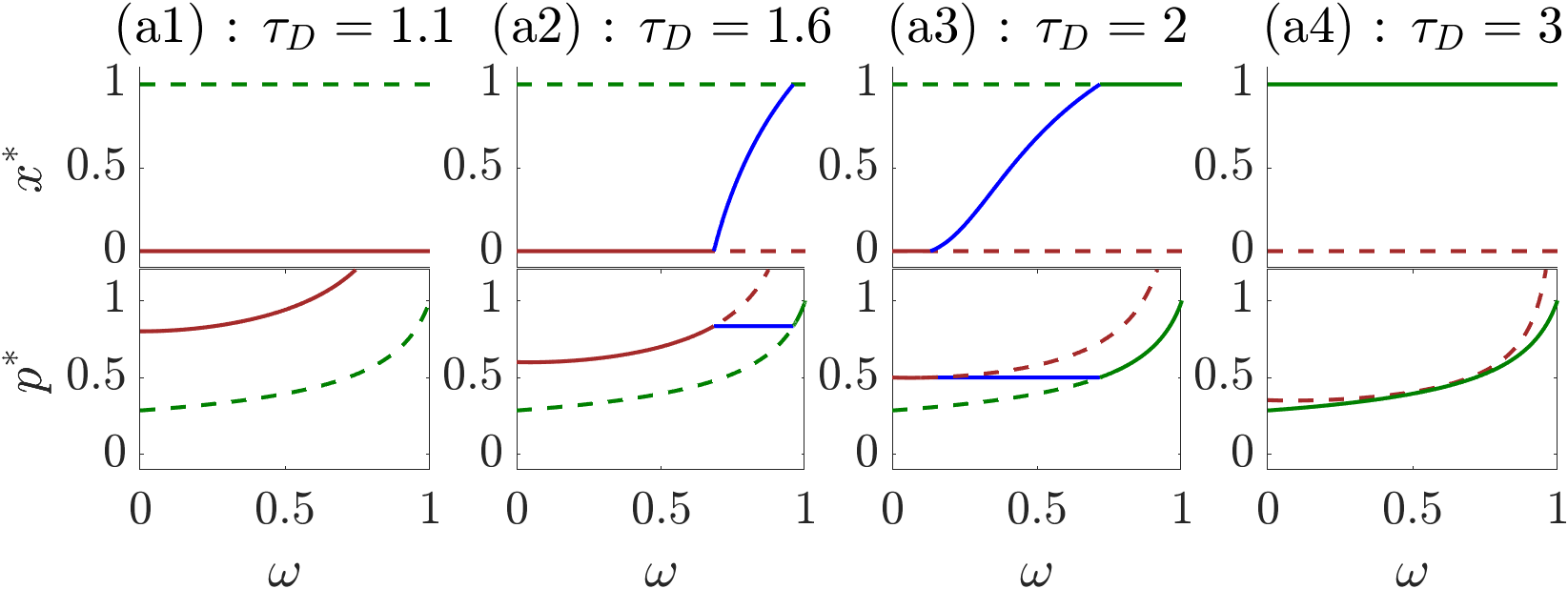}}\\
    \hline
    \rule{0pt}{55pt}
    \begin{tabular}{c}
    (b)\\    
    $c_1 = 0.07$\\
      $c_2 = 0.03$
    \end{tabular} &            
    \mline{\includegraphics[width=0.8\columnwidth]{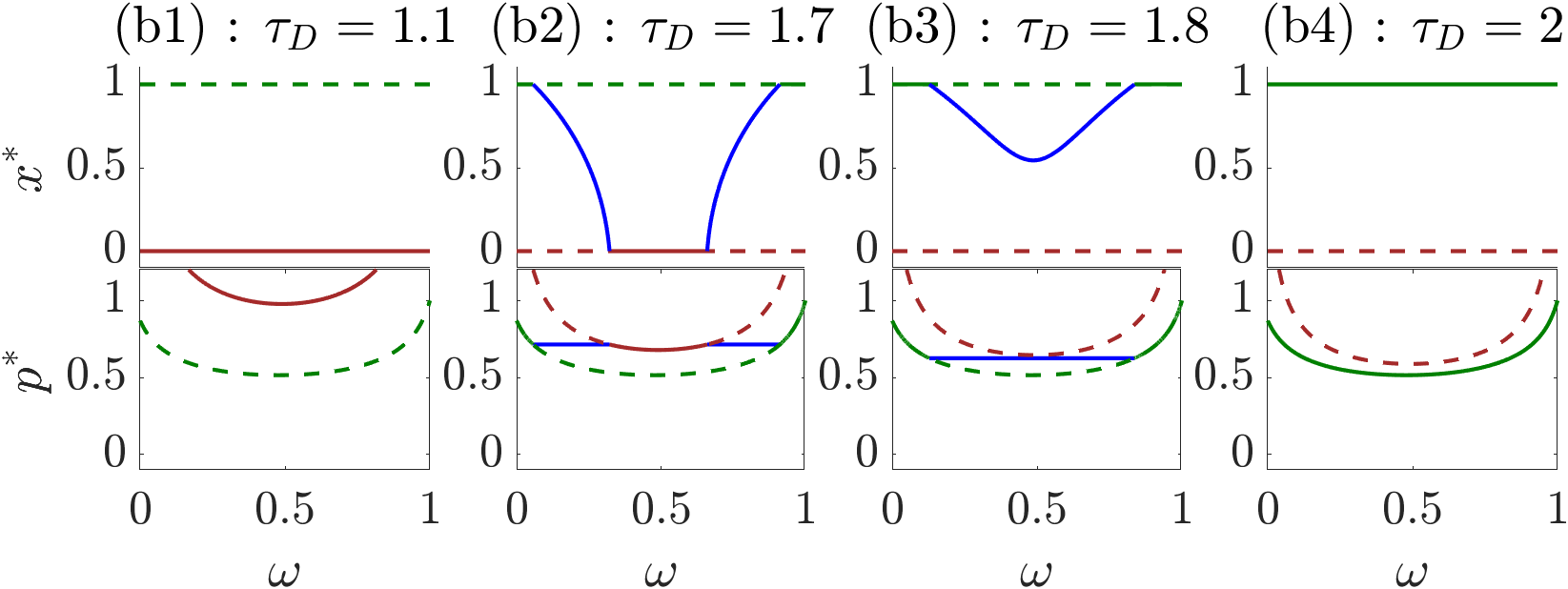}}\\
    \hline
    \rule{0pt}{55pt}
    \begin{tabular}{c}
    (c)\\
    $c_1 = 0.13$\\
      $c_2 = 0.1$
    \end{tabular} &
    \mline{\includegraphics[width=0.8\columnwidth]{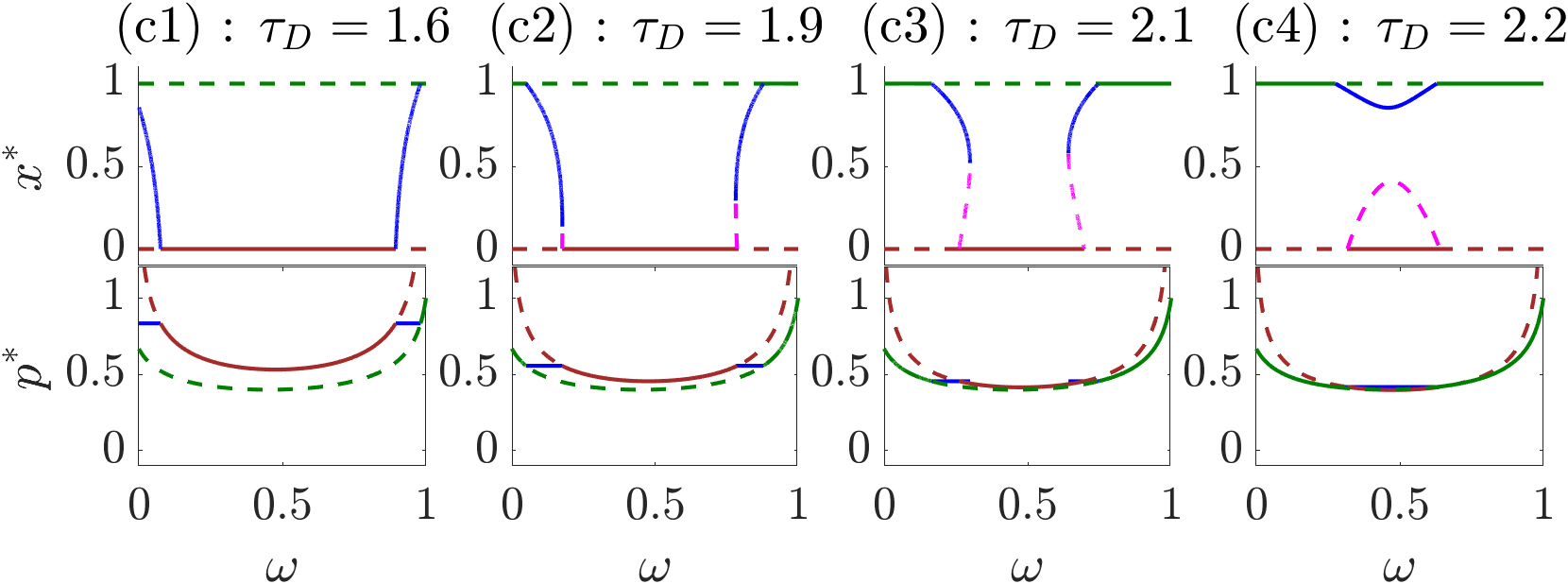}}\\
    \hline
    \rule{0pt}{55pt}
    \begin{tabular}{c}
    (d)\\    
    $c_1 = 0.3$\\    
      $c_2 = 0.2$
    \end{tabular} &
    \mline{\includegraphics[width=0.8\columnwidth]{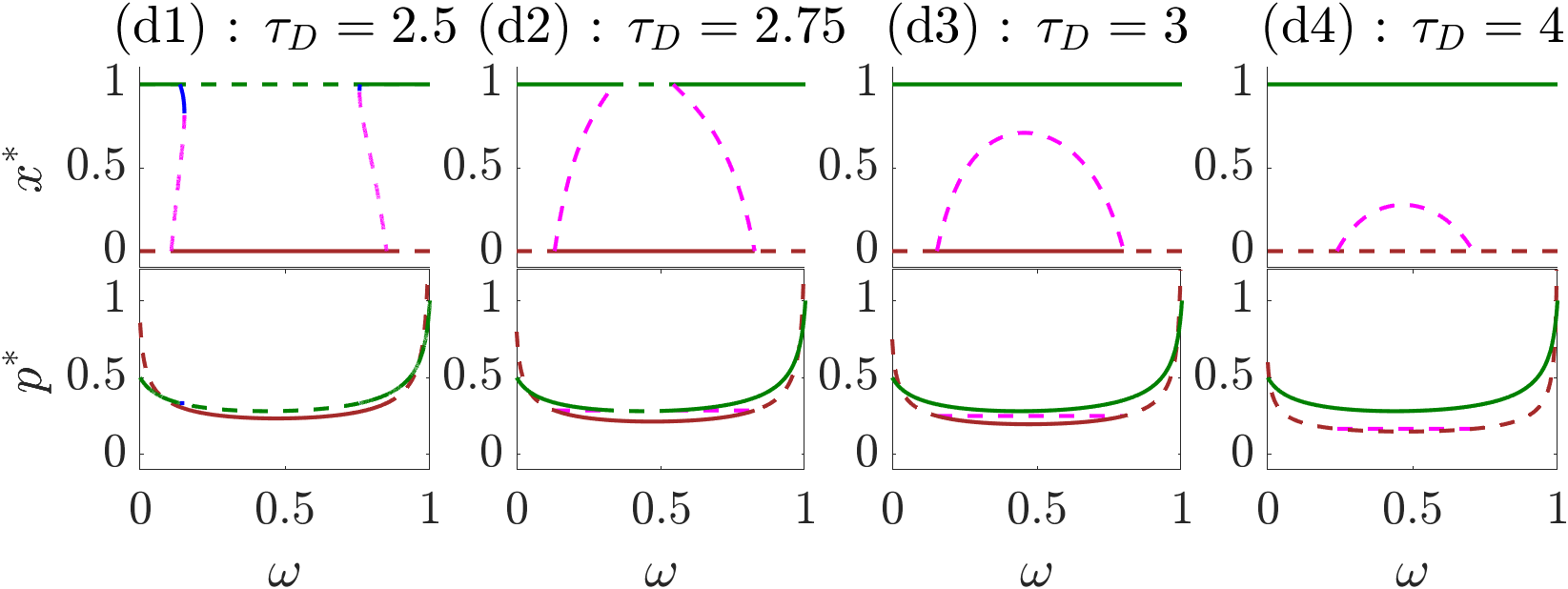}}\\
    \hline
  \end{tabular}
  \caption{Shares $x^{\ast}$and pollution level $p^{\ast}$ as $\omega$
    increases and for different values of $\tau_D$ in the case of
    different ex ante and marginal effectiveness of abatement, for intermediate
    clean technology emission level $\varepsilon_C = 0.2.$. Brown,
    green, blue and magenta are respectively related to steady states
    $\tmmathbf{\xi}_0^{\ast}, \tmmathbf{\xi}_1^{\ast},
    \tmmathbf{\xi}_b^{\ast}$ and $\tmmathbf{\xi}_a^{\ast}$. Solid and
    dashed lines respectively represent stable and unstable steady
    states. \label{fig:sc_med} }
\end{figure}

To this end, we introduce, for $i \in \{ C, D \}$, 


{\begin{equation}
  \begin{array}{l}
    z_{2, i} = \dfrac{2 \varepsilon_i}{\alpha + c_2 \tau_i  (1 - \omega)},\\
    z_{3, i} = \dfrac{2 \varepsilon_i  (\gamma + \sigma (1 - \gamma))}{2
    (\gamma + 1 + (1 - \gamma) \sigma) - (1 - \gamma)  (1 - \sigma)  (\alpha +
    c_2 \tau_i (1 - \omega))}, \;\\
    z_{4, i} = \dfrac{\varepsilon_i  (1 - 2 (\gamma + \sigma (1 - \gamma)))}{(1
    - \gamma)  (1 - \sigma)  (1 + \alpha + c_2 \tau_i (1 - \omega))}
  \end{array} 
  \label{eq:z24}
\end{equation}
and function
\begin{equation}\label{eq:c1i}
  c_{1, i} (z, \omega) = \frac{\varepsilon_i - \tau_i^2  (\alpha + c_2 (1 -
   \omega) \tau_i) z}{\chi \tau_i^4 \omega (1 - \omega) z^2}, \; i \in \{ C, D
   \}. \end{equation}

\begin{proposition}
  \label{th:stabxi0}Steady state $\tmmathbf{\xi}_0^{\ast} = (0, p_0^{\ast},
  k_0^{\ast})$ is locally asymptotically stable when
  \begin{equation}
    \omega > \frac{(1 - \gamma)  (1 - \sigma)  (\alpha + c_2 \tau_D + 2) -
    4}{c_2 \tau_D  (1 - \gamma)  (1 - \sigma)} \label{eq:stabp0w}
  \end{equation}
  provided that
  \begin{equation}
    \left\{ \begin{array}{l}
      c_1 > c_{1, D} (\bar{r}, \omega)\\
      c_1 > c_{1, D} (z_{2, D}, \omega)\\
      c_1 < c_{1, D} (z_{3, D}, \omega)\\
      c_1 < c_{1, D} (z_{4, D}, \omega)
    \end{array} \right. \label{eq:stab_xi0}
  \end{equation}
  The third condition is related to the possible emergence of a flip
  bifurcation and the fourth one to the possible emergence of a Neimark-Sacker
  bifurcation. For a given parameter configuration, the first two conditions
  provide a solution of the form $\omega \in (\omega_a, \omega_b) \cap [0,
  1]$, while the last two 
  provide a solution of the form $\omega
  \in ((- \infty, \omega_a) \cup (\omega_b, + \infty)) \cap [0, 1]$, where $\omega_a$ and $\omega_b$ have peculiar 
  expressions for each condition; intervals can be empty.
\end{proposition}

We have similar stability conditions for $\tmmathbf{\xi}_1^{\ast}$.

\begin{proposition}
  \label{th:stabxi1}Steady state $\tmmathbf{\xi}_1^{\ast} = (1, p_1^{\ast},
  k_1^{\ast})$ is locally asymptotically stable when
  \begin{equation}
    \omega > \frac{(1 - \gamma)  (1 - \sigma)  (\alpha + c_2 \tau_C + 2) -
    4}{c_2 \tau_C  (1 - \gamma)  (1 - \sigma)}
  \end{equation}
  provided that
  \begin{equation}
    \left\{ \begin{array}{l}
      c_1 < c_{1, C} (\bar{r}, \omega)\\
      c_1 > c_{1, C} (z_{2, C}, \omega)\\
      c_1 < c_{1, C} (z_{3, C}, \omega)\\
      c_1 < c_{1, C} (z_{4, C}, \omega)
    \end{array} \right. \label{eq:stab_xi1}
  \end{equation}
  The third condition is related to the possible emergence of a flip
  bifurcation and the fourth one to the possible emergence of a Neimark-Sacker
  bifurcation. For a given parameter configuration, the second condition
  provide a solution of the form $\omega \in (\omega_a, \omega_b) \cap [0,
  1]$, while the first and the last two conditions provide a solution of the
  form $\omega \in ((- \infty, \omega_a) \cup (\omega_b, + \infty)) \cap [0,
  1]$, where $\omega_a$ and $\omega_b$ have particular expressions for each
  condition; intervals can be empty.
\end{proposition}



\begin{figure}[ht!]
     \begin{tabular}{|b{1.8cm} b{13cm}|}
    \hline
       \multicolumn{2}{|c|}{High emission levels of clean
       producers }  \\
       \hline
       \rule{0pt}{55pt}
       \begin{tabular}{c}
         (a)\\         
         $c_1 = 0.03$\\
         $c_2 = 0.5$
       \end{tabular} &
       \mline{\includegraphics[width=0.8\columnwidth]{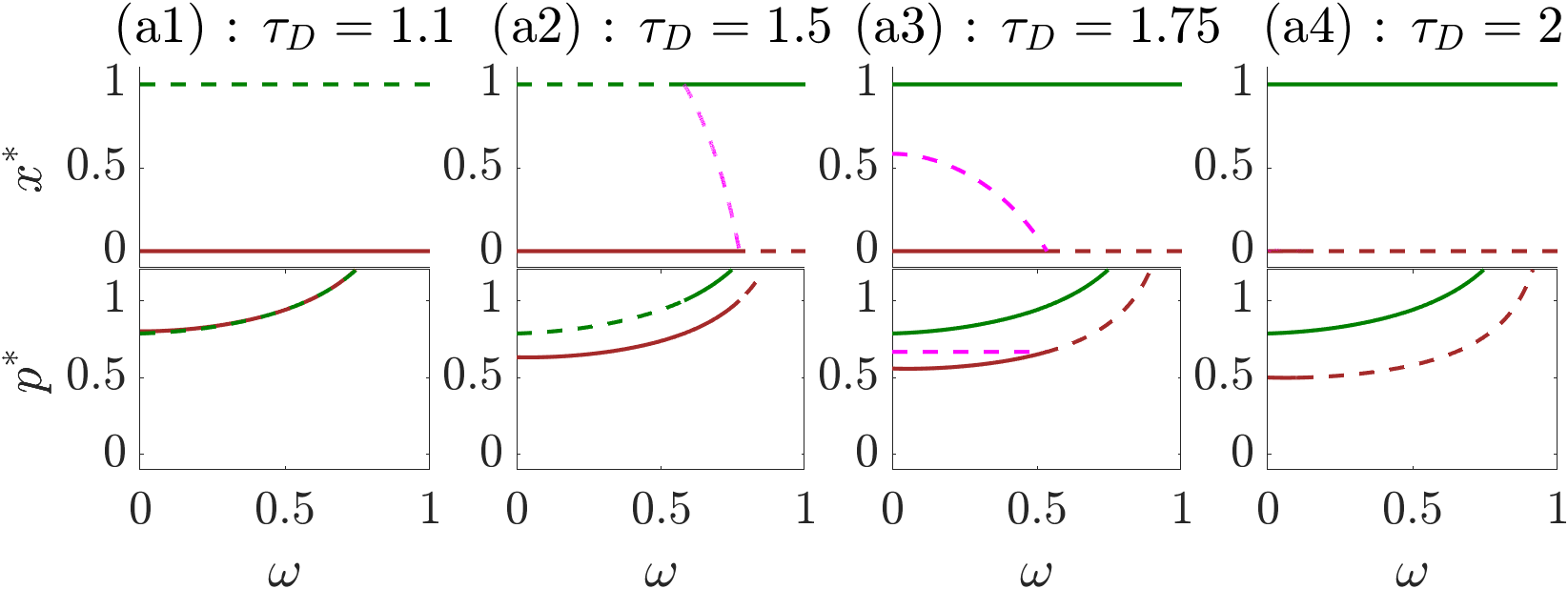}}\\
       \hline
       \rule{0pt}{55pt}
       \begin{tabular}{c}
         (b)\\   
         $c_1 = 0.07$\\
         $c_2 = 0.03$
         \end{tabular} &
    \mline{\includegraphics[width=0.8\columnwidth]{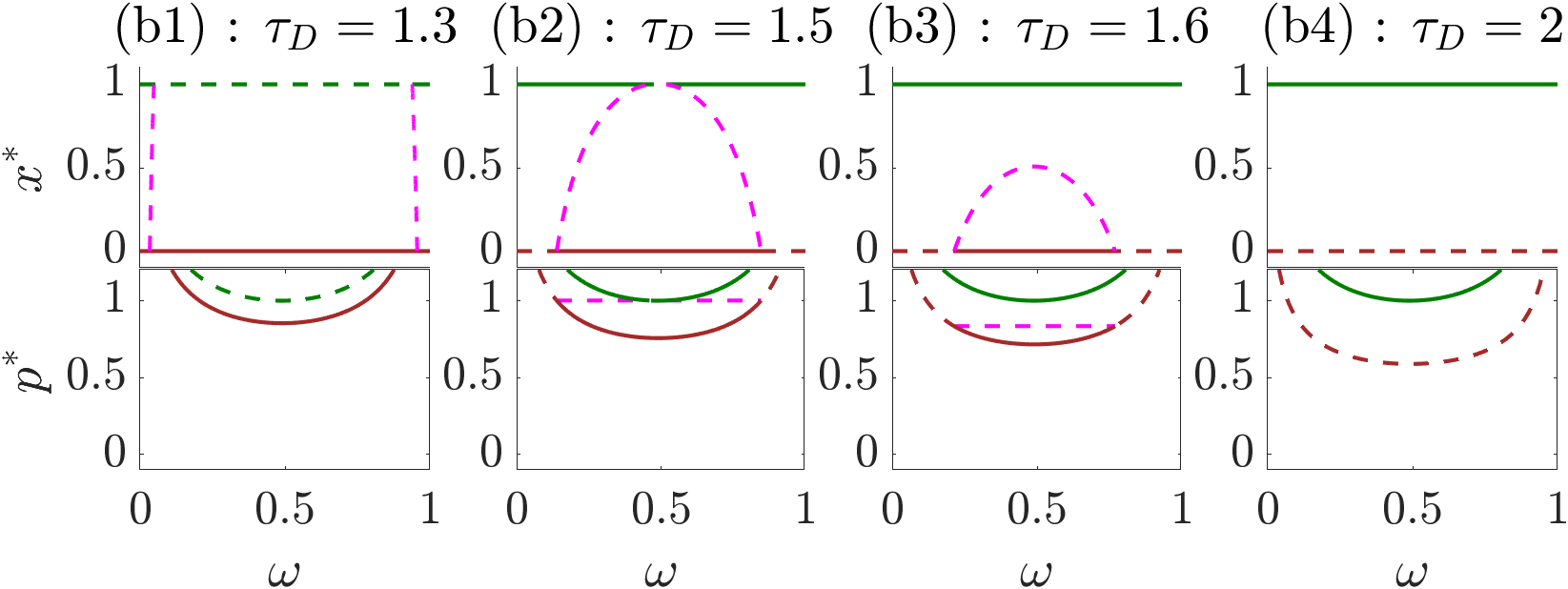}}\\
       \hline
       \rule{0pt}{55pt}
       \begin{tabular}{c}
         (c)\\ 
         $c_1 = 0.13$\\
         $c_2 = 0.1$
                       \end{tabular} &
    \mline{\includegraphics[width=0.8\columnwidth]{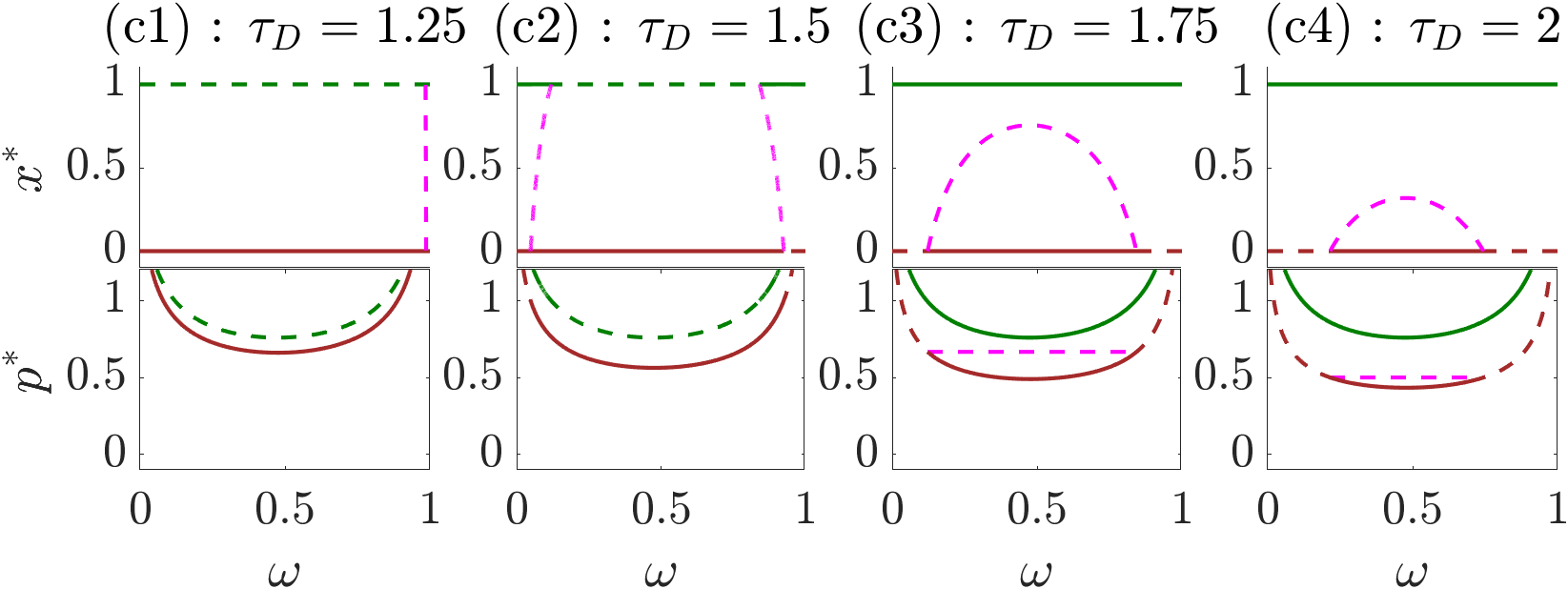}}\\
       \hline
       \rule{0pt}{55pt}
       \begin{tabular}{c}
    (d)\\
    $c_1 = 0.3$\\
    $c_2 = 0.2$
       \end{tabular}
                     &
    \mline{\includegraphics[width=0.8\columnwidth]{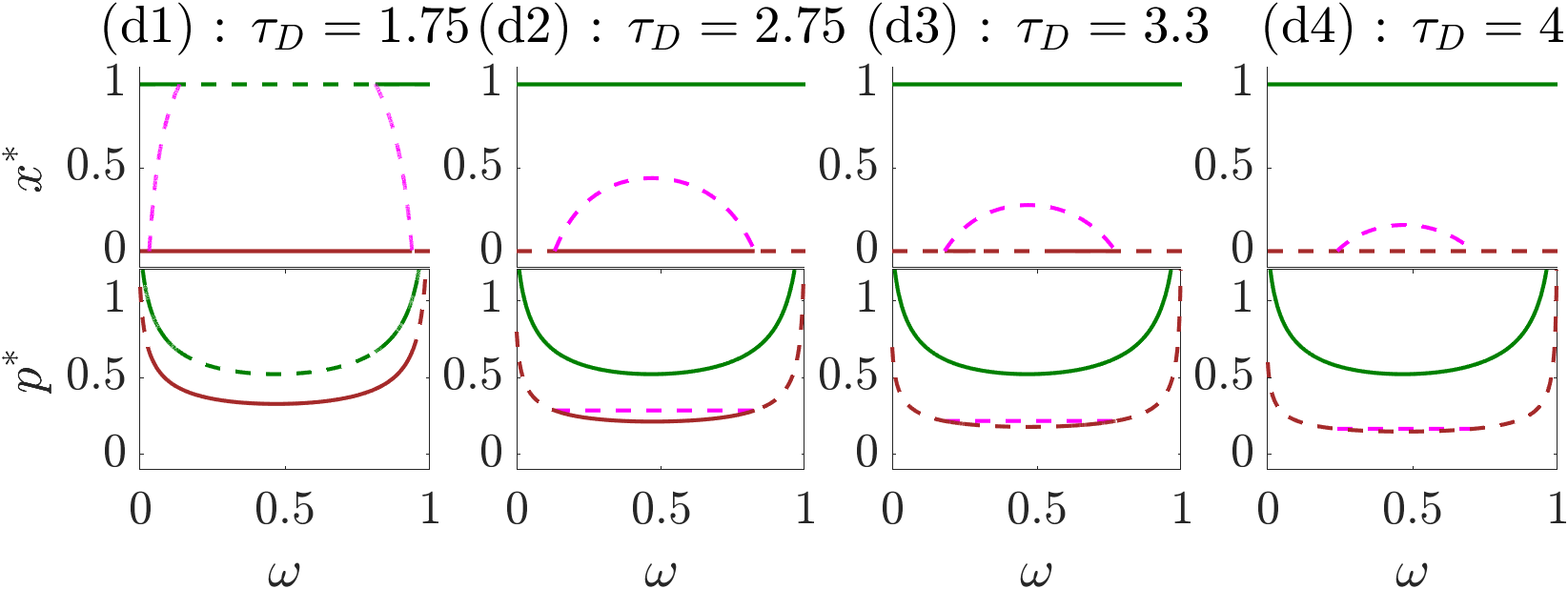}}\\
    \hline
  \end{tabular}
  \caption{Shares $x^{\ast}$and pollution level $p^{\ast}$ as $\omega$
    increases and for different values of $\tau_D$ in the case of
    different ex-ante and marginal effectiveness of abatement, for low
    clean technology emission level $\varepsilon_C = 0.55.$ Brown,
    green, blue and magenta are respectively related to steady states
    $\tmmathbf{\xi}_0^{\ast}, \tmmathbf{\xi}_1^{\ast},
    \tmmathbf{\xi}_b^{\ast}$ and $\tmmathbf{\xi}_a^{\ast}$. Solid and
    dashed lines respectively represent stable and unstable steady
    states. \label{fig:sc_large} }
\end{figure}

Stability conditions for $\tmmathbf{\xi}_0^{\ast}$ and
  $\tmmathbf{\xi}_1^{\ast}$ are expressed in terms of $c_1$, in
order to make them more explicit, thus providing information about the
stability regions with respect to $\omega$. We note that both $c_1$
and $\omega$ have an ambiguous effect on stability,  since
  they could both have stabilizing and destabilizing
role.  Moreover, stability of $\tmmathbf{\xi}_0^{\ast}$
  and $\tmmathbf{\xi}_1^{\ast}$ is not affected by the evolutionary
  pressure $\beta$.

Before discussing the results in Proposition \ref{th:stabxi0} and
\ref{th:stabxi1}, we infer some stability insights about
$\tmmathbf{\xi}_a^{\ast}$ and $\tmmathbf{\xi}_b^{\ast}$, referring to
simulations in Figures \ref{fig:sc_small}-\ref{fig:sc_large}. All the
simulations we performed for model \eqref{eq:model}, even those not
reported in the present contribution, show that
$\tmmathbf{\xi}_a^{\ast}$ is always unstable. Although such result is
not analytically achievable, this behavior can be justified based on
the characterization of $\tmmathbf{\xi}_a^{\ast}$ provided by
Proposition \ref{th:g}. In fact, if we consider a share of clean
producers slightly smaller than $x_a^{\ast}$, the pollution abatement
is stronger than emissions. This  leads pollution to
decrease, and can drive more agents to adopt the dirty technology,
 due to the reduced overall tax burden, so $x_a^{\ast}$
increases. Conversely, a share of clean
  producers slightly larger than $x_a^{\ast}$ results in a pollution
  abatement weaker than emissions. This leads pollution
to increase, and can drive more agents to adopt the clean technology,
as $\tau_D p$ increases, so $x_a^{\ast}$  increases. The
conclusion is that $\tmmathbf{\xi}_a^{\ast}$ represents a situation
 from  which agents would 
always divert
, and it does not play a relevant dynamical role in attracting
trajectories. The opposite role is played by
$\tmmathbf{\xi}_b^{\ast}$, toward which trajectories can
converge\footnote{Similarly, trajectories can converge toward an
  attractor arisen from the loss of stability of
  $\tmmathbf{\xi}_b^{\ast}$.}. For this reason, in what follows, we
focus our comments and interpretation on $\tmmathbf{\xi}_b^{\ast}$.

Now we come back to the discussion of the results in Proposition
\ref{th:stabxi0} and \ref{th:stabxi1}.  We recall that the
  choice of parameter configuration for the case studies aims at
  reducing
the possible emergence of complex dynamical behaviors for
$\tmmathbf{\xi}_0^{\ast}$ and $\tmmathbf{\xi}_1^{\ast}$, in particular
in terms of flip and Neimark-Sacker bifurcations, which is instead
investigated in Section \ref{sec:dyn_compl} by considering different
parameter settings. So now we focus on what happens when the first
conditions\footnote{We stress that a direct check using the parameters
  chosen for any reported simulation shows that the second conditions
  in \eqref{eq:stab_xi0} and \eqref{eq:stab_xi1} are never violated
  for $\omega\in [0, 1]$ and $\tau_D > \tau_C$ when the first ones
  hold true. Actually, in none of the simulations conducted did the
  second set of conditions appear to play a role in the emergence of
  bifurcations, suggesting that one can prove analytically that they
  cannot be satisfied with equality without simultaneously violating
  another condition. } in \eqref{eq:stab_xi0} and \eqref{eq:stab_xi1}
are violated.

We note that (see the proofs of Propositions \ref{th:stabxi0} and
\ref{th:stabxi1}) they correspond to, respectively,
\begin{equation}
  p_0^{\ast} <   \bar r\quad \text{and} \quad
  p_1^{\ast} >   \bar r . \label{eq:stab01}
\end{equation}
Their violation occurs for a pollution level equal to $\bar r$, which is the
  steady state pollution level characterizing
  $\tmmathbf{\xi}_b^{\ast}$. 
This suggests that,
  under the occurrence $p^{\ast}_0 = \bar r$ or $p^{\ast}_1 = \bar r$
  consequent to an increase of $\omega$, $\tmmathbf{\xi}_b^{\ast}$ coincides respectively with
$\tmmathbf{\xi}_0^{\ast}$ or $\tmmathbf{\xi}_1^{\ast}$, with a
possible switch in their stability through a transcritical bifurcation
mechanism. This is confirmed by the diagrams reported in Figures
\ref{fig:sc_med}-\ref{fig:sc_large}, 
from which becomes evident that when
$\tmmathbf{\xi}_a^{\ast}$ (magenta line) enters or leaves the feasible
region from $x = 0$ or $x = 1$, $\tmmathbf{\xi}_0^{\ast}$ (brown line)
or $\tmmathbf{\xi}_1^{\ast}$ (green line) lose or recover stability. \
We note that also in the case of $\tmmathbf{\xi}_a^{\ast}$ entering
the feasible region, we have a stability loss or gain for
$\tmmathbf{\xi}_0^{\ast}$ or $\tmmathbf{\xi}_1^{\ast}$\footnote{The
  stability recover may not take place as a consequence of the other
  stability conditions, as for example in panels (d3),(d4) of Figure
  \ref{fig:sc_small}, panel (d4) of Figure \ref{fig:sc_med}, panels
  (d3),(d4) of Figure \ref{fig:sc_large}. In all these simulations,
  this is due to the not fulfillment of the flip stability
  condition.}, while consistently with what already remarked,
$\tmmathbf{\xi}_a^{\ast}$ is unstable. The interpretation of these
stability changes is essentially tied to the underlying evolutionary
mechanism. For example, if $p_0^{\ast}$ is low (i.e., the first
condition in \eqref{eq:stab_xi0} is satisfied), the burden of
environmental taxation on the dirty technology is small, and this may
lead all agents to adopt the more profitable dirty technology. If an
increase in $\omega$ further reduces pollution, the first condition in
\eqref{eq:stab01} would still hold, so that $\tmmathbf{\xi}_0^{\ast}$
remains stable. Hence, its violation requires pollution to rise with
$\omega$ (and exceed the threshold
$\bar r = \frac{\lambda_0}{\tau_D - \tau_C}$, 
which in addition increases as $\tau_D$ approaches $\tau_C$), suggesting that a green transition is fostered by innovation only at the cost of worsening environmental conditions. Conversely, if $p_1^{\ast}$ is high (i.e., the first condition in \eqref{eq:stab_xi1} is satisfied), the high taxation burden may drive all agents to adopt the clean technology instead, leading $\tmmathbf{\xi}_1^{\ast}$ to become stable. The consequence is that if higher values of $\omega$ lower pollution  the second condition in \eqref{eq:stab01} could be violated, making $\tmmathbf{\xi}_1^{\ast}$ unstable. This indicates that research and innovation, while improving environmental quality, may instead trigger a transition towards dirty technologies, as they become more profitable 
  because of  the improved environmental quality. We can summarize the previous findings in the next outcome.

\begin{outcome}
  \label{out:gteq}As $\omega$ increases, a loss of stability of
  $\tmmathbf{\xi}_0^{\ast}$ in favor of the appearance of
  $\tmmathbf{\xi}_b^{\ast}$, or the disappearance of $\tmmathbf{\xi}_b^{\ast}$
  in favor of the stability gain for $\tmmathbf{\xi}_1^{\ast}$ can take place
  only when steady state pollution increases.
\end{outcome}

We stress that the previous outcome actually completes
the static results related to cases a), b), d) and e) of Corollary
\ref{th:static_trans} in light of the dynamical properties of the
involved steady states. Reappraising cases c) and f) of Corollary
\ref{th:static_trans} by looking at panels (d1)-(d3) of Figure
\ref{fig:sc_small} and (c2)-(c3) of Figure \ref{fig:sc_med}, we can
link the dynamical mechanism through which $\tmmathbf{\xi}_a^{\ast}$
and $\tmmathbf{\xi}_b^{\ast}$ enter/leave the feasible region with
$x^{\ast} \in (0, 1)$ to the occurrence of a fold bifurcation. This
opens  to the possibility of coexistence between stable
steady states $\tmmathbf{\xi}_b^{\ast}$ and either
$\tmmathbf{\xi}_0^{\ast}$ or $\tmmathbf{\xi}_b^{\ast}$. However, this
is not the only possible coexistence between stable steady states, as
conditions \eqref{eq:stab01} may hold simultaneously. In order
 for this to occur, from \eqref{eq:stab01}, we need
$p_0^{\ast} > p_1^{\ast}$, namely that the pollution level associated
with an all-green population must be higher than that of an all-dirty
one. Note that a similar counterintuitive occurrence  is
  in {\citet{CMN24}}.  In the present setting it also
  make the coexistence between $\tmmathbf{\xi}_0^{\ast}$ and
  $\tmmathbf{\xi}_1^{\ast}$ possible.  Looking at the
    first conditions in \eqref{eq:stab_xi0} and \eqref{eq:stab_xi1} in
    more details, we infer that such coexistence is more
    likely to occur when $\varepsilon_C$ is close enough to
  $\varepsilon_D$, namely in presence of ``dirty'' clean producers,
  or when values of $c_1$ and $c_2$ are both large enough
    and pretty similar. Both conditions imply that pollution with
  green agents can exceed that with dirty ones, since clean technology
  still produces significant emissions and, when per-unit taxation is
  inadequately small, this diverts resources from abatement innovation
  and implementation. 
  In the first case,   low per-unit taxation favors clean
  adoption as long as environmental quality is not too degraded; in
  the second, a better environmental quality makes
  dirty technology economically  reachable  despite heavier taxation.

\begin{outcome}
  \label{out:coex}Stable steady states $\tmmathbf{\xi}_0^{\ast}$ and
  $\tmmathbf{\xi}_1^{\ast}$ can coexist provided that the pollution
  level associated   to  a homogeneous population of
  clean producers is greater than that   of  a homogeneous
  population of dirty producers.
\end{outcome}

The aim of the remainder of this section is to examine, based on the
analytical findings and the related outcomes, with the help of the simulations
reported in Figures \ref{fig:sc_small}--\ref{fig:sc_large}, \ how increased
investment in innovation may (or may not) alter agents' choices, and to assess
these changes in terms of the regulator's other two targets, namely their
implications for environmental quality and economic sustainability.

Regarding taxation, we can distinguish environmental per-unit tax
regimes based on how much tax $\tau_D$ for dirty agents exceeds
$\tau_C$ for clean agents. Broadly speaking, we can arbitrarily define
three regimes,  characterized by   low
($\tau_C < 1 < \tau_D < 1.75$), medium ($1.75 < \tau_D < 3.5$), and
high ($\tau_D > 3.5$) environmental per-unit taxation. Under the
assumptions of the present model, the higher per-unit taxation is, the
more agents adopt the clean technology,  and this results
  in  an evolutionary selection (see the comments before Proposition
\ref{th:csab}).  However, this may occur for
economically unsustainable values of $\tau_D$, in particular when
clean emissions are very low (first case study).   Panels
  (a4), (b4), (c4) and (d4) show   that a small share of dirty agents
can persist even in presence of per-unit taxation, which
  is huge but not sufficient   to rule out the possibility  
  that some agents find the choice of a dirty technology still
  profitable.     This is a consequence of a  low steady
state pollution level,  due to the extremely  low
emissions of clean producers, which keeps the tax burden on dirty
agents sustainable even in the presence of a high
 level of $\tau_D$, thereby allowing a certain share of
dirty technology adoption to be still profitable. Even if these
scenarios are characterized by a very reduced pollution stock, the
economic sustainability goal is by far missed, given the high
$\tau_D$, so we do not discuss them further.

The effect  of an increase in $\omega$  on environmental
quality can be inferred looking at the lower graphs in the panels of
Figures \ref{fig:sc_small}--\ref{fig:sc_large}, from which we can
distinguish scenarios characterized by an improvement of the
environmental quality, entailing a reduction in pollution (brown or
green decreasing solid lines), by deterioration of the environmental
quality, associated with a raise in the pollution level (brown or
green increasing solid lines), or a neutral effect on the
environmental quality, always occurring in the presence of a
heterogeneous population of producers (constant solid blue line).

Note that, in some scenarios such as those reported in panels (a2) and (c2) of Figure \ref{fig:sc_small} or panel (a3) of Figure \ref{fig:ss_changes_med}, changes in pollution are negligible. Although these cases are formally included in those of improvement or deterioration, they can be embedded in the situation in which we observe a null effect of an increase in $\omega$.}

We now discuss the  possible  behaviors
  of technological choices evolution making reference   to
the upper graphs  (concerning shares evolution) in each
panel of Figures \ref{fig:sc_small}--\ref{fig:sc_large}. As $\omega$
increases, we can identify the following behaviors.
\subsubsection*{Technological transition}

We confirm the existence of a technological transition whenever the
share $x^{\ast}$ of clean producers at a unique stable steady state
changes smoothly with $\omega$. A green transition occurs when
$x^{\ast}$ increases (displayed by an upward-sloping blue line), while
a regression to dirty technologies occurs when $x^{\ast}$ decreases (
displayed by a downward-sloping blue line).

Concerning the green transition, we can identify three phases,
consisting of the onset, when 
  we pass from a homogeneous population of dirty producers (solid brown line) to a heterogeneous one having some green producers (blue line), in its progression, identified by a rising blue line and finally in its fulfillment, when all producers adopt the green technology (passing from solid blue to green lines). Conversely, a complete regression to dirty technology occurs
when in a homogeneous population of clean producers (solid green line)
agents increasingly adopt the dirty technology (solid blue line) until
a homogeneous population of brown producers (solid brown line) emerge.

We refer to panels (a2),(a3),(b1)-(b3),(c1)-(c3),(d1)-(d4) of Figure
\ref{fig:sc_small} and (a2),(a3),(b2),(b3),(c1)-(c4) of Figure
\ref{fig:sc_med} for several examples of technological transitions,
some of which are discussed in what follows.  Note that the 
 beginning and fulfillment of the transition can be
associated to a transcritical bifurcation.

We start noting that a technological transition is feasible only if
clean and dirty producers are characterized by suitably different
emission levels. In the third case study (Figure \ref{fig:sc_large}),
since $\varepsilon_C$ is close to $\varepsilon_D$, we actually never
observe technological transitions.  This can be explained as, being
the two technologies very similar in terms of emissions, it is more
likely that agents all converge compactly and very fastly  
  to  one technology or to  the other one. Conversely,
when a transition occurs, we can either have a green transition
(e.g. panels (a2) and (a3) in both Figures \ref{fig:sc_small} and
\ref{fig:sc_med}) or an initial  regression  to dirty
technologies, when $\omega$ is small, followed by a green transition,
as $\omega$ further increases (e.g. panels
(b1)-(b3),(c1)-(c3),(d1)-(d4) of Figure \ref{fig:sc_small} and
(b2),(b3),(c1)-(c4) of Figure \ref{fig:sc_med}).

The green transition scenarios occur when the ex-ante abatement level
of technology is already suitably effective even without investing in
innovation, because, in this case, increasing investment in innovation
is not an effective strategy. Doing so, from Outcome \ref{out:gteq},
we already know that the onset and fulfillment of green transition is
not compatible with a decrease in the pollution level, as well as from
Outcome \ref{out:pconstw} we know that, during the transition
progression, the environmental quality does not change.  Even under
these circumstances, raising $\omega$ can still be effective for the
regulator's objectives, as in the simulations reported in panels (a2)
and (a3) in both Figures \ref{fig:sc_small} and \ref{fig:sc_med} the
number of clean agents rises but little or no environmental
deterioration occurs, even keeping moderate the level of $\tau_D$.

If, instead, the marginal effect of new knowledge is significant when
compared to the ex-ante effectiveness in abatement (panels (b)-(d)),
the scenario becomes more complicated. In this case, increasing low
level investments in innovation may lead either to an increase of
dirty agents in an already heterogeneous population (leftmost parts of
graphs in panels (b1)-(b3),(c1)-(c3),(d1)-(d4) of Figure
\ref{fig:sc_small}) or to the   backsliding  from a
homogeneous green population to a heterogeneous one (leftmost parts of
graphs in panels (b2),(b3),(c1)-(c4) of Figure \ref{fig:sc_med}). In
both cases, after the share of green agents reaches a minimum (which
may trigger a reverse transition to a homogeneous population of dirty
agents, as shown in panels (b1),(c1),(c2),(d1) of Figure
\ref{fig:sc_small} and panels (b2),(c1) of Figure \ref{fig:sc_med})
the trend reverses as $\omega$ rises. These behaviors can be explained
in light of the emission levels of clean agents and the parameters
$c_1$ and $c_2$ that characterize abatement effectiveness. In both the
first and second case studies, emissions remain low and abatement can
still improve through innovation. Thus, higher innovation investment
allows similar pollution levels with more dirty agents, making an
all-dirty population sustainable in some cases, in which increasing
$\omega$ still enhances environmental conditions (particularly evident
for small per-unit taxation, as in panels (c1) of both Figures
\ref{fig:sc_small} and \ref{fig:sc_med}). When investments in
innovation further increase, they result ineffective for the
improvement of the environmental quality (Outcome \ref{out:w_ss}), but
this, fostering an increase of pollution, paradoxically promotes the
increase of the share of clean producers. In these scenarios, aligning
green transition and environmental improvement through taxation and
innovation policy proves challenging.

\subsubsection*{No technological transition}

These scenarios are indicated by the persistence of uniquely either
brown or green solid lines, showing that the production technology
choice is unaffected by the allocation of investments for
innovation. As predictable, the persistence of the only (stable)
homogeneous population of dirty agents occurs under low per-unit
taxation in all three case studies (e.g panels (a1)), while that of
green agents is observed at medium-to-high $\tau_D$ (e.g. panels
(a4),(b4) of Figure \ref{fig:sc_med} and (b4),(d4) of Figure
\ref{fig:sc_large})\footnote{These examples refer to cases where
  homogeneous populations arise for all values of $\omega$. However,
  as noted in the presence of transitions, homogeneous populations may
  also occur only within certain ranges of $\omega$, for which no
  transition takes place.  The discussion holds true for both
  situations.}. 

As noted in Outcome \ref{out:xbchange}, when the clean technology has
low emissions (first case study), dirty agents can survive even under
higher per-unit taxation, thanks to the low pollution level
characterizing a predominantly green population. Conversely, as
$\varepsilon_C$ increases (second and third case studies), a lower
per-unit taxation level is sufficient to stabilize
$\tmmathbf{\xi}_1^{\ast}$, since the high pollution levels, even in
presence of clean agents,  make dirty technologies
economically unsustainable.
 
In both cases,  changing $\omega$ does not foster a green
  transition, but  can improve abatement efficiency and reduce
pollution, in line with the environmental quality aim of the
regulator. The optimal policy corresponds to the distribution
$\tilde{\omega}$ from Propositions \ref{th:csxi0} and \ref{th:csxi1},
which, for the case study parameters, is slightly below $1 / 2$,
implying a roughly even allocation of resources between innovation and
implementation.

\subsubsection*{Coexistence}

\begin{figure}[h]
  \includegraphics[width=0.33\textwidth]{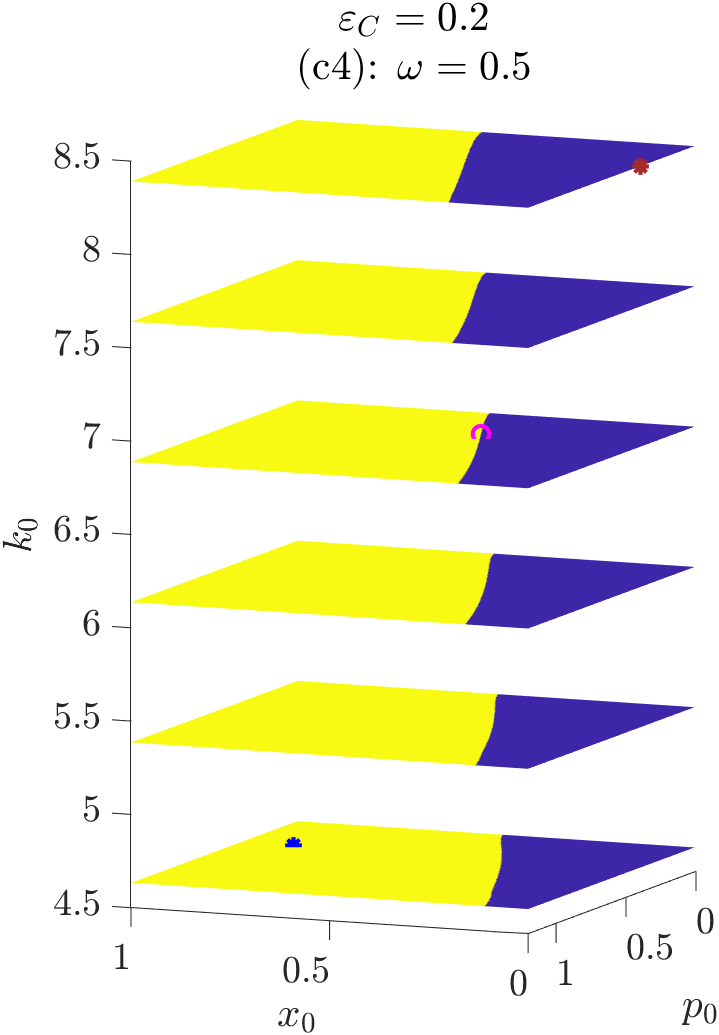}
  \includegraphics[width=0.33\textwidth]{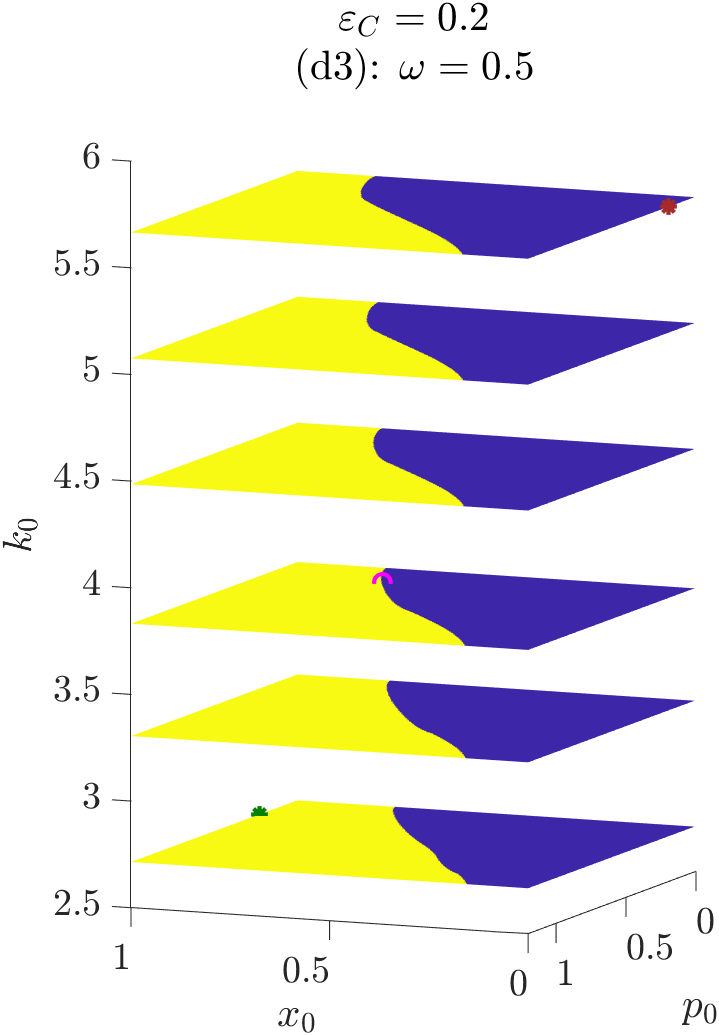}
  \includegraphics[width=0.33\textwidth]{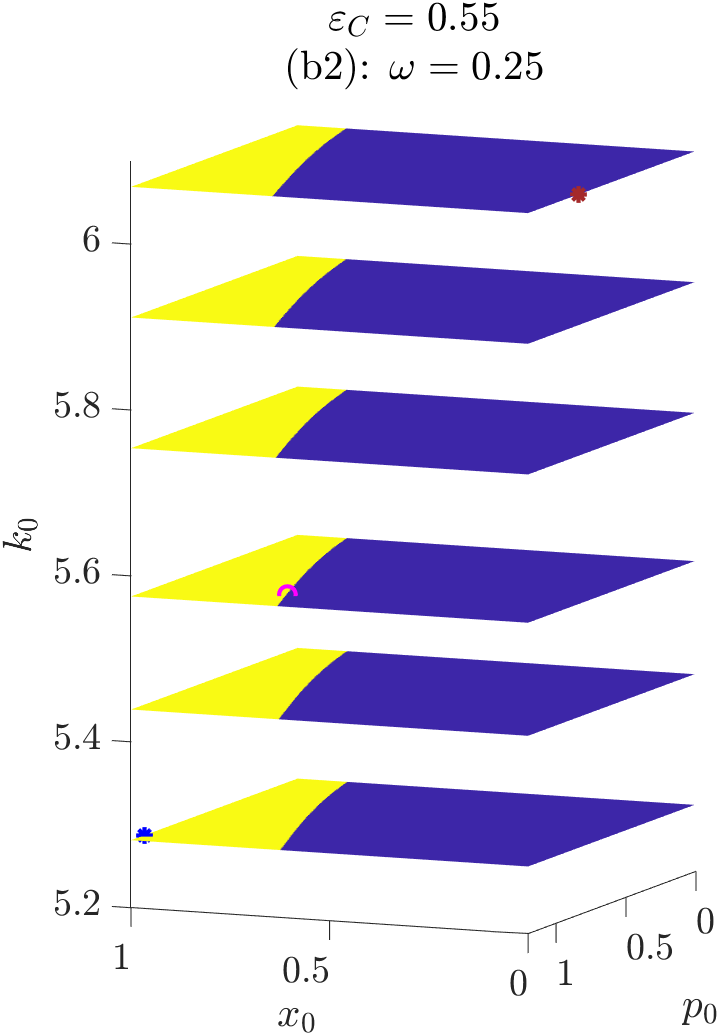}
  \caption{The blue regions depict sections of the basins of attraction of
  $\tmmathbf{\xi}_0^{\ast}$ (brown) and the yellow regions \ depict sections
  of the basins of attraction of $\tmmathbf{\xi}_1^{\ast}$ (green) or
  $\tmmathbf{\xi}_b^{\ast}$ (blue). Unstable steady state
  $\tmmathbf{\xi}_a^{\ast}$ is marked by a magenta circle. Left and middle
  panels refer to the simulations reported in panels (c4) and (d3) of Figure
  \ref{fig:sc_med}, right panel refers to the simulation reported in panel
  (b2) of Figure \ref{fig:sc_large}.\label{fig:basins}}
\end{figure}

The last framework we focus on is particularly interesting, as it
confirms the possibility of Outcome \ref{out:coex}, showing that
homogeneous populations of green and dirty producers can actually
coexist, both being simultaneously stable (green and brown lines,
solid for the same values of $\omega$, as in panels (d1)-(d3) of
Figure \ref{fig:sc_med} and (a2)-(a4),(b2),(b3),(c2)-(c4),(d2),(d3) of
Figure \ref{fig:sc_large}).  Additionally, some panels also highlight
coexistence between a heterogeneous population with a homogeneous one
of dirty producers (blue and brown lines, solid for the same values of
$\omega$, as in panels (c2)-(c4) of Figure \ref{fig:sc_med}). In this
latter case, we observe a scenario similar to that related to
coexistence between $\tmmathbf{\xi}_1^{\ast}$ and
$\tmmathbf{\xi}_0^{\ast}$, but now arising when the pollution level of
$\tmmathbf{\xi}_1^{\ast}$ is lower than that of
$\tmmathbf{\xi}_0^{\ast}$, making the latter unstable. This latter
phenomenon can be linked to the occurrence of a fold
bifurcation. Looking at the basins of attraction reported in Figure
\ref{fig:basins}, we can see that for the convergence toward either
$\tmmathbf{\xi}_1^{\ast} /\tmmathbf{\xi}_b^{\ast}$ or
$\tmmathbf{\xi}_0^{\ast}$ the initial pollution level must be
sufficiently close to that steady, as well as the population share,
and, albeit to a much lesser extent, the initial stock of
knowledge. The intuition is clear, and the behavior is relevant, as it
points out how a correct policy may be effective or not also depending
on the current situation. In particular, for its success the
environmental situation needs not to be excessively
compromised. Moreover, since the basins of attraction of the stable
attractors change with $\omega$, we may have that varying the policy
can affect the basin which the same initial state belong to. This can
lead trajectories to divert from an attractor to another, fostering
either an abrupt `green jump' or a `dirty fall'. The coexistence of
steady states or attractors makes policy choices particularly
delicate, as outcomes depend not only on the
system's intricate dynamics but also on path
dependence. Simulations indicate that this coexistence mainly arises
when clean technology has high emissions, underscoring how challenging
it is for regulators to achieve their objectives in such a
framework. Note that coexisting steady states always occur in presence
of feasible steady state $\tmmathbf{\xi}_a^{\ast}$, which is unstable
but denotes the presence of a surface, on which it lies, which
delimits the basins of attraction of the other steady states.

\subsection{Complex dynamics}\label{sec:dyn_compl}
\begin{figure}[t!]
  \begin{center}
    $\varepsilon_C = 0.002$\\
    \includegraphics[width=0.95\textwidth]{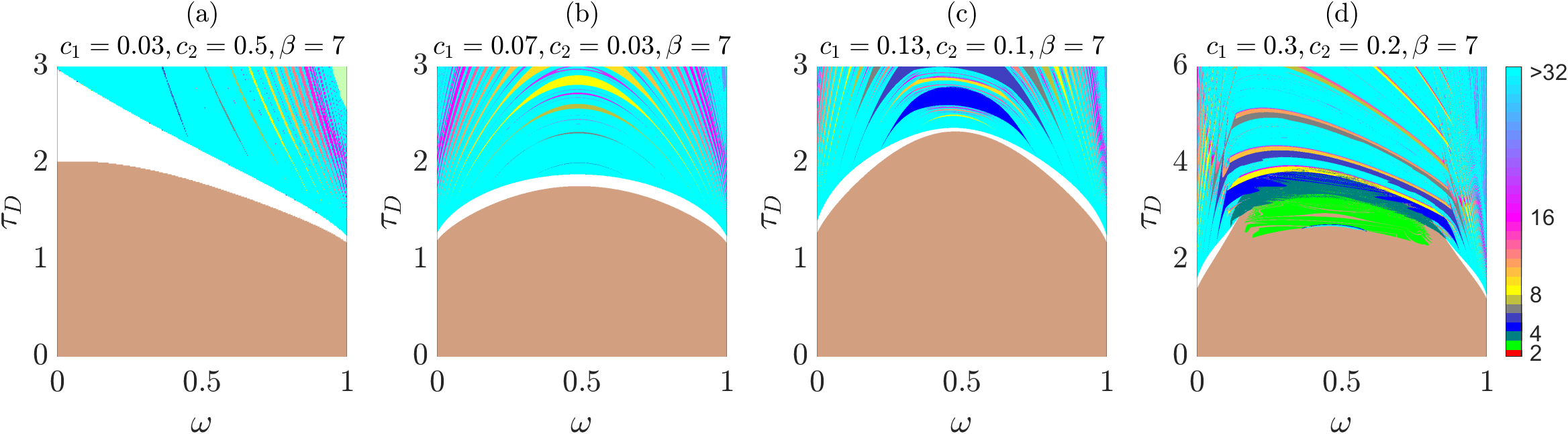}
  \end{center}
  \caption{Two dimensional bifurcation diagrams for low emission levels of
  the clean producers.\label{fig:bibif_small}}
\end{figure}

\begin{figure}[t!]
  \begin{center}
    $\varepsilon_C = 0.2$\\
 \includegraphics[width=0.95\textwidth]{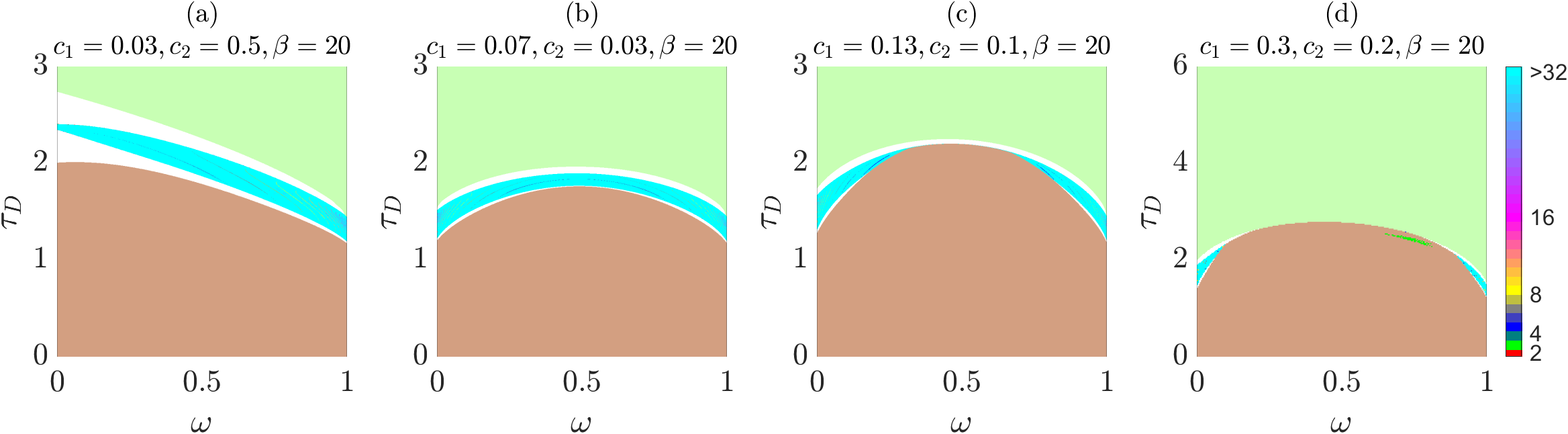}  
\end{center}  
  
  \caption{Two dimensional bifurcation diagrams for intermediate emission
  levels of the clean producers.\label{fig:bibif_med}}
\end{figure}

\begin{figure}[t!]
  \begin{center}
    $\varepsilon_C = 0.55$\\
 \includegraphics[width=0.95\textwidth]{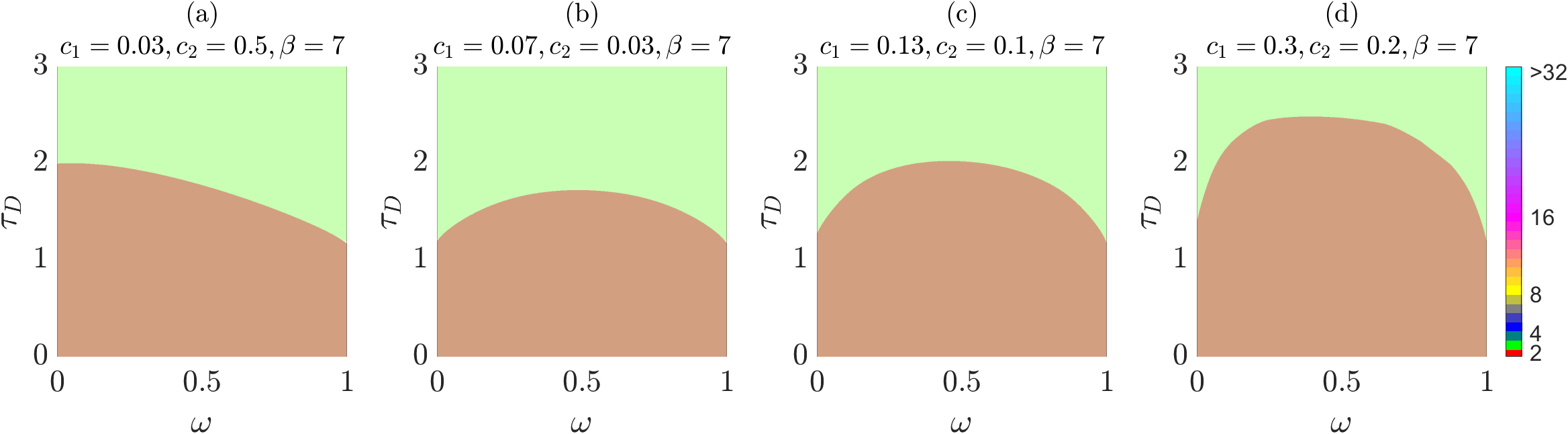}  
  \end{center}
  \caption{Two dimensional bifurcation diagrams for large emission levels of
  the clean producers.\label{fig:bibif_large}}
\end{figure}

\begin{figure}[t!]
   \includegraphics[width=0.95\textwidth]{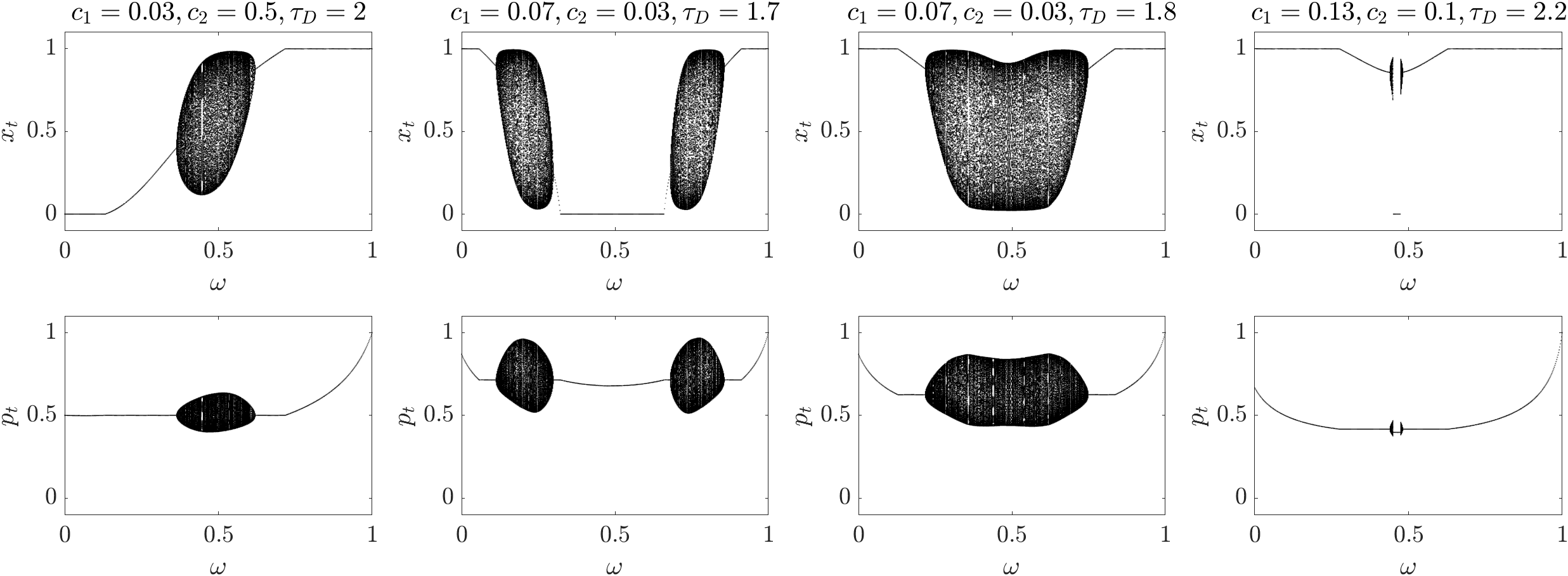}
  \caption{Bifurcation diagrams on increasing $\omega$ for some parameter
  configurations of the three case studies.\label{fig:bif_cs}}
\end{figure}

So far, we have mostly focused on the possibility of stability swapping
between stable steady states. However, both $\tmmathbf{\xi}_0^{\ast},
\tmmathbf{\xi}_1^{\ast}$ and $\tmmathbf{\xi}_b^{\ast}$ may lose stability,
giving rise to out-of-equilibrium dynamics. As we noted, in the case studies we considered parameter settings that protected $\tmmathbf{\xi}_0^{\ast}$ and
$\tmmathbf{\xi}_1^{\ast}$ from flip or Neimark--Sacker bifurcations, whereas the stability of $\tmmathbf{\xi}_b^{\ast}$ is affected by the evolutionary pressure $\beta$. In Figures \ref{fig:bibif_small}--\ref{fig:bibif_large}, we present two-dimensional bifurcation diagrams in the $(\omega, \tau_D)$ parameter space, obtained by setting $\beta = 7$. Brown, light green, and white denote the stable steady states $\tmmathbf{\xi}_0^{\ast}, \tmmathbf{\xi}_1^{\ast}$ and $\tmmathbf{\xi}_b^{\ast}$, respectively, while other colors represent attractors consisting of more than a single point (e.g., red for period-two cycles, green for period-three cycles, and so on,
with cyan indicating attractors with more than 32 points, namely, large-period
cycles, quasi-periodic, and chaotic attractors). We can highlight the direct
transitions, as $\omega$ or $\tau_D$ increase, between white and cyan regions, corresponding to Neimark--Sacker bifurcations. This is evident from the bifurcation diagrams in $\omega$ reported in Figure \ref{fig:bif_cs}, which reveal multiple bubbling phenomena, with stability lost and regained through Neimark--Sacker bifurcations as the share of resources devoted to innovation increases. We stress that this occurs most clearly in the first case study, namely for low emission levels of clean producers. For interpretation, let us assume a situation characterized by high pollution levels, consistent with a population predominantly composed of dirty producers. The evolutionary
mechanism gradually induces part of the population to adopt clean technology, leading to a reduction in pollution. If evolutionary selection is highly responsive, this decrease may result in very low pollution levels, achieved thanks to the presence of many clean agents. At this point, the opposite process is triggered, since low pollution levels make the widespread adoption of dirty technology economically sustainable. Finally, we note that the green regions within the brown ones in panel (d) of Figures \ref{fig:bibif_small}
and \ref{fig:bibif_med} denote coexistence between the stable state
$\tmmathbf{\xi}_0^{\ast}$ and a periodic attractor, which is not linked to any steady state of the model.

\begin{figure}[t!]
   \includegraphics[width=0.95\textwidth]{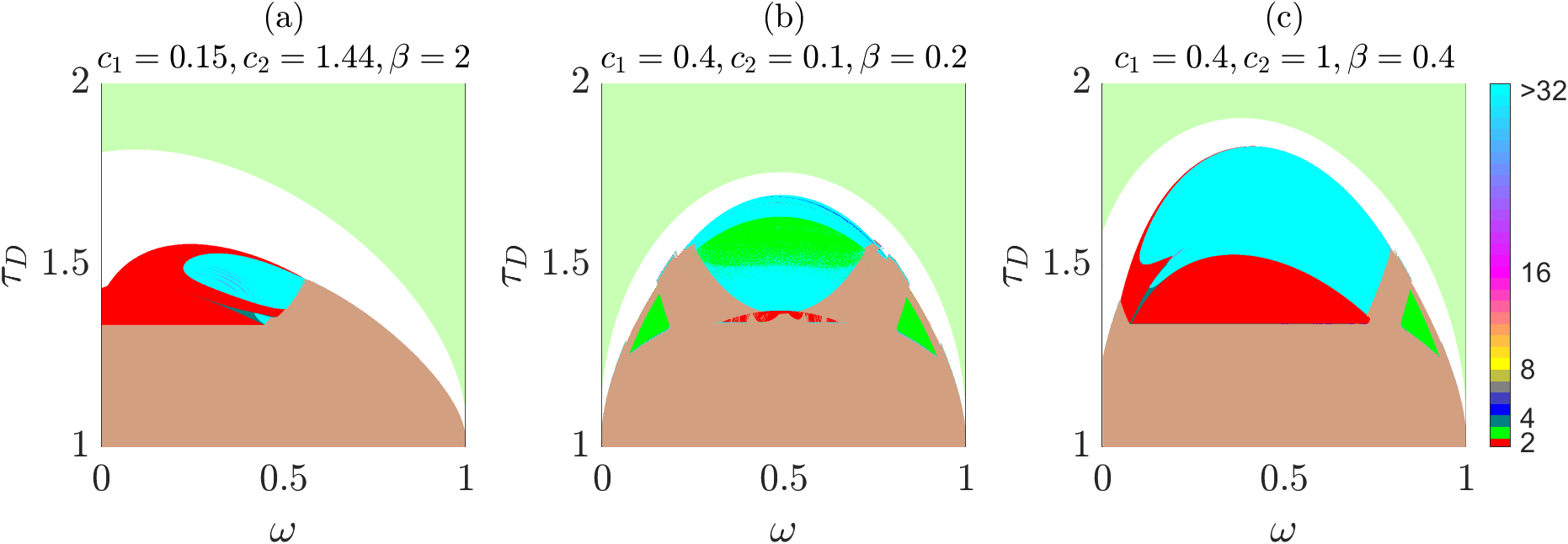}
  \caption{Two dimensional bifurcation diagrams for large emission levels of
  the clean producers.\label{fig:bibif_unstable}}
\end{figure}

\begin{figure}[t!]
   \includegraphics[width=0.95\textwidth]{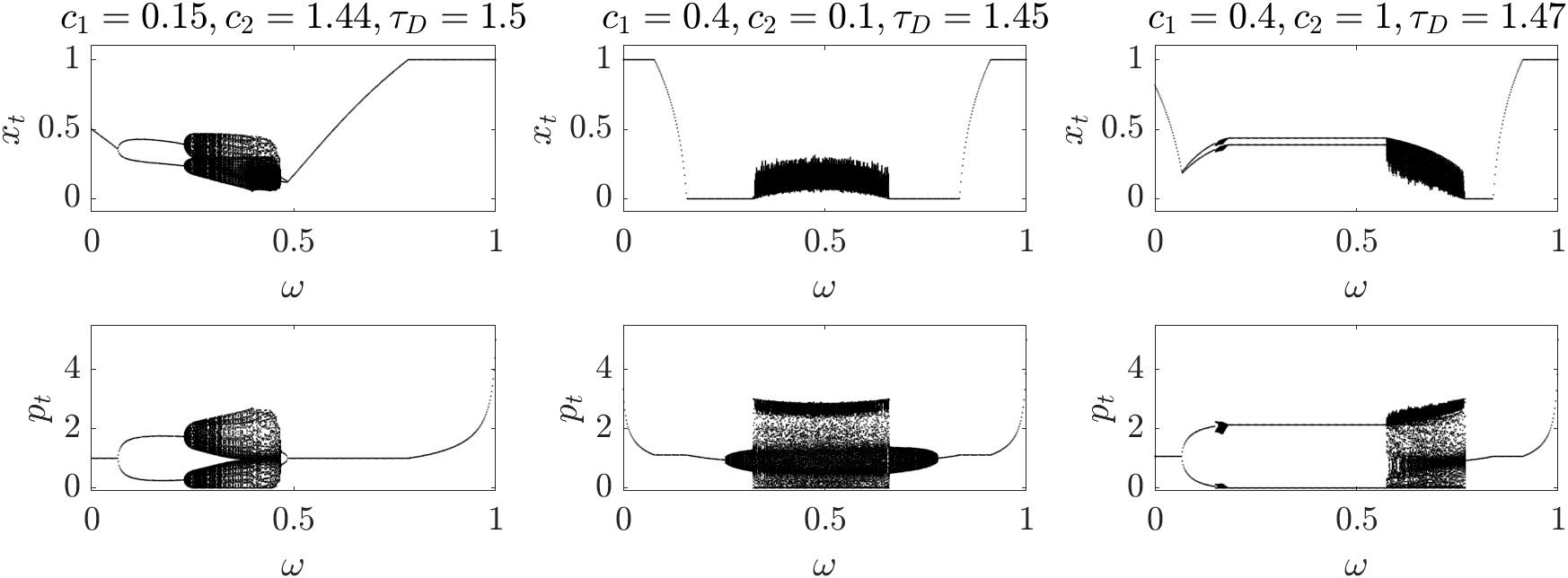}
  \caption{Bifurcation diagrams on increasing $\omega$ for some parameter
  configurations of the three case studies.\label{fig:bif_unstable}}
\end{figure}

To illustrate the possible emergence of complex dynamics related to the
destabilization of steady states characterized by homogeneous distributions of technology adoption, we must to some extent depart from the parameter settings used in the case studies. In particular, in what follows we focus on $\tmmathbf{\xi}_0^{\ast}$, since similar results and considerations can be extended to $\tmmathbf{\xi}_1^{\ast}$. We consider higher emission levels for both clean and dirty producers, setting $\varepsilon_C = 1$ and $\varepsilon_D = 3$, and we explore different configurations with respect to $c_1, c_2$, and $\beta$, as reported in the two-dimensional bifurcation diagrams of Figure 
\ref{fig:bibif_unstable}. Panel (a) provides evidence of a flip bifurcation for $\tmmathbf{\xi}_b^{\ast}$ (transition between red and white regions), which is also reported in the bifurcation diagram in panel (a) of Figure \ref{fig:bif_unstable}. The stable steady state $\tmmathbf{\xi}_b^{\ast}$ loses stability, giving rise to a period-two cycle for small values of $\omega$, which then undergoes a secondary Neimark--Sacker bifurcation. As $\omega$ further increases, stability is restored through a period-halving bifurcation. Panels (b) and (c) of Figure \ref{fig:bibif_unstable} show stability loss and/or recovery for $\tmmathbf{\xi}_0^{\ast}$ via Neimark--Sacker and flip bifurcations, respectively. In particular, looking at panel (b) of Figure \ref{fig:bif_unstable}, we observe a pair of Neimark--Sacker bifurcations affecting $\tmmathbf{\xi}_0^{\ast}$. We also
remark the occurrence of complex dynamics for intermediate values of $\omega$, with large oscillations in both the share of clean producers and the pollution levels. Finally, in panel (c) of Figure \ref{fig:bif_unstable}, we can note the loss of stability by means of a flip bifurcation of $\tmmathbf{\xi}_b^{\ast}$ for small values of $\omega$. As $\omega$ increases, the chaotic attractor arising is than replaced by stable steady state $\tmmathbf{\xi}_0^{\ast}$, which incurs a flip bifurcation for $\omega \approx 0.768.$ This shows how the transition from a heterogeneous to a homogeneous population of producers may pass from the occurrence of complex dynamics.

It is worth noting that the reported simulations employ moderate or even small values of the intensity of choice $\beta$. Indeed, the evolutionary selection mechanism plays a key role in sustaining these erratic trajectories, but they cannot be ascribed to simple overreaction phenomena.

Finally, we remark that for these parameter settings as well, we find evidence of additional coexistence phenomena, as indicated by the small green and red regions in panels (b) and (c) of Figure \ref{fig:bibif_unstable}.

\

\section{Concluding insights and outlook}\label{sec:concl}
The proposed model 
and its analysis have been widely as well as thoroughly 
discussed and interpreted within the paper, so here we limit ourselves to highlighting a few distinctive elements that underscore its complexity. If the goal is to foster a green transition through environmental taxation and its use for innovation and the implementation of systems aimed at improving environmental quality, some crucial outcomes must be taken into consideration. Firstly, investments in innovation can be effective in fostering a transition toward green technologies only in the presence of appropriate  
 fiscal choices and provided that clean technology indeed entails low emission levels. Otherwise, the element of path dependency, also emphasized in the empirical literature
({\citet{AHTZ19}}), becomes crucial for the coexistence of situations
that, in the long run, may lead either to a complete green transition
or to a full backsliding toward dirty technologies. However, even in
the case of clean technologies with low or virtually zero emission
levels, several factors may hinder the achievement of the targets. The
cleaner the low-impact technology is, the more the reduced levels of
pollution may allow for the sustainable presence of a share of dirty
producers. This possibility can become more concrete when innovation
investments improve abatement efficiency, thereby postponing the need
for structural changes toward the adoption of green
technologies. Moreover, in line with the observations of the European
Environment Agency, the reduction of environmental taxation that
occurs as the green transition progresses may lead to a decrease in
the resources allocated to the environment, including those devoted to
innovation. This, in turn, can give rise to phenomena characterized by
oscillations in the diffusion of green technologies and in
environmental quality.

The results highlight the importance of a dynamical approach to the
problem, and show how nonlinearities are crucial for understanding the
phenomena documented in the empirical literature. The present research
can be enriched in several ways. First, the economic dimension is
highly stylized. From this perspective, one could introduce a dynamic
description of the market in which agents operate, whose choices are
both influenced by and exert effects on the environmental and
evolutionary dimensions. This would make it possible to allow
producers to invest in technologies leading to a structural
improvement of production processes, and to allocate resources from
environmental taxation that provide incentives for innovation and its implementation into 
 circular economics, as stressed by   \citet{SABRNOM20}. Another relevant aspect that could be incorporated is the social dimension, since, as the literature shows, social interactions play a crucial role in either facilitating or hindering green transition policies.

\bibliography{Biblio}{}	

\begin{thebibliography}{23}
\providecommand{\natexlab}[1]{#1}
\providecommand{\url}[1]{\texttt{#1}}
\expandafter\ifx\csname urlstyle\endcsname\relax
  \providecommand{\doi}[1]{doi: #1}\else
  \providecommand{\doi}{doi: \begingroup \urlstyle{rm}\Url}\fi

\bibitem[Acemoglu et~al.(2012)Acemoglu, Aghion, Bursztyn, and Hemous]{AABH12}
D.~Acemoglu, P.~Aghion, L.~Bursztyn, and D.~Hemous.
\newblock The environment and directed technical change.
\newblock \emph{American Economic Review}, 102:\penalty0 131--166, 2012.

\bibitem[Aghion et~al.(2019)Aghion, Hepburn, Teytelboym, and Zenghelis]{AHTZ19}
P.~Aghion, C.~Hepburn, A.~Teytelboym, and D.~Zenghelis.
\newblock \emph{Handbook on Green Growth}, chapter Path dependence, innovation
  and the economics of climate change.
\newblock Edward Elgar Publishing, 2019.
\newblock \doi{10.4337/9781788110686.00011}.

\bibitem[Arilla-Llorente et~al.(2024)Arilla-Llorente, Gavurova, Rigelsky, and
  Ribeiro-Soriano]{arilla2024quantifying}
Ramon Arilla-Llorente, Beata Gavurova, Martin Rigelsky, and Domingo
  Ribeiro-Soriano.
\newblock Quantifying the dynamics of relationships between eco-innovations and
  sdg 8.
\newblock \emph{Energy Economics}, 130:\penalty0 107280, 2024.

\bibitem[Brock and Hommes(1997)]{brock1997rational}
William~A Brock and Cars~H Hommes.
\newblock A rational route to randomness.
\newblock \emph{Econometrica: Journal of the Econometric Society}, pages
  1059--1095, 1997.

\bibitem[Cavalli et~al.(2024)Cavalli, Moretto, and Naimzada]{CMN24}
Fausto Cavalli, Enrico Moretto, and Ahmad Naimzada.
\newblock Green transition and environmental quality: an evolutionary approach.
\newblock \emph{Annals of Operations Research}, pages 1009--1035, 2024.

\bibitem[Cressman(2003)]{cressman2003evolutionary}
Ross Cressman.
\newblock \emph{Evolutionary dynamics and extensive form games}, volume~5.
\newblock MIT Press, 2003.

\bibitem[D{\'\i}az-Garc{\'\i}a et~al.(2015)D{\'\i}az-Garc{\'\i}a,
  Gonz{\'a}lez-Moreno, and S{\'a}ez-Mart{\'\i}nez]{diaz2015eco}
Cristina D{\'\i}az-Garc{\'\i}a, {\'A}ngela Gonz{\'a}lez-Moreno, and Francisco~J
  S{\'a}ez-Mart{\'\i}nez.
\newblock Eco-innovation: insights from a literature review.
\newblock \emph{Innovation}, 17\penalty0 (1):\penalty0 6--23, 2015.

\bibitem[Ding et~al.(2025)Ding, Feng, and Chen]{ding2025supply}
Hongjie Ding, Muzi Feng, and Qian Chen.
\newblock How supply chain disruptions, renewable energy consumption, and
  eco-innovation mitigate environmental degradation? a path towards sustainable
  development in france.
\newblock \emph{Energy Economics}, 145:\penalty0 108346, 2025.

\bibitem[Ebaidalla(2024)]{ebaidalla2024impact}
Ebaidalla~M Ebaidalla.
\newblock The impact of taxation, technological innovation and trade openness
  on renewable energy investment: Evidence from the top renewable energy
  producing countries.
\newblock \emph{Energy}, 306:\penalty0 132539, 2024.

\bibitem[Fischer and Newell(2008)]{FN08}
C.~Fischer and R.~G. Newell.
\newblock Environmental and technology policies for climate mitigation.
\newblock \emph{Journal of Environmental Economics and Management},
  55:\penalty0 142--162, 2008.

\bibitem[Goolsbee and Jones(2021)]{GoolsbeeJones+2021}
Austan Goolsbee and Benjamin Jones.
\newblock \emph{Innovation and Public Policy}.
\newblock University of Chicago Press, Chicago, 2021.
\newblock ISBN 9780226805597.
\newblock \doi{doi:10.7208/chicago/9780226805597}.
\newblock URL \url{https://doi.org/10.7208/chicago/9780226805597}.

\bibitem[Hall(2019)]{hall2019tax}
Bronwyn~H Hall.
\newblock \emph{Tax policy for innovation}.
\newblock Number w25773. National Bureau of Economic Research, 2019.

\bibitem[Jevons(1865)]{Jev865}
W.~S. Jevons.
\newblock \emph{The {C}oal {Q}uestion: An Inquiry Concerning the Progress of
  the Nation, and the Probable Exhaustion of Our Coal-mines}.
\newblock 1865.

\bibitem[La~Torre and Marsiglio(2010)]{la2010endogenous}
Davide La~Torre and Simone Marsiglio.
\newblock Endogenous technological progress in a multi-sector growth model.
\newblock \emph{Economic Modelling}, 27\penalty0 (5):\penalty0 1017--1028,
  2010.

\bibitem[Li et~al.(2025)Li, Sun, and Zhang]{LI2025107215}
Xiaoliang Li, Shuie Sun, and Ally~Quan Zhang.
\newblock Modeling green reputation decisions in a nonlinear cournot duopoly of
  carbon emission abatement.
\newblock \emph{Economic Modelling}, 151:\penalty0 107215, 2025.
\newblock ISSN 0264-9993.
\newblock \doi{https://doi.org/10.1016/j.econmod.2025.107215}.
\newblock URL
  \url{https://www.sciencedirect.com/science/article/pii/S026499932500210X}.

\bibitem[Lyon and Maxwell(2011)]{LM11}
T.~P. Lyon and J.~W. Maxwell.
\newblock Greenwash: Corporate environmental disclosure under threat of audit.
\newblock \emph{Journal of Economics \& Management Strategy}, 20:\penalty0
  3--41, 2011.
\newblock \doi{10.1111/j.1530-9134.2010.00282.x}.

\bibitem[Rodrik(2014)]{rodrik2014green}
Dani Rodrik.
\newblock Green industrial policy.
\newblock \emph{Oxford review of economic policy}, 30\penalty0 (3):\penalty0
  469--491, 2014.

\bibitem[Roseland(2000)]{Ros00}
M.~Roseland.
\newblock Sustainable community development: integrating environmental,
  economic, and social objectives.
\newblock \emph{Progress in Planning}, 54:\penalty0 73--132, 2000.

\bibitem[Saunders(2013)]{Sau13}
H.~D. Saunders.
\newblock Historical evidence for energy efficiency rebound in 30 {U}{S}
  sectors and a toolkit for rebound analysts.
\newblock \emph{Technological Forecasting and Social Change}, 80:\penalty0
  1317--1330, 2013.
\newblock \doi{10.1016/j.techfore.2012.12.007}.

\bibitem[Sharif et~al.(2023)Sharif, Kocak, Khan, Uzuner, and
  Tiwari]{sharif2023demystifying}
Arshian Sharif, Sinem Kocak, Hafizah Hammad~Ahmad Khan, Gizem Uzuner, and Sunil
  Tiwari.
\newblock Demystifying the links between green technology innovation, economic
  growth, and environmental tax in asean-6 countries: The dynamic role of green
  energy and green investment.
\newblock \emph{Gondwana Research}, 115:\penalty0 98--106, 2023.

\bibitem[Sorrell(2009)]{Sor09}
S.~Sorrell.
\newblock Jevons’ paradox revisited: The evidence for backfire from improved
  energy efficiency.
\newblock \emph{Energy Policy}, 37:\penalty0 1456--1469, 2009.

\bibitem[Sovacool et~al.(2020)Sovacool, Ali, Bazilian, Radley, Nemery, Okatz,
  and Mulvaney]{SABRNOM20}
B.~K. Sovacool, S.~H. Ali, M.~Bazilian, B.~Radley, B.~Nemery, J.~Okatz, and
  D.~Mulvaney.
\newblock Sustainable minerals and metals for a low-carbon future.
\newblock \emph{Science}, 367:\penalty0 30--33, 2020.
\newblock \doi{10.1126/science.aaz6003}.

\bibitem[Zeppini(2015)]{zeppini2015discrete}
Paolo Zeppini.
\newblock A discrete choice model of transitions to sustainable technologies.
\newblock \emph{Journal of economic behavior \& organization}, 112:\penalty0
  187--203, 2015.

\end{thebibliography}
\bibliographystyle{plainnat}

\section*{Appendix}
\begin{proof}[Prop. \ref{th:ss}]
  To find steady states we set $x_{t + 1} = x_t = x, p_{t + 1} = p_t = p$ and
  $k_{t + 1} = k_t = k$ in \eqref{eq:model}. It is straightforward to see that
  $x = 0$ and $x = 1$ provide identities in the first equation in
  \eqref{eq:model}.
  
  If $x = 0$, we have $\bar{\tau}_t = \tau_D$, so from the third equation in
  \eqref{eq:model} we find $k = \chi \omega \tau_D p$,which replaced in the
  second equation in \eqref{eq:model} provides
  \begin{equation}
    h_0 (p, \omega) = c_1 \chi \omega (1 - \omega) \tau_D^2 p^2 + (\alpha +
    c_2 (1 - \omega) \tau_D) p - \varepsilon_D = 0. \label{eq:impxi0}
  \end{equation}
  For $\omega \neq 0, 1$, the left hand side in \eqref{eq:impxi0} represents a
  convex parabola with respect to $p$, strictly negative for $p = 0$, we have
  that \eqref{eq:impxi0} has a unique feasible solution $p_0^{\ast} \in (0, +
  \infty)$, whose expression is
  \begin{equation}
    p_0^{\ast} = \frac{- (\alpha + c_2 (1 - \omega) \tau_D) +
    \sqrt{\Delta_0}}{2 c_1 \chi \omega (1 - \omega) \tau_D^2}, \label{eq:p0}
  \end{equation}
  in which we set
  \begin{equation}
    \Delta_0 = (\alpha + c_2 (1 - \omega) \tau_D)^2 + 4 \varepsilon_D c_1 \chi
    \omega (1 - \omega) . \label{eq:Delta0}
  \end{equation}
  If $\omega = 0$ or $\omega = 1$, the left hand side in \eqref{eq:impxi0}
  represent a straight, increasing line with respect to $p$, strictly negative
  for $p = 0$, so we again have that \eqref{eq:impxi0} has a unique solution,
  whose expression is now
  \[ p_0^{\ast} = \frac{\varepsilon_D}{\alpha + c_2  (1 - \omega) \tau_D} . \]
  Similarly, when $x = 1$, we have $\bar{\tau}_t = \tau_C$, so from the third
  equation in \eqref{eq:model} we find $k = \chi \omega \tau_C p$, which
  replaced in the second equation in \eqref{eq:model} provides
  \begin{equation}
    h_1 (p, \omega) = c_1 \chi \omega (1 - \omega) \tau_C^2 p^2 + (\alpha +
    c_2 (1 - \omega) \tau_C) p - \varepsilon_C = 0 \label{eq:impxi1}
  \end{equation}
  What we said about the solution of \eqref{eq:impxi0} still applies to
  equation \eqref{eq:impxi1}, so we again have a unique feasible solution
  $p_1^{\ast} \in (0, + \infty)$, whose expression when $\omega \neq 0, 1$ is
  \[ p_1^{\ast} = \frac{- (\alpha + c_2 (1 - \omega) \tau_C) +
     \sqrt{\Delta_1}}{2 c_1 \chi \omega (1 - \omega) \tau_C^2}, \label{eq:p1}
  \]
  in which we set
  \[ \Delta_1 = (\alpha + c_2 (1 - \omega) \tau_C)^2 + 4 \varepsilon_C c_1
     \chi \omega (1 - \omega) \]
  For $\omega = 0, 1$ $p_1^{\ast}$ becomes
  \[ p_1^{\ast} = \frac{\varepsilon_C}{\alpha + c_2  (1 - \omega) \tau_C} . \]
  Assume now that $x \neq 0, 1$. The replicator mechanism is at a steady state
  provided that $p =\bar r$, while from the third equation in \eqref{eq:model} we have
  \begin{equation}
    k = \bar r(\omega (\tau_D (1 - x) + \tau_C x)) \chi \label{eq:ksint}
  \end{equation}
  which replaced in the second equation in \eqref{eq:model} provides
  \begin{equation}
    \begin{array}{l}
      - c_1 \lambda_0^2 \omega (1 - \omega) \chi x^2 + \bar r\left( 
      (1 - \omega) (c_2 (\tau_D - \tau_C) + 2 c_1 \lambda_0 \tau_D \omega
      \chi) - (\varepsilon_D - \varepsilon_C) \right) x\\
      + \varepsilon_D - \bar r\alpha -
      \bar r\tau_D  \left( c_2 + c_1\bar r  \tau_D
      \omega \chi \right)  (1 - \omega)
      = 0
    \end{array}
  \end{equation}
  The second degree equation \eqref{eq:impxi1} can be then solved by no, one
  or two values $x \in (0, 1)$.
\end{proof}

Let us introduce function
\begin{equation}
  \begin{array}{lll}
    f (x) & = & - c_1 \lambda_0^2 \omega (1 - \omega) \chi x^2 + \bar r\left( 
      (1 - \omega) (c_2 (\tau_D - \tau_C) + 2 c_1 \lambda_0 \tau_D \omega
      \chi) - (\varepsilon_D - \varepsilon_C) \right) x\\
    &  &  + \varepsilon_D - \bar r\alpha -
      \bar r\tau_D  \left( c_2 + c_1\bar r  \tau_D
      \omega \chi \right)  (1 - \omega)
      = 0 \label{eq:f}
  \end{array}
\end{equation}
To stress the relevance of the parameter under investigation, in the next
results we use notation $f (x, \tau_D)$ and $f (x, \omega)$ to highlight the
dependence of $f$ from $\tau_D$ and $\omega$, respectively.

We state the following result.

\begin{lemma}
  For each given value $x \in [0, 1]$, function $f (x, \tau_D)$ is increasing
  with respect to $\tau_D$.
\end{lemma}

\begin{proof}
  We have
  \[ \frac{\partial f}{\partial \tau_D} = - \bar r  \frac{2 c_1 \lambda_0
     \tau_C \omega \chi (\tau_D - \tau_C)  (1 - \omega) x - (\tau_D - \tau_C) 
     (\alpha + c_2 \tau_C (1 - \omega)) - 2 c_1 \lambda_0 \tau_C \tau_D \omega
     \chi (1 - \omega)}{(\tau_D - \tau_C)^2} \]
  in which the numerator is an increasing line, negative for $x = 0$. Since at
  $x = 1$ we have
  \[ - (\tau_D - \tau_C)  (\alpha + c_2 \tau_C (1 - \omega)) - 2 c_1 \lambda_0
     \tau_C^2 \omega \chi (1 - \omega) < 0 \]
  we can conclude that $\frac{\partial f}{\partial \tau_D}$ is positive for
  any $x$.
\end{proof}

\begin{proof}[Prop. \ref{th:g}]
  The shape of functions on varying $\omega$ or $x$ is evident, as well as its
  convexity for $\omega \in (0, 1)$ and concavity for $x \in [0, 1]$. In this
  latter case, a direct computation shows that the parabola with respect to
  $x$ described the right-hand-side of \eqref{eq:bal} attains its vertex at
  \[ \frac{c_2}{2 c_1 \lambda_0 \omega \chi} + \frac{\tau_D}{\tau_D - \tau_C}
     > 1, \]
  which provides the monotonicity of $x \mapsto g (x, \omega)$. Setting $c_2 =
  0$, the expression of $g$ depends on $\omega (1 - \omega)$, and hence,
  considered as a function of $\omega$, it represents a parabola with vertex
  at $\omega = \frac{1}{2}$.
\end{proof}

\begin{proof}[Prop. \ref{th:ss0}]
  If $\omega = 0$, the lhs of equation \eqref{eq:f} can be written as
  \begin{equation}
    f (x) = (\lambda_0 c_2 - \varepsilon_D + \varepsilon_C) x + \varepsilon_D   - \bar r(\alpha + \tau_D c_2)
    \label{eq:impxi2}
  \end{equation}
  Let
  \[ f (0) = f_0 = \varepsilon_D - \bar r(\alpha + \tau_D
     c_2)  \text{, {\hspace{0.17em}}} f (1) = f_1 =
     \varepsilon_C - \bar r (\alpha + \tau_C c_2)
     \text{.} \]
  Linear equation $f (x) = 0$ is impossible when $f_0 = 0$ and $f_1 \ne 0$
  while it becomes an identity if $f_0 = 0$ and $f_1 = 0$. In all other cases, equation $f (x) = 0$ has unique solution
  \begin{equation}
       \hat x =  \frac{\bar r(\alpha + \tau_D c_2) - \varepsilon_D }{\lambda_0 c_2 - \varepsilon_D + \varepsilon_C}. \label{x-hat}
  \end{equation}
  
  Four cases are now to be investigated.
  \begin{itemize}
    \item $0 < \hat{x} < 1$ when $f_0 < 0$ and $f_1 > 0$. This occurs under
    the first condition in case a). Recalling that $\tau_D > \tau_C$, this
    condition cannot be empty.
    
    These conditions provide a lower bound for $\varepsilon_C$ and an upper
    one for $\varepsilon_D$. Further, function $f (x)$ is upward sloping.
    
    \item $0 < \hat{x} < 1$ when $f_0 > 0$ and $f_1 < 0$. This occurs under
    the second condition in case a). As in the previous case, this interval
    cannot be empty.
    
    These conditions provide a upper bound for $\varepsilon_C$ and a lower one
    for $\varepsilon_D$. Further, function $f (x)$ is downward sloping.
    
    \item $\hat{x} \not\in (0 ; 1)$ when $f_0 \geq 0$ and $f_1 \geq 0$. This
    occurs under the first condition in case b). Note that these two
    inequalities permits to write condition $\varepsilon_D - \varepsilon_C >
    c_2$.
    
    Further, if $f_0 > f_1$ then $\hat{x} > 1$ while if $f_0 < f_1$ then
    $\hat{x} < 0$.
    
    \item $\hat{x} \not\in (0 ; 1)$ when $f_0 \leq 0$ and $f_1 \leq 0$. This
    occurs under the second condition in case b). Note that these two
    inequalities permits to write condition $\varepsilon_D - \varepsilon_C <
    c_2$.
    
    Further, if $f_0 > f_1$ then $\hat{x} < 0$ while if $f_0 < f_1$ then
    $\hat{x} > 0$.
  \end{itemize}
\end{proof}

\begin{proof}[Prop. \ref{th:incw}]
  From \eqref{eq:dgw} the sign of $\frac{\partial g}{\partial
  \omega}$ is determined by that of
  \begin{equation}
    \tilde{g} (\omega, x) = c_1 \chi (1 - 2 \omega)  \left[
    \bar r (\tau_D (1 - x) + \tau_C x) \right] -
    c_2 \label{eq:dgrel}
  \end{equation}
  Note that $\tilde g (\omega, x)$ is decreasing in $\omega$ and $x$ and
  $\tilde g(\omega, x) < 0$ for any $x$ when $\omega > 1 / 2$.
  
  If $\tilde{g} (0, 0)$ is negative, namely if
  \begin{equation}
    c_1 \chi \bar r\tau_D - c_2 <
    0, \label{eq:g00}
  \end{equation}
  $\tilde{g} (\omega, x)$ is negative for any $x$ and $\omega$. Recalling
  \eqref{eq:g_sign}, the last inequality is equivalent to $R_{\tau} (0, 0) <
  R_c$, from which we obtain case a).
  
  Conversely, if $\tilde{g} (0, 1)$ is positive, namely if
  \begin{equation}
    c_1 \chi \bar r\tau_C -  c_2 > 0, \label{eq:g01}
  \end{equation}
  $\tilde{g} (0, x)$ is positive for any $x$. \ Recalling \eqref{eq:g_sign},
  the last inequality is equivalent to $R_{\tau} (0, 1) > R_c$, which provides
  condition of case c). As $\omega$ increases, recalling that for $\omega > 1
  / 2$ expression $\tilde{g} (\omega, x)$ must be negative for any $x$,
  recalling the monotonicity of $\tilde{g} (\omega, x)$ in $x$, there is
  $\omega_C$ such that $\tilde{g} (\omega_C, 1) = 0$ and $\omega_D > \omega_C$
  such that $\tilde{g} (\omega_D, 0) = 0$. This means that for $\omega \in (0,
  \omega_C)$ we have $\tilde{g} (\omega, x) > 0$ for any $x$, for $\omega \in
  (\omega_C, \omega_D)$ we have that there exists $\tilde{x} (\omega)$ for
  which $\tilde{g} (\omega, \tilde{x} (\omega)) = 0$ and such that $\tilde{g}
  (\omega, x) > 0$ for $x \in (0, \tilde{x} (\omega))$ and $\tilde{g} (\omega,
  x) < 0$ for $x \in (\tilde{x} (\omega), 1)$ Moreover, for $\omega \in
  (\omega_D, 1)$, we have $\tilde{g} (\omega, x) < 0$ for any $x \in (0, 1)$.
  Solving the last three equalities provides
  \[ \omega_C = \frac{1}{2} - \frac{c_2  }{2 c_1 \bar r
     \tau_C \chi}, \hspace{0.27em} \omega_D = \frac{1}{2} - \frac{c_2  }{2 c_1\bar r \tau_D \chi} \]
  and
  \[ \tilde{x} (\omega) = \frac{\tau_D}{\tau_D - \tau_C} - \frac{c_2}{c_1
     \lambda_0 \chi (1 - 2 \omega)} \]
  which allows concluding case c).
  
  Finally, let us consider the case in which $\tilde{g} (0, 0)$ is positive
  but $\tilde{g} (0, 1)$ is negative. Based on \eqref{eq:g00} and
  \eqref{eq:g01}, this corresponds to
  \[ \left\{\begin{array}{l}
       c_1 \chi  \bar r\tau_D - c_2
       > 0\\
       c_1 \chi \bar r \tau_C - c_2
       < 0
     \end{array}\right. \quad \Leftrightarrow \quad R_{\tau} (0, 1) < R_c <
     R_{\tau} (0, 0) \]
  This provides the condition of case b), under which we have that $\tilde{g}
  (\omega, \tilde{x} (\omega)) = 0$ and $\tilde{g} (\omega, x) > 0$ for $x \in
  (0, \tilde{x} (\omega))$ and $\tilde{g} (\omega, x) < 0$ for $x \in
  (\tilde{x} (\omega), 1)$. As $\omega$ increases, we find $\omega_C$ such
  that $\tilde{g} (\omega_D, 0) = 0$, and hence we have $\tilde{g} (\omega, x)
  < 0$ for $\omega \in (\omega_C, 1)$ and for any $x \in (0, 1)$. This
  provides cases b1) and b2) and allows concluding the proof.
\end{proof}

\begin{proof}[Proof of Prop. \ref{th:csxi0}]
  The steady state pollution is a solution to equation \eqref{eq:impxi0}. If
  $c_1 = 0$ we have
  \[ p_0^{\ast} = \frac{\varepsilon_D}{\alpha + c_2  (1 - \omega) \tau_D} \]
  in which case $p_0^{\ast}$ is an increasing function of $\omega$.
  
  If $c_1 > 0$, the derivative of $p_0^{\ast}$ with respect to $\omega$ can be
  obtained by applying the implicit function theorem to $h_0 (p, \omega) = 0$,
  (function $h_0$ is defined in equation \eqref{eq:impxi0}), which provides
  \[ \frac{dp_0^{\ast}}{d \omega} = - \frac{\frac{\partial h_0}{\partial
     \omega}}{\frac{\partial h_0}{\partial p}} = - \frac{(c_1 (1 - 2 \omega)
     \tau_D \chi p_0^{\ast} - c_2) p_0^{\ast} \tau_D}{2 c_1 \tau_D^2 \omega (1
     - \omega) \chi p_0^{\ast} + \alpha + c_2 \tau_D  (1 - \omega)}, \]
  whose sign is determined by the sign of
  \[ - c_1  (1 - 2 \omega) \tau_D \chi p_0^{\ast} + c_2 . \]
  If $\omega \geq 1 / 2$, we have $\frac{dp_0^{\ast}}{d \omega} > 0$,
  conversely, if $\omega < 1 / 2$ we have
  \[ \left\{ \begin{array}{lll}
       p_0^{\ast} > \tilde{p} = \frac{c_2}{c_1 (1 - 2 \omega) \tau_D \chi} &
       \Rightarrow & \frac{dp_0^{\ast}}{d \omega} < 0\\
       p_0^{\ast} < \tilde{p} = \frac{c_2}{c_1 (1 - 2 \omega) \tau_D \chi} &
       \Rightarrow & \frac{dp_0^{\ast}}{d \omega} > 0
     \end{array} \right. \]
  If $h_0 (\tilde{p}, \omega) < 0$, we have $\tilde{p} < p_0^{\ast}$ and hence
  $\frac{dp_0^{\ast}}{d \omega} < 0$, while if $h_0 (\tilde{p}, \omega) > 0$,
  we have $\tilde{p} > p_0^{\ast}$ and hence $\frac{dp_0^{\ast}}{d \omega} >
  0$. So the sign of $h_0 (\tilde{p}, \omega)$ provides the sign of
  $dp_0^{\ast} / d \omega$. We have
  \[ \begin{array}{lll}
       h_0 (\tilde{p}, \omega) & = & [\omega^2 \tau_D  (c_2^2 - 4 \chi c_1
       \varepsilon_D) - \omega (2 c_2^2 \tau_D - 4 \chi c_1 \varepsilon_D
       \tau_D + 2 \alpha c_2)\\
       &  & + c_2^2 \tau_D - c_1 \chi \varepsilon_D \tau_D + \alpha c_2]
       \cdot \frac{1}{c_1 \tau_D  (1 - 2 \omega)^2 \chi}
     \end{array} \]
  whose sign is determined by the numerator. Let us introduce function
  $\tilde{h}_0 : [0, 1 / 2] \rightarrow \mathbb{R}, \omega \mapsto \tilde{h}_0
  (\omega)$, defined by the numerator of $h_0$, i.e.
  \[ \tilde{h}_0 (\omega) = \xi_2 \omega^2 + \xi_1 \omega + \xi_0 \]
  where
  \[ \begin{array}{lll}
       \xi_2 > 0 & \Leftrightarrow & \varepsilon_D < \frac{c_2^2}{4 \chi} =
       \varepsilon_{D, 2}\\
       \xi_1 > 0 & \Leftrightarrow & \varepsilon_D > \frac{c_2^2 \tau_D +
       \alpha c_2}{2 \chi c_1 \tau_D} = \varepsilon_{D, 1}\\
       \xi_0 > 0 & \Leftrightarrow & \varepsilon_D < \frac{c_2^2 \tau_D +
       \alpha c_2}{c_1 \chi \tau_D} = \varepsilon_{D, 0}
     \end{array} \]
  Note that $\varepsilon_{D, 2} < \varepsilon_{D, 1} < \varepsilon_{D, 0}$.
  The rightmost inequality is straightforward, while the leftmost is a
  consequence of
  \[ \varepsilon_{D, 1} - \varepsilon_{D, 2} = \frac{c_2  (2 \alpha + c_2
     \tau_D)}{4 c_1 \tau_D \chi} > 0 \]
  Moreover, also recalling that for $\omega \geq 1 / 2$ we have
  $\frac{dp_0^{\ast}}{d \omega} > 0$, we have $\tilde{h}_0 \left( \frac{1}{2}
  \right) = \frac{c_2^2 \tau_D}{4 \chi} > 0$.\\
  We summarize in the following table the possible cases depending on the
  value of $\varepsilon_D$, highlighting that $\tilde{h}_0 (0) = \xi_0$ and
  $\tilde{h}'_0 (0) = \xi_1$.
  
  \begin{minipage}{0.6\columnwidth}
    \[ \begin{array}{llcccc}
         &  & \xi_2 & \xi_1 & \xi_0 & \tilde{h}_0 (\omega) \text{on } [0, 1 /
         2]\\
         a) & \varepsilon_D < \varepsilon_{D, 2} < \varepsilon_{D, 1} <
         \varepsilon_{D, 0} & > 0 & < 0 & > 0 & \text{convex parabola}\\
         b) & \varepsilon_D = \varepsilon_{D, 2} < \varepsilon_{D, 1} <
         \varepsilon_{D, 0} & 0 & < 0 & > 0 & \text{decreasing line}\\
         c) & \varepsilon_{D, 2} < \varepsilon_D \leq \varepsilon_{D, 1} <
         \varepsilon_{D, 0} & < 0 & \leq 0 & > 0 & \text{concave parabola}\\
         d) & \varepsilon_{D, 2} < \varepsilon_{D, 1} < \varepsilon_D \leq
         \varepsilon_{D, 0} & < 0 & > 0 & \geq 0 & \text{concave parabola}\\
         e) & \varepsilon_{D, 2} < \varepsilon_{D, 1} < \varepsilon_{D, 0} <
         \varepsilon_D & < 0 & > 0 & < 0 & \text{concave parabola}
       \end{array} \]
  \end{minipage}%
  {\hspace{6em}}\begin{minipage}{0.3\columnwidth}
    \includegraphics[width=0.8 \textwidth]{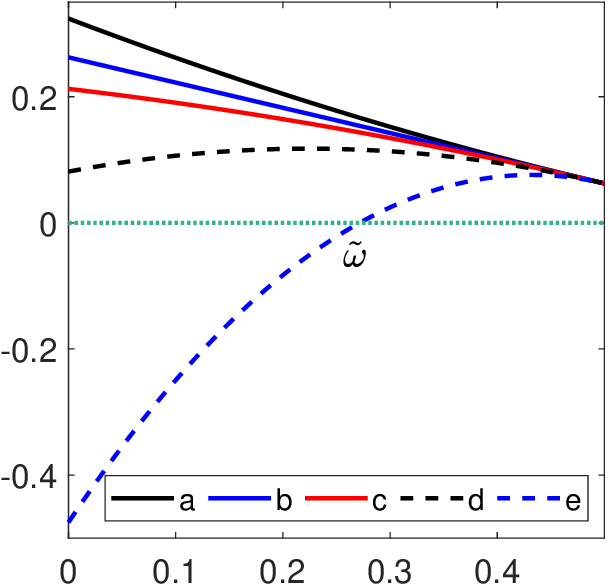}
  \end{minipage}
  
 Note that all the curves have the same positive value for $\omega = 1 / 2$.
  
  Case a)
  
  Function $\tilde{h}_0$ is a convex parabola, positive and decreasing at
  $\omega = 0$. It attains its minimum at
  \[ \omega_V = \frac{c_2^2 \tau_D - 2 c_1 \varepsilon_D \tau_D \chi + \alpha
     c_2}{c_2^2 \tau_D - 4 c_1 \varepsilon_D \tau_D \chi} \]
  Since $\omega_V > 1 / 2$ can be rewritten as
  \[ \frac{c_2  (2 \alpha + c_2 \tau_D)}{2 \tau_D  (c_2^2 - 4 \chi c_1
     \varepsilon_D)} = \frac{c_2  (2 \alpha + c_2 \tau_D)}{2 \tau_D \xi_2} > 0.
  \]
  which is true since $\xi_2 > 0$, we have that $\tilde{h}_0 (\omega) > 0$.
  
  Case b)
  
  Function $\tilde{h}_0$ is a decreasing straight line, strictly positive at
  $\omega = 0$ and $\omega = 1 / 2$, and hence $\tilde{h}_0 (\omega) > 0$.
  
  Cases c,d)
  
  Function $\tilde{h}_0$ is a concave parabola, positive at $\omega = 0$.
  Independently of the monotonicity of $\tilde{h}_0$ at $\omega = 0$, we have
  $\tilde{h}_0 (\omega) \geq 0$
  
  Case e)
  
  Function $\tilde{h}_0$ is a concave parabola, strictly negative at $\omega =
  0$. Recalling that $\tilde{h}_0  (1 / 2) > 0$, there exists a unique
  $\tilde{\omega} \in (0, 1 / 2)$ such that $\tilde{h}_0 (\omega) < 0$ for
  $\omega \in (0, \tilde{\omega})$ and $\tilde{h}_0 (\omega) > 0$ for $\omega
  \in (\tilde{\omega}, 1 / 2)$.
\end{proof}

\begin{proof}[Proof of Prop. \ref{th:csxi1}]
  The steady state pollution is a solution to equation \eqref{eq:impxi1}. If
  $c_1 = 0$ we have
  \[ p_1^{\ast} = \frac{\varepsilon_C}{\alpha + c_2  (1 - \omega) \tau_C} \]
  in which case $p_1^{\ast}$ is an increasing function of $\omega$.
  
  If $c_1 > 0$, the derivative of $p_1^{\ast}$ with respect to $\omega$ can be
  obtained by applying the implicit function theorem to $h_1 (p, \omega) = 0$,
  (function $h_1$ is defined in equation \eqref{eq:impxi1}), which provides
  \[ \frac{dp_1^{\ast}}{d \omega} = - \frac{\frac{\partial h_1}{\partial
     \omega}}{\frac{\partial h_1}{\partial p}} = - \frac{(c_1 (1 - 2 \omega)
     \tau_C \chi p_1^{\ast} - c_2) p_1^{\ast} \tau_D}{2 c_1 \tau_C^2 \omega (1
     - \omega) \chi p_1^{\ast} + \alpha + c_2 \tau_C  (1 - \omega)}, \]
  whose sign is determined by the sign of
  \[ - c_1  (1 - 2 \omega) \tau_C \chi p_1^{\ast} + c_2 \]
  If $\omega \geq 1 / 2$, we have $\frac{dp_1^{\ast}}{d \omega} > 0$,
  conversely, if $\omega < 1 / 2$ we have
  \[ \left\{ \begin{array}{lll}
       p_1^{\ast} > \tilde{p} = \frac{c_2}{c_1 (1 - 2 \omega) \tau_C \chi} &
       \Rightarrow & \frac{dp_1^{\ast}}{d \omega} < 0\\
       p_1^{\ast} < \tilde{p} = \frac{c_2}{c_1 (1 - 2 \omega) \tau_C \chi} &
       \Rightarrow & \frac{dp_1^{\ast}}{d \omega} > 0
     \end{array} \right. \]
  If $h_1 (\tilde{p}, \omega) < 0$, we have $\tilde{p} < p_1^{\ast}$ and hence
  $\frac{dp_1^{\ast}}{d \omega} < 0$, while if $h_1 (\tilde{p}, \omega) > 0$,
  we have $\tilde{p} > p_1^{\ast}$ and hence $\frac{dp_1^{\ast}}{d \omega} >
  0$. So the sign of $h_1 (\tilde{p}, \omega)$ provides the sign of
  $dp_1^{\ast} / d \omega$. We have
  \[ \begin{array}{lll}
       h_1 (\tilde{p}, \omega) & = & [\omega^2 \tau_C  (c_2^2 - 4 \chi c_1
       \varepsilon_C) - \omega (2 c_2^2 \tau_C - 4 \chi c_1 \varepsilon_C
       \tau_C + 2 \alpha c_2)\\
       &  & + c_2^2 \tau_C - \chi \varepsilon_C \tau_C + \alpha c_2] \cdot
       \frac{1}{c_1 \tau_C  (1 - 2 \omega)^2 \chi}
     \end{array} \]
  whose sign is determined by the numerator. Let us introduce function
  $\widetilde{h_1} : [0, 1 / 2] \rightarrow \mathbb{R}, \omega \mapsto
  \tilde{h}_1 (\omega)$, defined by the numerator of $h_1$, i.e.
  \[ \tilde{h}_1 (\omega) = \xi_2 \omega^2 + \xi_1 \omega + \xi_0 \]
  where
  \[ \begin{array}{lll}
       \xi_2 > 0 & \Leftrightarrow & \varepsilon_C < \frac{c_2^2}{4 \chi c_1}
       = \varepsilon_{C, 2}\\
       \xi_1 > 0 & \Leftrightarrow & \varepsilon_C > \frac{c_2^2 \tau_C +
       \alpha c_2}{2 \chi c_1 \tau_C} = \varepsilon_{C, 1}\\
       \xi_0 > 0 & \Leftrightarrow & \varepsilon_C < \frac{c_2^2 \tau_C +
       \alpha c_2}{c_1 \chi \tau_C} = \varepsilon_{C, 0}
     \end{array} \]
  Note that $\varepsilon_{C, 2} < \varepsilon_{C, 1} < \varepsilon_{C, 0}$.
  The rightmost inequality is straightforward, while the leftmost is a
  consequence of
  \[ \varepsilon_{C, 1} - \varepsilon_{C, 2} = \frac{c_2  (2 \alpha + c_2
     \tau_C)}{4 c_1 \tau_C \chi} > 0 \]
  Moreover, also recalling that for $\omega \geq 1 / 2$ we have
  $\frac{dp_1^{\ast}}{d \omega} > 0$, we have $\tilde{h}_1 \left( \frac{1}{2}
  \right) = \frac{c_2^2 \tau_C}{4 \chi} > 0$. We summarize in the following
  table the possible cases depending on the value of $\varepsilon_C$,
  highlighting that $\tilde{h}_1 (0) = \xi_0$ and $\tilde{h}_1' (0) = \xi_1$.
  
  \begin{minipage}{0.6\columnwidth}
    \[ \begin{array}{llcccc}
         &  & \xi_2 & \xi_1 & \xi_0 & \tilde{h}_1 (\omega) \text{on } [0, 1 /
         2]\\
         a) & \varepsilon_C < \varepsilon_{C, 2} < \varepsilon_{C, 1} <
         \varepsilon_{C, 0} & > 0 & < 0 & > 0 & \text{convex parabola}\\
         b) & \varepsilon_C = \varepsilon_{C, 2} < \varepsilon_{C, 1} <
         \varepsilon_{C, 0} & 0 & < 0 & > 0 & \text{decreasing line}\\
         c) & \varepsilon_{C, 2} < \varepsilon_C \leq \varepsilon_{C, 1} <
         \varepsilon_{C, 0} & < 0 & \leq 0 & > 0 & \text{concave parabola}\\
         d) & \varepsilon_{C, 2} < \varepsilon_{C, 1} < \varepsilon_D \leq
         \varepsilon_{C, 0} & < 0 & > 0 & \geq 0 & \text{concave parabola}\\
         e) & \varepsilon_{C, 2} < \varepsilon_{C, 1} < \varepsilon_{C, 0} <
         \varepsilon_C & < 0 & > 0 & < 0 & \text{concave parabola}
       \end{array} \]
  \end{minipage}%
  {\hspace{6em}} \begin{minipage}{0.3\columnwidth}
    \includegraphics[width=0.8\textwidth]{proof_p0.eps}
  \end{minipage}
  
  Note that all the curves have the same positive value for $\omega = 1 / 2$.
  
  Case a)
  
  Function $\tilde{h}_1$ is a convex parabola, positive and decreasing at
  $\omega = 0$. It attains its minimum at
  \[ \omega_V = \frac{c_2^2 \tau_C - 2 c_1 \varepsilon_C \tau_C \chi + \alpha
     c_2}{c_2^2 \tau_C - 4 c_1 \varepsilon_C \tau_C \chi} \]
  Since $\omega_V > 1 / 2$ can be rewritten as
  \[ \frac{c_2  (2 \alpha + c_2 \tau_C)}{2 \tau_C  (c_2^2 - 4 \chi c_1
     \varepsilon_D)} = \frac{c_2  (2 \alpha + c_2 \tau_C)}{2 \tau_C \xi_2} > 0
  \]
  which is true since $\xi_2 > 0$, we have that $\tilde{h}_1 (\omega) > 0$.
  
  Case b)
  
  Function $\tilde{h}_1$ is a decreasing straight line, strictly positive at
  $\omega = 0$ and $\omega = 1 / 2$, and hence $\tilde{h}_1 (\omega) > 0$.
  
  Cases c,d)
  
  Function $g$ is a concave parabola, positive at $\omega = 0$. Independently
  of the monotonicity of $\tilde{h}_1$ at $\omega = 0$, we have $\tilde{h}_1
  (\omega) \geq 0$.
  
  Case e)
  
  Function $g$ is a concave parabola, strictly negative at $\omega = 0$.
  Recalling that $\tilde{h}_1  (1 / 2) > 0$, there exists a unique
  $\tilde{\omega} \in (0, 1 / 2)$ such that $\tilde{h}_1 (\omega) < 0$ for
  $\omega \in (0, \tilde{\omega})$ and $\tilde{h}_1 (\omega) > 0$ for $\omega
  \in (\tilde{\omega}, 1 / 2)$.
\end{proof}

\begin{proof}[Prop. \ref{th:csab}]
  Recalling \eqref{eq:bal0} and that $x_a^{\ast} \leq x_b^{\ast}$, the
  behavior of $x_a^{\ast}$ and $x_b^{\ast}$ can be obtained by simple
  geometrical considerations based on the possible monotonicity behavior of
  parabolic convex function $g$ reported in Proposition \ref{th:incw}.
  Concerning the role of $\tau_D$, a direct check shows that $\frac{\partial
  g}{\partial \tau_D}  (x, \omega) > 0$, and hence simple geometrical
  considerations again allows concluding.
\end{proof}

To prove Propositions \ref{th:stabxi0} and \ref{th:stabxi1} we compute the
Jacobian matrix of \eqref{eq:model}
\[ J = \left( \begin{array}{ccc}
     J_{11} & J_{12} & 0\\
     J_{21} & J_{22} & J_{23}\\
     J_{31} & J_{32} & J_{33}
   \end{array} \right) \]

where
\[ J_{11} = \frac{e^{\beta (\lambda_0 - p(\tau_D - \tau_C))}}{(x + (1-x)e^{\beta (\lambda_0 - p(\tau_D - \tau_C))})^2}, \hspace{0.27em} J_{12} = \frac{\beta (\tau_D
   - \tau_C)x(1-x)e^{\beta (\lambda_0 - p( \tau_D - \tau_C))}}{(x + (1 - x)e^{\beta (\lambda_0 - p (\tau_D - \tau_C))} )^2}\]
\[ J_{21} = \varepsilon_C - \varepsilon_D + p (\tau_D - \tau_C)  (1 - \omega) 
   (c_2 + c_1 k), \hspace{0.27em} J_{22} = 1 - \alpha - (1 - \omega)\bar{\tau}(x)  (c_2 + c_1 k) \]
\[ J_{23} = - c_1 (1 - \omega)p \bar{\tau}(x) \]
\[ J_{31} = - \frac{dk^{\gamma}  (p \omega)^{1 - \gamma}  (\tau_D - \tau_C) 
   (1 - \gamma)}{(\bar{\tau}(x))^{\gamma}}, \hspace{0.27em} J_{32}
   = \frac{dk^{\gamma} \omega^{1 - \gamma}  (\bar{\tau}(x))^{1 -
   \gamma}  (1 - \gamma)}{p^{\gamma}} \]
\[ J_{33} = \sigma + d \gamma \left( \frac{p \bar{\tau}(x)\omega}{k}  \right)^{1 - \gamma} \]

\[ J_{11} = \frac{e^{\beta p (\tau_C + \tau_D + \lambda_0)}}{(e^{\beta
   (\lambda_0 + p \tau_C)} (1 - x) + xe^{\beta p \tau_D})^2}, \hspace{0.27em}
   J_{12} = \frac{\beta xe^{\beta (\lambda_0 + p \tau_C - p \tau_D)}  (\tau_D
   - \tau_C)  (1 - x)}{(x + e^{\beta (\lambda_0 + p \tau_C - p \tau_D)} (1 -
   x))^2} \]
\[ J_{21} = \varepsilon_C - \varepsilon_D + p (\tau_D - \tau_C)  (1 - \omega) 
   (c_2 + c_1 k), \hspace{0.27em} J_{22} = 1 - \alpha - (\tau_C x + \tau_D (1
   - x))  (1 - \omega)  (c_2 + c_1 k) \]
\[ J_{23} = - c_1 p (\tau_C x + \tau_D (1 - x))  (1 - \omega) \]
\[ J_{31} = - \frac{dk^{\gamma}  (p \omega)^{1 - \gamma}  (\tau_D - \tau_C) 
   (1 - \gamma)}{(\tau_C x + \tau_D (1 - x))^{\gamma}}, \hspace{0.27em} J_{32}
   = \frac{dk^{\gamma} \omega^{1 - \gamma}  (\tau_C x + \tau_D (1 - x))^{1 -
   \gamma}  (1 - \gamma)}{p^{\gamma}} \]
\[ J_{33} = \sigma + d \gamma \left( \frac{p \omega}{k} (\tau_C x + \tau_D (1
   - x)) \right)^{1 - \gamma} \]
Based on the expression of $J$, we study local asymptotic stability.

\begin{proof}[Proof of Prop. \ref{th:stabxi0}]
  We recall that at $\tmmathbf{\xi}_0^{\ast}$ we have $x^{\ast}_0 = 0,
  k^{\ast}_0 = \chi \omega \tau_D p^{\ast}_0$, with $p^{\ast}_0$ solution to
  \eqref{eq:impxi0}, i.e.
  \[ h_0 (p^{\ast}_0, \omega) = c_1 \chi \omega (1 - \omega) \tau_D^2
     (p^{\ast}_0)^2 + (\alpha + c_2 (1 - \omega) \tau_D) p^{\ast}_0 -
     \varepsilon_D = 0 \]
  Using the expression of $k_0^{\ast}$ in the previous equation we can write
  \[ h_0 (p^{\ast}_0, \omega) = c_1  (1 - \omega) \tau_D p^{\ast}_0 k_0^{\ast}
     + (\alpha + c_2 (1 - \omega) \tau_D) p^{\ast}_0 - \varepsilon_D = 0 \]
  from which we can obtain
  \begin{equation}
    c_1 k_0^{\ast} + c_2 = \frac{1}{(1 - \omega) \tau_D}  \left(
    \frac{\varepsilon_D}{p_0^{\ast}} - \alpha \right) \label{eq:th0astp}
  \end{equation}
  We have
  \[ J_0^{\ast} = \left( \begin{array}{ccc}
       \frac{1}{e^{\beta (\lambda_0 - p_0^{\ast} (\tau_D - \tau_C))}} & 0 &
       0\\
       \varepsilon_C - \varepsilon_D + p_0^{\ast} (\tau_D - \tau_C) (1 -
       \omega) (c_2 + c_1 k^{\ast}_0) & 1 - \alpha - \tau_D (1 - \omega) (c_2
       + c_1 k^{\ast}_0) & - c_1 p^{\ast}_0 \tau_D (1 - \omega)\\
       - \frac{d (k^{\ast}_0)^{\gamma} (p^{\ast}_0)^{1 - \gamma} \omega^{1 -
       \gamma} (\tau_D - \tau_C) (1 - \gamma)}{\tau_D^{\gamma}} & \frac{d
       (k^{\ast}_0)^{\gamma} \omega^{1 - \gamma} \tau_D^{1 - \gamma} (1 -
       \gamma)}{(p^{\ast}_0)^{\gamma}} & \sigma + d \gamma \left(
       \frac{p^{\ast}_0 \omega \tau_D}{k^{\ast}_0} \right)^{1 - \gamma}
     \end{array} \right) \]
  Using \eqref{eq:th0astp}, we have
  \[ (J_0^{\ast})_{2, 1} = \varepsilon_C - \varepsilon_D + \frac{(\tau_D -
     \tau_C)}{\tau_D}  (\varepsilon_D - \alpha p_0^{\ast}) \]
  and
  \[ (J_0^{\ast})_{2, 2} = 1 - \frac{\varepsilon_D}{p_0^{\ast}} \]
  Using the value of $k_0^*$, the definition of $\chi$, and the
  expressions of $(J_0^{\ast})_{2, 1}$ and $(J_0^{\ast})_{2, 1}$ as
  written above, $J_0^{\ast}$ can be simplified as follows:
  \[ \left( \begin{array}{ccc}
       \frac{1}{e^{\beta (\lambda_0 - p_0^{\ast} (\tau_D - \tau_C))}} & 0 &
       0\\
       \varepsilon_C - \varepsilon_D + \frac{(\tau_D - \tau_C)}{\tau_D}
       (\varepsilon_D - \alpha p_0^{\ast}) & 1 -
       \frac{\varepsilon_D}{p_0^{\ast}} & - c_1 p^{\ast}_0 \tau_D (1 -
       \omega)\\
       - d \chi^{\gamma} p^{\ast}_0 \omega (\tau_D - \tau_C) (1 - \gamma) & d
       \chi^{\gamma} \omega \tau_D (1 - \gamma) & \sigma + \gamma (1 - \sigma)
     \end{array} \right) \]
  One eigenvalue is $\frac{1}{e^{\beta (\lambda_0 - p_0^{\ast} (\tau_D -
  \tau_C))}}$, so since it is indeed greater than $- 1$ it requires
  \[ \frac{1}{e^{\beta (\lambda_0 - p_0^{\ast} (\tau_D - \tau_C))}} < 1 \]
  from which we find
  \begin{equation}
    p_0^{\ast} < \bar r. \label{eq:stabp01}
  \end{equation}
  The two remaining eigenvalues are those of
  \[ \tilde{J}_0^{\ast} = \left( \begin{array}{cc}
       1 - \frac{\varepsilon_D}{p_0^{\ast}} & - c_1 p^{\ast}_0 \tau_D (1 -
       \omega)\\
       d \chi^{\gamma} \omega \tau_D (1 - \gamma) & \sigma + \gamma (1 -
       \sigma)
     \end{array} \right) \]
  and they lie in the unit circle provided that
  \begin{equation}
    \left\{ \begin{array}{l}
      1 - \text{tr} (\tilde{J}_0^{\ast}) + \det (\tilde{J}_0^{\ast}) > 0\\
      1 + \text{tr} (\tilde{J}_0^{\ast}) + \det (\tilde{J}_0^{\ast}) > 0\\
      1 - \det (\tilde{J}_0^{\ast}) > 0
    \end{array} \right. \label{eq:sc2d}
  \end{equation}
  We have
  \[ \left\{ \begin{array}{l}
       \text{tr} (\tilde{J}_0^{\ast}) = 1 - \frac{\varepsilon_D}{p_0^{\ast}} +
       \sigma + \gamma (1 - \sigma)\\
       \det (\tilde{J}_0^{\ast}) = \frac{dc_1 \tau_D^2 \omega (1 - \gamma) (1
       - \omega) \chi^{\gamma} (p_0^{\ast})^2 + (\gamma + \sigma (1 - \gamma))
       p_0^{\ast} - \varepsilon_D (\gamma + \sigma (1 - \gamma))}{p_0^{\ast}}
     \end{array} \right. \]
  From \eqref{eq:impxi0} we find
  \[ (p_0^{\ast})^2 = \frac{\varepsilon_D - (\alpha + c_2 (1 - \omega) \tau_D)
     p_0^{\ast}}{c_1 \chi \omega (1 - \omega) \tau_D^2} \]
  which used in the expression of $\det (\tilde{J}_0^{\ast})$ provides
  \[ \det (\tilde{J}_0^{\ast}) = \frac{[\gamma + \sigma (1 - \gamma) - (1 -
     \gamma) (1 - \sigma) (\alpha + c_2 (1 - \omega) \tau_D)] p_0^{\ast} +
     \varepsilon_D  [(1 - \gamma) (1 - \sigma) - (\gamma + \sigma (1 -
     \gamma))]}{p_0^{\ast}} \]
  so we have
  \[ \left\{ \begin{array}{l}
       \text{tr} (\tilde{J}_0^{\ast}) = 1 - \frac{\varepsilon_D}{p_0^{\ast}} +
       \sigma + \gamma (1 - \sigma)\\
       \det (\tilde{J}_0^{\ast}) = \frac{\varepsilon_D [(1 - \gamma) (1 -
       \sigma) - (\gamma + \sigma (1 - \gamma))]}{p_0^{\ast}} + \gamma +
       \sigma (1 - \gamma) - (1 - \gamma) (1 - \sigma) (\alpha + c_2 (1 -
       \omega) \tau_D)
     \end{array} \right. \]
Conditions in \eqref{eq:sc2d} are then equivalent to inequalities 
  \begin{equation}
    \left\{ \begin{array}{l}
      p_0^{\ast} < \frac{2 \varepsilon_D}{\alpha + c_2 \tau_D (1 - \omega)}\\
      p_0^{\ast} > \frac{2 \varepsilon_D (\gamma + \sigma (1 - \gamma))}{2
      (\gamma + 1 + (1 - \gamma) \sigma) - (1 - \gamma) (1 - \sigma) (\alpha +
      c_2 \tau_D (1 - \omega))}\\
      p_0^{\ast} > \frac{\varepsilon_D (1 - 2 (\gamma + \sigma (1 -
      \gamma)))}{(1 - \gamma) (1 - \sigma) (1 + \alpha + c_2 \tau_D (1 -
      \omega))}
    \end{array} \right. \label{eq:stabp0234}
  \end{equation}
  in which the second one requires 
  $$2 (\gamma + 1 + (1 -
  \gamma) \sigma) - (1 - \gamma)  (1 - \sigma)  (\alpha + c_2 \tau_D (1 -
  \omega)) > 0$$ to hold, providing condition \eqref{eq:stabp0w}.
  
  Now we make more explicit solutions to \eqref{eq:stabp01} and
  \eqref{eq:stabp0234}, solving for $c_1$ and clarifying the behavior as
  $\omega$ increases. We start noting that all these inequalities are of the
  form $p_0^{\ast} > z$ or $p_0^{\ast} < z$, for a generic $z$. From
  \eqref{eq:p0}, solving $p_0^{\ast} > z$, we firstly obtain
  \[ \sqrt{\Delta_0} > 2 c_1 \chi \omega (1 - \omega) \tau_D^2 z + (\alpha +
     c_2 (1 - \omega) \tau_D) . \]
  Squaring both sides, using \eqref{eq:Delta0} and rearranging the resulting
  expression we find
  \[ 4 c_1 \chi \omega (1 - \omega)  (\varepsilon_D - c_1 \chi \tau_D^4 \omega
     (1 - \omega) z^2 + \tau_D^2 (\alpha + c_2 (1 - \omega) \tau_D) z) > 0 \]
  namely
  \begin{equation}
    \varepsilon_D - c_1 \chi \tau_D^4 \omega (1 - \omega) z^2 - \tau_D^2 
    (\alpha + c_2 (1 - \omega) \tau_D) z > 0 \label{eq:cond_z}
  \end{equation}
  We start discussing the possible solutions to \eqref{eq:cond_z} with respect
  to $\omega$, depending on the 4 possible expressions of $z$, corresponding
  to the right hand sides of \eqref{eq:stabp01} and \eqref{eq:stabp0234}. If
  $z = \bar r$, we have that the left hand side of
  \eqref{eq:cond_z} represents the graph of a convex parabola. Since
  \eqref{eq:cond_z} corresponds to $p_0^{\ast} > z$, while \eqref{eq:stabp01}
  has the form $p_0^{\ast} < z$, we consider \eqref{eq:cond_z} with inequality
  $<$ and hence its solutions are of the form $(\omega_a, \omega_b)$.
  
  Let us now focus on expressions for $z$ corresponding to the right hand
  sides of \eqref{eq:stabp0234}, which all can be written in the form 
  \begin{equation}
    z = \frac{B_1}{B_2 + B_3  (\alpha + c_2\tau_D (1 - \omega))} \label{eq:z}
  \end{equation}
  where $B_1, B_2, B_3$ are suitable constants, with $B_2 \geq 0$. In
  the first and second right hand sides in \eqref{eq:stabp0234} we
  have $B_1 > 0$, while the third condition is always fulfilled if
  $B_1 \leq 0$, so in what follows can assume $B_1 > 0$ for each right
  hand side in \eqref{eq:stabp0234}. For the first and third right
  hand sides in \eqref{eq:stabp0234} we have $B_3 > 0$, while for the
  second right hand side in \eqref{eq:stabp0234} we have $B_3 < 0$. In
  any case, recalling condition \eqref{eq:stabp0w}, we have
  $B_2 + B_3 (\alpha + c_2\tau_D (1 - \omega)) > 0$ and condition
  \eqref{eq:cond_z} can be rewritten as
  \begin{equation}
    \frac{\varepsilon_D}{\tau_D^2 z} - c_1 \chi \tau_D^2 \omega (1 - \omega)
    z - (\alpha + c_2\tau_D (1 - \omega)) > 0 \label{eq:cond_z2}
  \end{equation}
  in which, from \eqref{eq:z}, $\frac{\varepsilon_D}{\tau_D^2 z} - (\alpha + c_2\tau_D (1 - \omega))$ is a first
  degree polynomial with respect to $\omega$, so we can study the
  convexity/concavity of $S (\omega) = - c_1 \chi \tau_D^2 \omega (1 - \omega)z $, with $z$, depending on $\omega$, \
  given by \eqref{eq:z}. We have
  \[ S'' (\omega) = \frac{2 B_1 c_1 \chi \tau_D^2  (B_2 + \alpha B_3)  (B_2 + B_3
     (\alpha + c_2\tau_D))}{(B_2 + B_3 (\alpha + c_2\tau_D (1 - \omega)))^3} \]
  For the first and third condition in \eqref{eq:stabp0234} we have $S''
  (\omega) > 0$, so, since the first condition in \eqref{eq:stabp0234} comes
  from inequality $p_0^{\ast} < z$, its solution is of the form $(\omega_a,
  \omega_b)$, while since the last condition in \eqref{eq:stabp0234} comes
  from inequality $p_0^{\ast} > z$, its solution is of the form $(- \infty,
  \omega_a) \cup (\omega_b, + \infty)$.
  
  Now let us consider the second condition in \eqref{eq:stabp0234}. We
  recall that this condition requires $z > 0$ as otherwise it is not
  fulfilled. If $B_2 + B_3 (\alpha + c_2\tau_D (1 - \omega)) > 0$ is
  fulfilled for any $\omega \in [0, 1]$, we then have that it is
  fulfilled for both $\omega = 1$, for which we have
  $B_2 + \alpha B_3 > 0$, and for $\omega = 0$, for which we have
  $B_2 + B_3 (\alpha + c_2\tau_D) > 0$, we can conclude that
  $S'' (\omega) > 0$ and, since the second condition in
  \eqref{eq:stabp0234} comes from inequality $p_0^{\ast} > z$, its
  solution is of the form
  $(- \infty, \omega_a) \cup (\omega_b, + \infty)$. Conversely, if
  $ B_2 + B_3 (\alpha + c_2\tau_D (1 - \omega)) > 0$ is fulfilled only
  for $\omega \in (\omega_0, 1]$ with $\omega_0 \in [0, 1)$, we have
  that for $\omega \in (\omega_0, 1]$, recalling that $B_3 < 0$ and
  the other coefficients are positive, we can write
  \[ B_2 + \alpha B_3 > B_2 + B_3  (\alpha + c_2\tau_D ) > B_2 + B_3  (\alpha + c_2\tau_D (1 - \omega)) > B_2 + B_3  (\alpha + c_2\tau_D (1 - \omega_o)) = 0 \]
  and so $S'' (\omega) > 0$ and, since the second condition in
  \eqref{eq:stabp0234} comes from inequality $p_0^{\ast} > z$, its solution is
  of the form $[(- \infty, \omega_a) \cup (\omega_b, + \infty)] \cap
  (\omega_0, 1]$.
  
This concludes the analysis of the stability intervals with respect to $\omega$.
  
  We now focus on the stability conditions with respect to parameter $c_1$. To this end, solving equation \eqref{eq:cond_z} withe respect to $c_1$ provides
\begin{equation}
  p_0^{\ast} > z \Longleftrightarrow c_1 < c_{1, C}  (z, \omega)
  \label{eq:c1p0}
\end{equation}
where $c_{1, C}  (z, \omega)$ is defined in \eqref{eq:c1i}. Therefore, 
considering $z = \bar{r}, z = z_{2, C}, z = z_{3, C}$ and $z = z_{4,
  C}$ 
we find that conditions
\[ \left\{ \begin{array}{l}
     p_0^{\ast} < z_1\\
     p_0^{\ast} < z_2\\
     p_0^{\ast} > z_3\\
     p_0^{\ast} > z_4
   \end{array} \right. \]
are equivalent to the stability conditions \eqref{eq:z24}.
\end{proof}

\begin{proof}[Prop. \ref{th:stabxi1}]
  We recall that at $\tmmathbf{\xi}_1^{\ast}$ we have $x^{\ast}_1 = 1,
  k^{\ast}_1 = \chi \omega \tau_c p^{\ast}_1$, with $p^{\ast}_1$ solution to
  \eqref{eq:impxi1}, i.e.
  \[ h_1 (p^{\ast}_1, \omega) = c_1 \chi \omega (1 - \omega) \tau_c^2
     (p^{\ast}_1)^2 + (\alpha + c_2 (1 - \omega) \tau_c) p^{\ast}_1 -
     \varepsilon_c = 0 \]
  Using the expression of $k_1^{\ast}$ in the previous equation we can write
  \[ h_1 (p^{\ast}_1, \omega) = c_1  (1 - \omega) \tau_c p^{\ast}_1 k_1^{\ast}
     + (\alpha + c_2 (1 - \omega) \tau_c) p^{\ast}_1 - \varepsilon_c = 0 \]
  from which we can obtain
  \begin{equation}
    c_1 k_1^{\ast} + c_2 = \frac{1}{(1 - \omega) \tau_c}  \left(
    \frac{\varepsilon_c}{p_1^{\ast}} - \alpha \right) \label{eq:th1astp}
  \end{equation}
  We have
  \[ J_1^{\ast} = \left( \begin{array}{ccc}
       e^{\beta (\lambda_0 - p_1^{\ast} (\tau_d - \tau_c))} & 0 & 0\\
       \varepsilon_c - \varepsilon_d + p_1^{\ast} (\tau_d - \tau_c) (1 -
       \omega) (c_2 + c_1 k^{\ast}_1) & 1 - \alpha - \tau_c (1 - \omega) (c_2
       + c_1 k^{\ast}_1) & - c_1 p^{\ast}_1 \tau_c (1 - \omega)\\
       - \frac{d (k^{\ast}_1)^{\gamma} (p^{\ast}_1)^{1 - \gamma} \omega^{1 -
       \gamma} (\tau_d - \tau_c) (1 - \gamma)}{\tau_c^{\gamma}} & \frac{d
       (k^{\ast}_1)^{\gamma} \omega^{1 - \gamma} \tau_c^{1 - \gamma}
                                                                  (1 -
                                                                  
       \gamma)}{(p^{\ast}_1)^{\gamma}} & \sigma + d \gamma \left(
       \frac{p^{\ast}_1 \omega \tau_c}{k^{\ast}_1} \right)^{1 - \gamma}
     \end{array} \right) \]
  Using \eqref{eq:th1astp}, we have
  \[ (J_1^{\ast})_{2, 1} = \varepsilon_c - \varepsilon_d + \frac{(\tau_d -
     \tau_c)}{\tau_c}  (\varepsilon_c - \alpha p_1^{\ast}) \]
  and
  \[ (J_1^{\ast})_{2, 2} = 1 - \frac{\varepsilon_c}{p_1^{\ast}} \]
  Using the value of $k_1^{\ast}$, the definition of $\chi$, and the
  expressions of $(J_1^{\ast})_{2, 1}$ and $(J_1^{\ast})_{2, 1}$ as written
  above, $J_1^{\ast}$ can be simplified as follows:
  \[ \left( \begin{array}{ccc}
       e^{\beta (\lambda_0 - p_1^{\ast} (\tau_d - \tau_c))} & 0 & 0\\
       \varepsilon_c - \varepsilon_d + \frac{(\tau_d - \tau_c)}{\tau_c}
       (\varepsilon_c - \alpha p_1^{\ast}) & 1 -
       \frac{\varepsilon_c}{p_1^{\ast}} & - c_1 p^{\ast}_1 \tau_c (1 -
       \omega)\\
       - d \chi^{\gamma} p^{\ast}_1 \omega (\tau_d - \tau_c) (1 - \gamma) & d
       \chi^{\gamma} \omega \tau_c (1 - \gamma) & \sigma + \gamma (1 - \sigma)
     \end{array} \right) \]
  One eigenvalue is $e^{\beta (\lambda_0 - p_1^{\ast} (\tau_d - \tau_c))}$, so
  since it is indeed greater than $- 1$ it requires
  \[ e^{\beta (\lambda_0 - p_1^{\ast} (\tau_d - \tau_c))} < 1 \]
  from which we find
  \[ p_1^{\ast} > \bar{r} . \]
  The two remaining eigenvalues are those of
  \[ \tilde{J}_1^{\ast} = \left( \begin{array}{cc}
       1 - \frac{\varepsilon_c}{p_1^{\ast}} & - c_1 p^{\ast}_1 \tau_c (1 -
       \omega)\\
       d \chi^{\gamma} \omega \tau_c (1 - \gamma) & \sigma + \gamma (1 -
       \sigma)
     \end{array} \right) \]
  and they lie in the unit circle provided that \eqref{eq:sc2d} holds true at
  $\tilde{J}_1^{\ast}$. We have
  \[ \left\{ \begin{array}{l}
       \text{tr} (\tilde{J}_1^{\ast}) = 1 - \frac{\varepsilon_c}{p_1^{\ast}} +
       \sigma + \gamma (1 - \sigma)\\
       \det (\tilde{J}_1^{\ast}) = \frac{dc_1 \tau_c^2 \omega (1 - \gamma) (1
       - \omega) \chi^{\gamma} (p_1^{\ast})^2 + (\gamma + \sigma (1 - \gamma))
       p_1^{\ast} - \varepsilon_c (\gamma + \sigma (1 - \gamma))}{p_1^{\ast}}
     \end{array} \right. \]
  From \eqref{eq:impxi1} we find
  \[ (p_1^{\ast})^2 = \frac{\varepsilon_c - (\alpha + c_2 (1 - \omega) \tau_c)
     p_1^{\ast}}{c_1 \chi \omega (1 - \omega) \tau_c^2} \]
  which used in the expression of $\det (\tilde{J}_1^{\ast})$ provides
  \[ \det (\tilde{J}_1^{\ast}) = \frac{[\gamma + \sigma (1 - \gamma) - (1 -
     \gamma) (1 - \sigma) (\alpha + c_2 (1 - \omega) \tau_c)] p_1^{\ast} +
     \varepsilon_c  [(1 - \gamma) (1 - \sigma) - (\gamma + \sigma (1 -
     \gamma))]}{p_1^{\ast}} \]
  so we have
  \[ \left\{ \begin{array}{l}
       \text{tr} (\tilde{J}_1^{\ast}) = 1 - \frac{\varepsilon_c}{p_1^{\ast}} +
       \sigma + \gamma (1 - \sigma)\\
       \det (\tilde{J}_1^{\ast}) = \frac{\varepsilon_c [(1 - \gamma) (1 -
       \sigma) - (\gamma + \sigma (1 - \gamma))]}{p_1^{\ast}} + \gamma +
       \sigma (1 - \gamma) - (1 - \gamma) (1 - \sigma) (\alpha + c_2 (1 -
       \omega) \tau_c)
     \end{array} \right. \]
  The expressions of tr$(\tilde{J}_1^{\ast})$ and $\det (\tilde{J}_1^{\ast})$
  are analogous to those of tr$(\tilde{J}_0^{\ast})$ and $\det
  (\tilde{J}_0^{\ast})$, with $\tau_C$, $\varepsilon_C$ and $p_1^{\ast}$ in
  place of $\tau_D$, $\varepsilon_D$ and $p_0^{\ast}$, respectively.
  Accordingly, the stability conditions with respect to $c_1$ are analogous to
  those in Proposition \ref{th:stabxi0}, apart from that corresponding to
  $p_1^{\ast} > \bar{r}$, in which we have a change in the sign of the
  inequality.
\end{proof}

\end{document}